\pdfoutput=1
\documentclass[11pt, table]{article}

\usepackage{amsmath,amsfonts,bm}

\def\eqref#1{equation~\ref{#1}}

\def\1{\bm{1}}

\DeclareMathAlphabet{\mathsfit}{\encodingdefault}{\sfdefault}{m}{sl}
\SetMathAlphabet{\mathsfit}{bold}{\encodingdefault}{\sfdefault}{bx}{n}

\usepackage{hyperref}

\usepackage{acl} 

\usepackage{times}
\usepackage{latexsym}

\usepackage[T1]{fontenc}

\usepackage[utf8]{inputenc}

\usepackage{microtype}

\usepackage{inconsolata}

\usepackage{graphicx}
\usepackage[]{xcolor} 
\usepackage[]{tcolorbox} 

\usepackage{hyperref}
\usepackage{url}
\usepackage{subfigure}
\usepackage{multirow}
\usepackage{bbm}

\usepackage{enumitem}

\title{RoundTable: Investigating Group Decision-Making Mechanism in Multi-Agent Collaboration}

\author{Young-Min Cho\textsuperscript{1,2}\thanks{Work completed during employment at Amazon.}, Raphael Shu\textsuperscript{2}, Nilaksh Das\textsuperscript{2}, Tamer Alkhouli\textsuperscript{2}\\
\textbf{Yi-An Lai\textsuperscript{2}, Jason Cai\textsuperscript{2}, Monica Sunkara\textsuperscript{2}, Yi Zhang\textsuperscript{2}, Dan Roth\textsuperscript{1*}} \\
\textsuperscript{1}University of Pennsylvania, \textsuperscript{2}AWS GenAI \\
\texttt{jch0@seas.upenn.edu}\\
\texttt{\{zhongzhu, yizhngn\}@amazon.com}
}

\begin{document}
\maketitle
\begin{abstract}
Effective group decision-making is critical in Multi-Agent Systems (MAS). Yet, how different mechanisms for reaching consensus impact collaboration quality and efficiency remains understudied. We conduct a systematic study on group decision-making mechanisms in a decentralized setting. Through controlled experiments, we analyze how different voting rules affect decision quality and efficiency in a multi-round collaboration. Results reveal that majority voting often cause inefficient collaboration due to its strict acceptance criteria. At the extreme, unanimous voting gives 87\% lower initial performance than the best-performing method. Our qualitative analysis of cross-agent communication shows that messages become longer and more repetitive over time: while message length increases by 84\%, similarity to the previous round increases to 90\%. Based on these insights, language-based early stopping methods make the performance 13\% closer to oracle while reducing rounds by 50\%. Our findings highlight the crucial role of group decision-making in optimizing MAS collaboration.
\end{abstract}

\section{Introduction}

Collaboration is a fundamental aspect of human society. It enables individuals to address complex challenges that would be unmanageable otherwise \citep{de2005collaboration, graesser2018advancing, bittner2013shared}. Inspired by this principle, Large Language Model (LLM)-based Multi-Agent Systems (MAS) have emerged as a paradigm where autonomous LLM agents interact, reason, and coordinate to contribute diverse perspectives, fostering richer problem-solving through collective intelligence. This approach is valuable in complex domains such as software development \cite{hong2023metagpt}; financial trading, \cite{li2023tradinggpt}; and healthcare \cite{tang2023medagents}. These systems demonstrate the importance of MAS, which coordinates interactions among agents to enhance task performance and solution quality.

\begin{figure}[t]
    \centering
    \includegraphics[width=\linewidth]{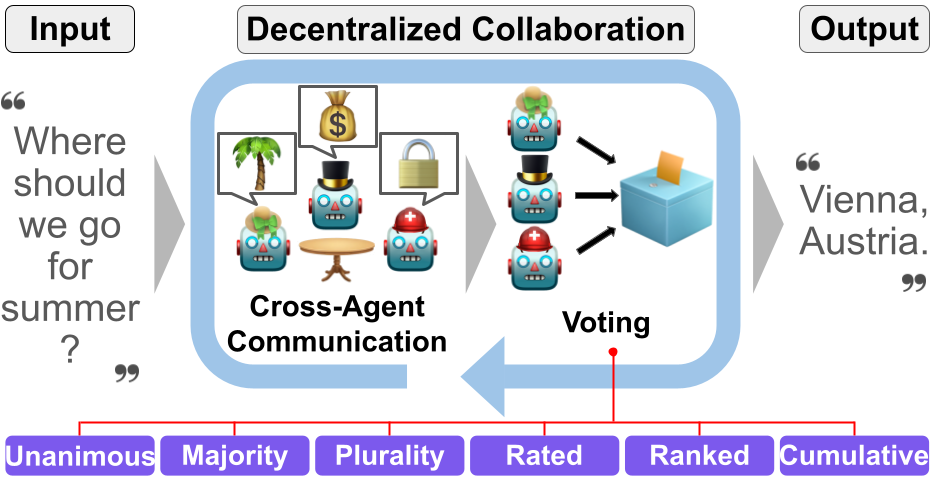}
    \caption{An example of decentralized multi-agent collaboration, in which agents reach consensus through multi-round communication and decision-making. }
    \label{fig:spirit}
\end{figure}

One fundamental component in MAS is decision-making mechanism, which defines how decisions are made in the system. It enables agents to coordinate their actions and achieve collective goals efficiently. Decision-making can be categorized into two levels: 1) \textit{individual decision-making}, where each agent independently evaluates its environment, sets objectives, and selects actions to optimize its own outcomes; and 2) \textit{group (collective) decision-making}, where a framework aggregates the opinions of all agents to produce a collective outcome, with implicit early stopping when consensus stabilizes. Since agents likely to have varying perspectives by the nature of MAS, aggregating different opinions to make a group level decision becomes especially important.

Although individual decision-making is mostly powered by LLM, group decision-making varies across systems. While centralized MAS resolving conflicts between agents through a leader or hierarchical decision-making pipeline \citep{hong2023metagpt, qian2023communicative, dong2023self}, decentralized MAS distributes power equally among agents \citep{mandi2024roco, zhang2023building, xu2023exploring, park2023generative}. Despite the recent success of decentralized MAS, such as Multi-Agent Debate \cite{du2023improving, chan2023chateval, liang2023encouraging}, existing studies predominantly rely on majority voting or average rating without thoroughly evaluating alternative mechanisms.

Indeed, the aggregation of diverse opinions has been extensively studied in economics and political science. As a theoretical framework for decision-making, social choice theory examines how collective decisions are made by aggregating individual preferences into a unified outcome \cite{arrow2012social, riker1988liberalism, endriss2008computational}. Within social choice theory, voting rules are the most representative and widely used functions that maps individual votes to a collective outcome. Examples of voting rules include unanimous, majority, plurality, rated, ranked, and cumulative voting.

While traditional studies in economics and computational social choice are extensive, investigating social choice in LLM-based multi-agent systems remains non-trivial. This is because the mere acknowledgment of using a specific voting rule can influence agents’ actions and strategies. In decentralized MAS, where agents are adaptive and capable of strategic behavior, the voting rule is not just an aggregation tool but also a factor that shapes the dynamics of interaction. As shown in Figure \ref{fig:spirit}, this feedback loop between the voting rule and agent behavior makes it essential to investigate how different voting rules affect not only the final outcomes but also the process of reaching consensus. Furthermore, the decision-making mechanism requires a stopping signal during collaboration to reduce communication costs and enhance collaboration quality. However, social choice theory does not account for communication dynamics or provide criteria for early stopping. Therefore, studying the decision-making mechanisms in MAS is critical for designing systems that ensure robust, fair, and efficient collective intelligence.

This study investigates group decision-making mechanisms in decentralized MAS. We develop an experimental platform, \textit{RoundTable}, which enables multi-round, multi-agent decentralized collaboration and allows controlled experiments to compare the effects of different decision-making mechanisms. By comparing various voting rules, we see that too strict voting threshold and format yield worse performance. Unanimous voting, which has the strictest threshold and format, shows 87\% lower initial round performance and 39\% lower final round performance. We further analyze cross-agent communications to characterize MAS interactions, finding that messages grow longer and more repetitive over time. Message length increases by 84\%, while similarity to the previous round reaches 90\%, indicating inefficiencies in prolonged collaboration. Based on these insights, we find language-based early stopping improves performance by 13\% while halving the number of rounds, enabling timely, efficient collaboration.

This paper provides the following research contributions:

\begin{itemize}
    \item We show that majority voting often fails in multi-agent systems because its strict acceptance criteria hinder decision-making.
    \item We identify that extended collaboration often leads to inefficiencies, characterized by longer and more repetitive messages, which result in redundant communication and reduced effectiveness.
    \item Our experiments show that  can be leveraged for early stopping, enhancing performance and reducing the number of rounds.
\end{itemize}


\section{Related Work}
\paragraph{Centralized vs Decentralized MAS}
Centralized MAS have been extensively studied, with systems like MetaGPT \citep{hong2023metagpt} and AutoGen \citep{wu2023autogen} using manager agents or hierarchical structures for decision-making. While effective for coordination, these systems face issues such as single points of failure and decision biases due to concentrated authority \citep{owens2024multi, jiang2019learning}.

Decentralized MAS, as seen in RoCo \citep{mandi2024roco} and Generative Agents \citep{park2023generative}, distribute decision power among agents, enhancing flexibility, robustness, and adaptability. Despite these strengths, current research often overlooks challenges like consensus-building and efficient cross-agent communication, especially in complex, dynamic environments \citep{xu2023exploring, zhang2023building}.

\paragraph{Social Choice Theory}
Social choice theory, rooted in economics, examines how individual preferences can be aggregated for collective decision-making, with foundational work like Arrow’s Impossibility Theorem highlighting the challenges of fairness \citep{arrow2012social}. It has since expanded into computational social choice for algorithmic decision-making and automated negotiation to resolve conflicts in multi-agent settings \citep{endriss2008computational, kraus2001strategic}. In LLM-based MAS, voting rules help consolidate diverse agent opinions, but most applications rely heavily on majority voting due to its simplicity \citep{li2022competition, chan2023chateval}. This over-reliance limits adaptability, overlooking more flexible voting rules like ranked or rated voting that could improve collaboration and decision efficiency in dynamic MAS environments.
\paragraph{Cross-Agent Communication}
Cross-agent communication is essential in decentralized MAS, enabling information sharing, negotiation, and consensus-building without central control \citep{jennings2001automated, park2023generative, han2024llm}. However, structured protocols like turn-taking and fixed roles can create power imbalances, undermining decentralization. While prior work optimizes message passing and protocols, how to use language patterns to understand collaboration remains underexplored. This study examines language cues as indicators of collaboration quality and introduces language-based early stopping to enhance decision-making efficiency.

\section{Decentralized MAS}
A decentralized MAS is a collection of autonomous agents collaborating towards a common goal without a central controller. Each agent in the system operates based on its individual context, defined by available information, historical interactions, and personalized objectives. The collaboration unfolds over multiple rounds, where agents communicate, propose solutions, and vote to reach a collective decision. This system is characterized by 1) The collective decision emerges from individual proposals and votes rather than being imposed by a single entity; 2) Agents update their strategies based on past interactions, ensuring adaptability to changing circumstances.

\begin{figure*}[t]
    \centering
    \includegraphics[width=\textwidth]{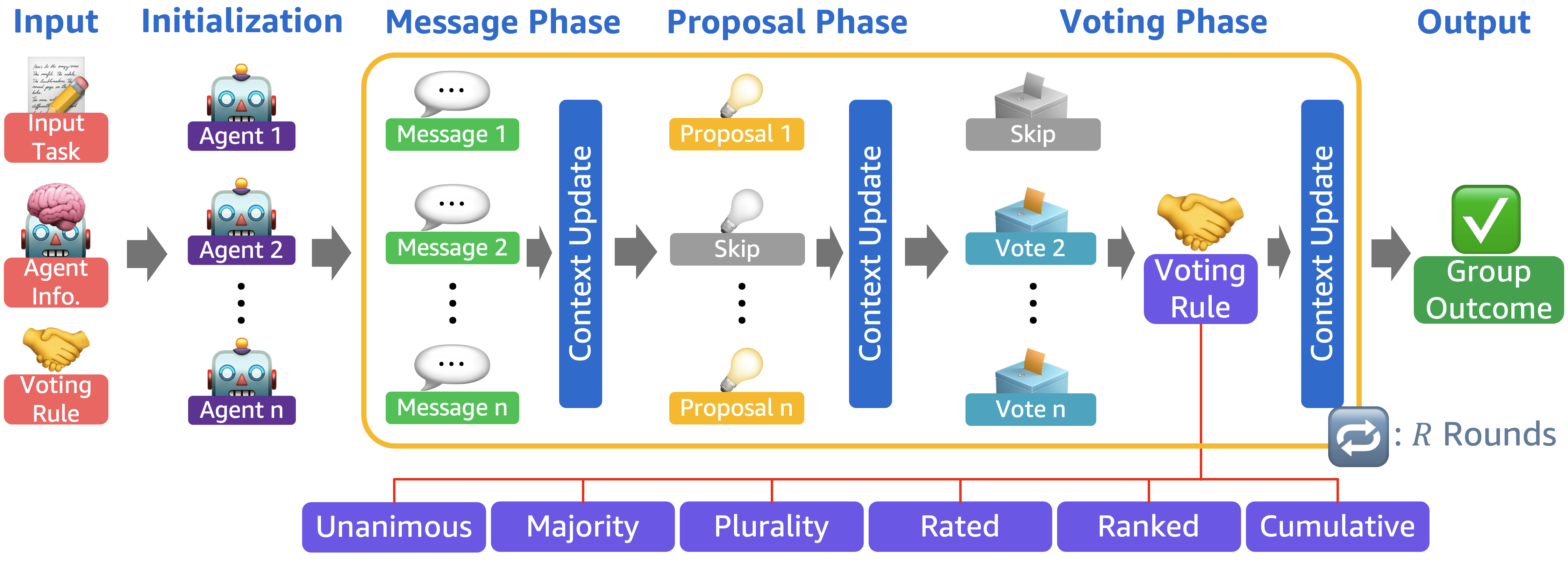}
    \caption{Overview of our multi-agent collaboration platform: RoundTable. It allows a multi-round collaboration where agents simultaneously send messages, propose solutions, and vote. Based on a voting rule, RoundTable selects the most preferred proposal for group decisions. Detailed introduction of RoundTable is in Appendix \ref{app:roundtable_platform}.}
    \label{fig:roundtable}
\end{figure*}

\subsection{Problem Definition}
We consider an environment where all possible states of the world are defined as $\mathcal{X}$ and a task $\Sigma$ is set as the global objective. Here, a decentralized MAS system is defined as a collaborative decision-making process over $R$ rounds, involving a set of agents $i \in \mathcal{I}$. 

\paragraph{Agent Definition} At each round $r$, an agent $i$ takes actions with LLM and its latest individual context, denoted as $\Omega_{i} = \{B_i, H\}$. 

\noindent• Background $B_i$ includes: \textit{Task $\Sigma$}, \textit{Collaboration Rules}, and \textit{Self Information} (persona, data access, and objective). 

\noindent• Latest History $H$ includes: \textit{Messages $M$}, \textit{Proposals $P$}, \textit{Votes $V$}, and \textit{Group Decision $x^*_r \in \mathcal{X}$}: The latest decision reached by the group at round $r$.

\paragraph{Action Definition}
The collaboration at each round $r$ is structured into three ordered phases, which defines action space for agents:\\\\
\noindent \textit{• Message Phase}: Each agent formulates a message based on its context: $m_{i,r} = f_m(\Omega_{i})$

\noindent \textit{• Proposal Phase}: Each agent generates a proposal for a potential group decision: $p_{i,r} = f_p(\mathcal{X}, \Omega_{i}), \quad p_{i,r} \in \mathcal{X}$

\noindent \textit{• Voting Phase}: Each agent evaluates and votes on proposals: $v_{i,p,r} = f_v(p, \Omega_i)$, where the format of vote is defined by the input voting rule $F$.

\paragraph{Voting Rules}
At the end of each round, a voting rule  \( F \) determines the collective decision for the round:  
\[
\hat{x^*}_r = F(\{(c, \{v_{i,c,r} \mid i \in \mathcal{I} \}) \mid c \in C \})
\]
where the candidate set \( C \) consists of all proposals and the previous group decision: $C = P \cup \{\hat{x^*}_{r-1}\}$.

To investigate the impact of different voting rules in multi-agent collaboration, we compare the following six voting rules: Unanimous, majority and plurality voting, where agents only choose the best candidate from the list (\textit{discrete voting rules}); rated, ranked and cumulative voting, which ask a nuanced, gradient preference over candidates (\textit{preference voting rules}).\footnote{Mathematical definitions and prompt texts are presented in Appendix \ref{app:social_choice}.}

    \noindent\textit{• Unanimous Voting} \citep{arrow2012social}: The proposal that receives votes from all agents will be selected.
    
    \noindent\textit{• Majority Voting} \citep{arrow2012social}: The proposal that receives votes from more than half of all agents will be selected.
    
    \noindent\textit{• Plurality Voting} \citep{arrow2012social}: The proposal that receives the most votes will be selected.
    
    \noindent\textit{• Rated Voting} \citep{baujard2018voters}: Each agent rates proposals on a 5-point scale, selecting the one with the highest total score.
    
    \noindent\textit{• Ranked Voting} \citep{arrow2012social}: Agents rank proposals by preference, with points assigned as 1, 1/2, 1/3, etc., and the highest-scoring proposal is selected.
    
    \noindent\textit{• Cumulative Voting} \citep{black1958theory}: Each agent distributes $|\mathcal{I}|$ points among $|\mathcal{I}|$ proposals, selecting the one with the highest total.\newline

The goal of the decentralized MAS is to find the solution $x^* \in \mathcal{X}$ with inputs: the task $\Sigma$, all agent information $\{\phi_i|\forall i\in \mathcal{I}\}$, and a voting rule $F$. Based on the definitions above, we propose \textbf{RoundTable}, an experiment platform for decentralized multi-agent collaboration that can compare various voting rules through controlled experiments. The overview of RoundTable is shown in Figure \ref{fig:roundtable}.\footnote{details of the platform design and the LLM prompts are shown in Appendix \ref{app:roundtable_platform} and \ref{app:roundtable_prompt}.}

\section{Experiments}
Using RoundTable, we analyze multi-agent behavior patterns over $R=10$ rounds per collaboration. Since disagreement among agents creates tension between cooperation and competition, we evaluate RoundTable in two environments: competitive-focused and cooperative-focused. In both, we report performance as the average of 100 independent collaborations. 

\subsection{Competitive Environment - Exchange Economy}
\paragraph{Introduction} 
We use an exchange economy for the competitive environment as the advantages make it ideal for evaluating a decentralized MAS \citep{varian1992microeconomic}. First, it is a plus-sum game where agents seek equilibrium, meaning collaboration opportunities exist if allocation is not yet optimal. Second, multiple equilibria in the market allow dynamic collaboration. Third, while equilibrium doesn't guarantee maximum utility, it must lie within one of the equilibria, helping to assess whether conflicts between agents hinder the group's progress toward the ultimate goal.

Here, $K$ types of goods and the same number of agents participate in a market, with each good having a quantity of 100. Each agent has a Cobb-Douglas utility function $u_i = \prod a_k^{\theta_k}, \sum \theta_k = 1$, where $a_k$ represents the amount of good $k$ for agent $i$, and $\theta_k$ reflects the relative preference for each good \citep{cobb1928theory}. Agents' objective is set to maximize their individual utility by finding an optimal allocation of goods. From a group perspective, the hidden goal is to maximize the total utility $U=\sum u_i$, which is not revealed to agents.\footnote{For more on the exchange economy environment, see Appendix \ref{app:exchange_economy}.} 

To reflect common MAS collaboration patterns, we use an asymmetric utility function setup with $K=3$ types of goods and agents, where each agent prefers different goods, mimicking real-world scenarios where agents specialize in different areas.\footnote{Other types of utility function sets are shared in Appendix \ref{app:other_utility_sets}.}  Each agent's utility function is $u_i = a_i^{0.8}\prod_{k \neq i} a_k^{\tilde{\theta}}$, where $\tilde{\theta} = \frac{0.2}{K-1}$.

\paragraph{Metrics}
We use various metrics to analyze multi-agent collaboration. For quality, we report the \textit{group total utility} $U=\frac{\sum u_i}{U_{max}}$, where $U_{max}$ is the largest possible utility achievable in the environment. \textit{Efficiency} is measured by the area under the curve (AUC), $AUC@n = \sum_{r = 1}^{n} \frac{U_r}{U_{max}}$. \textit{Fairness} is assessed by $\frac{u_{min}}{u_{max}}$, which compares the smallest individual utility to the largest at the last round. \textit{Rationality} is the ratio of rounds where the individual utility of the proposed allocation exceeds the current one.\footnote{This differs from compromise, where proposing lower utility than the previous proposal is natural, but a lower utility than the current allocation is irrational.} Finally, rigidity measures how often an old allocation is kept, $rigidity = \frac{1}{10}\sum_{r=1}^{10}\mathbbm{1}(U_r = U_{r-1})$.

\subsection{Cooperative Environment - Recommendation System}
\paragraph{Introduction}
For the cooperative environment, we focus on the recommendation system. It reveals unique collaborative behaviors absent in exchange economies. First, strong information asymmetry exists, as no partial information can accurately predict. Second, group decision-making fosters diverse problem-solving approaches, with agents both consuming and contributing information in a feedback loop that enhances recommendations.

The task is to predict user ratings for specific items by analyzing historical interactions and detecting trends and correlations in the data. This is challenging for a single-agent system due to the overwhelming amount of data, but essential information is scattered and limited. In our setting, we divide the background information into three parts and assign each to different agents. The agents must then collaborate by exploring, analyzing, exchanging, and synthesizing their information to reach a comprehensive conclusion.

\paragraph{Dataset}
We evaluate MAS performance using the MovieLens-100k dataset \citep{harper2015movielens}. The dataset consists of information of all users and movies, and the rating history of users on movies. Due to resource constraints, we randomly selected 100 examples from the u1.test split for evaluation. To simplify the data structure, we pre-processed it into three tables: \textit{basicInfo}, \textit{userHistory}, and \textit{movieHistory}. Each table was assigned to a different agent, creating a 3-agent collaboration system.\footnote{The definition and example of each table can be found in Appendix \ref{app:recommendation_system}.}

\paragraph{Metrics}
We use Mean Absolute Error (MAE) and Root Mean Squared Error (RMSE) to represent utility $U_r$, assessing the accuracy of predictions compared to the gold rating at round $r$. To evaluate MAS performance in the recommendation task, we compare it against two baselines: a simple \textit{Always Guess 4}, which predicts the median regardless of input, and a strong State of the Art (\textit{SoTA}) model from \citet{behera2023collaborative}, which employs collaborative filtering with temporal features.

\section{Results}
\begin{figure}[t]
    \centering
    \includegraphics[width=\linewidth]{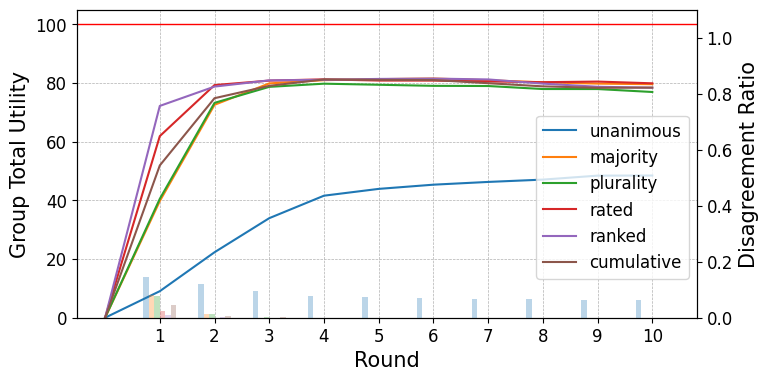}
    \caption{Voting rules comparison in exchange economy, using 3 agent, asymmetric, \textit{gpt-4o-mini} setting. Line plots (left y-axis) show the group total utility achieved over rounds; bar plots (right y-axis) represent the ratio of cases where participants yet to reach an agreement until a certain round. The red horizontal line indicates the maximum achievable group total utility.}
    \label{fig:economics}
\end{figure}

\begin{table*}[t]
\vspace{0.3cm}
\resizebox{\textwidth}{!}{%
\begin{tabular}{llllllllll}
\textbf{}  & \multicolumn{1}{c}{\textbf{\begin{tabular}[c]{@{}c@{}}GROUP\\TOTAL\\UTILITY@3\end{tabular}}} & \multicolumn{1}{c}{\textbf{\begin{tabular}[c]{@{}c@{}}GROUP\\TOTAL\\UTILITY@5\end{tabular}}} & \multicolumn{1}{c}{\textbf{\begin{tabular}[c]{@{}c@{}}GROUP\\TOTAL\\UTILITY@10\end{tabular}}} & \multicolumn{1}{c}{\textbf{AUC@3}} & \multicolumn{1}{c}{\textbf{AUC@5}} & \multicolumn{1}{c}{\textbf{AUC@10}} & \multicolumn{1}{c}{\textbf{RATIONALITY}} & \multicolumn{1}{c}{\textbf{FAIRNESS}} & \multicolumn{1}{c}{\textbf{RIGIDITY}} \\ \hline
\multicolumn{10}{c}{\textbf{Model: gpt-4o-mini, \# Agent: $\mathbf{K=3}$}}\\
Unanimous & \textbf{\textcolor{blue}{33.94 (3.96)}} & \textbf{\textcolor{blue}{43.92 (3.96)}} & \textbf{\textcolor{blue}{48.48 (3.86)}} & \textbf{\textcolor{blue}{21.77 (2.86)}} & \textbf{\textcolor{blue}{30.17 (3.05)}} & \textbf{\textcolor{blue}{38.65 (3.32)}} & \textbf{\textcolor{red}{35.00 (1.73)}} & \textbf{\textcolor{red}{87.96 (1.85)}} & \textbf{\textcolor{red}{93.00 (0.61)}} \\
Majority & 79.88 (0.80) & 81.33 (0.72) & 79.61 (0.88) & 64.08 (1.81) & 70.91 (1.23) & 75.67 (0.84) & 23.80 (0.94) & 64.00 (3.07) & 71.30 (1.32) \\
Plurality & 78.70 (1.32) & 79.42 (1.12) & 76.95 (1.24) & 64.17 (1.85) & 70.34 (1.37) & 74.26 (1.11) & 26.50 (1.10) & \textbf{\textcolor{blue}{58.11 (3.29)}} & 69.10 (1.43) \\
Rated & 80.83 (0.49) & 80.85 (0.97) & \textbf{\textcolor{red}{79.91 (0.83)}} & 74.02 (0.92) & 76.83 (0.68) & 78.63 (0.59) & 19.80 (0.89) & 67.02 (2.97) & 66.70 (1.56) \\
Ranked & \textbf{\textcolor{red}{80.92 (0.62)}} & \textbf{\textcolor{red}{81.41 (0.73)}} & 78.40 (1.46) & \textbf{\textcolor{red}{77.31 (1.00)}} & \textbf{\textcolor{red}{78.89 (0.74)}} & \textbf{\textcolor{red}{79.41 (0.68)}} & \textbf{\textcolor{blue}{19.03 (0.85)}} & 65.10 (3.05) & \textbf{\textcolor{blue}{61.40 (1.93)}} \\
Cumulative & 79.05 (1.11) & 81.10 (0.89) & 78.45 (1.15) & 68.60 (1.62) & 73.63 (1.11) & 76.48 (0.86) & 23.13 (1.05) & 59.24 (3.36) & 65.50 (1.60) \\
\end{tabular}
}
\caption{Exchange economics environment results across different voting rules. The reported numbers are an average of all simulations, and the numbers in parentheses are standard errors. The smallest and largest value in a category is colored in \textcolor{blue}{blue} and \textcolor{red}{red}. We see that discrete voting rules show lower and less stable early performance, due to the higher probability of disagreement.}
\label{tbl:economics}
\end{table*}
\paragraph{Exchange Economy}
The experiment results show distinct performance differences across voting rules in Table \ref{tbl:economics} and Figure \ref{fig:economics}.\footnote{We explore more details with ablation studies, including scalability test with different number of agents and generalizability experiment with varying LLM models. For the results, please refer to Appendix \ref{app:ablation}.} Preference voting rules achieve higher performance early in the collaboration process compared to discrete voting rules, indicating that nuanced evaluations better aggregate agent preferences. Discrete voting rules struggle with lower and less stable early performance due to greater disagreement, which impairs decision-making. Despite differing early trajectories, all voting rules reach high-quality outcomes by the end. In most cases, performance peaks in the middle rounds, suggesting that early stopping could be advantageous. Rationality scores show that agents often make suboptimal choices, but decentralized MAS still performs well overall. Preference voting rules allow more dynamic decision updates, while discrete voting rules exhibit greater rigidity, particularly with unanimous voting.

When comparing discrete voting rules, unanimous voting performs the worst due to its rigidity and requirement for total agreement, resulting in an 87\% lower initial round performance and a 39\% lower final round performance compared to the best-performing models. Moreover, fairness progressively declines from unanimous to plurality voting. These findings suggest that stricter voting thresholds, such as unanimity, enforce greater fairness among agents but lead to significantly worse outcomes.

\begin{figure}[t]
    \centering
    \includegraphics[width=\linewidth]{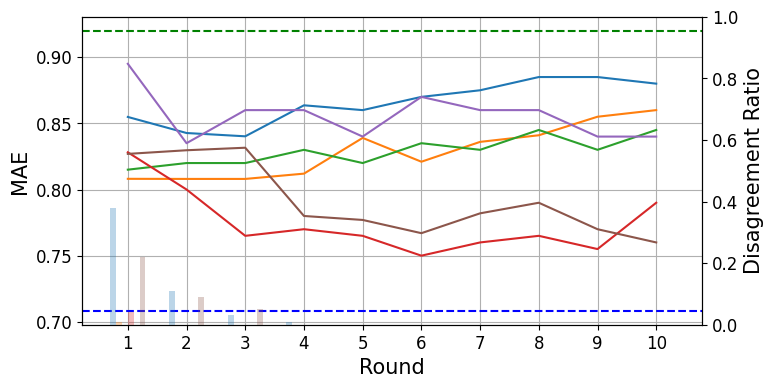}
    \caption{Comparison between voting rules in recommendation system environment. Line plots (left y-axis) show the comparison of MAE achieved over rounds; bar plots (right y-axis) represent the ratio of cases where participants yet to reach an agreement until a certain round. The green horizontal line is baseline (\textit{Always Guess 4}), and the blue line is SoTA performance \cite{behera2023collaborative}. }
    \label{fig:recommendation_mae}
\end{figure}

\paragraph{Recommendation System}
We illustrate the performance of recommendation system in Figure \ref{fig:recommendation_mae}, \ref{fig:recommendation_rmse} and Table \ref{tbl:recommendation}. The results show that multi-agent collaboration achieves moderate performance with distributed information, with the decentralized MAS performing comparably to SoTA approaches. While the SoTA method was purely machine-learning-driven, the MAS approach showed reasonable outcomes, indicating that collaboration between agents can explore and produce meaningful signals that help to make decision. 
Similar to the exchange economy, multiple results in the recommendation system follow a V-shaped curve. This suggests that while agent collaboration was efficient early on, it declined in later rounds, highlighting diminishing returns from repeated interactions. The trend underscores the need to balance sustained collaboration within MAS over time.

When comparing discrete to preference voting rules, we find that discrete voting rules generally worsened or plateaued after several rounds of collaboration. In contrast, preference voting rules show consistent improvement over time. This suggests that adaptive and flexible voting formats enhance collaboration by allowing for more nuanced input and iterative refinement across multiple rounds.

\section{Cross-Agent Communication Analysis}
\label{sec:conversation_analysis}
\paragraph{Method}
From the experiment results, we notice the performance degraded throughout the prolonged collaboration. To understand agent collaboration behaviors and communication, we perform a language analysis focused on three key features.\footnote{Detailed definition of four features is shown in Appendix \ref{app:language_features}. All observations are collected in 3 agents, gpt-4o-mini setting.} \textit{Message Length}: The length of the message. \textit{Message Complexity}: The complexity of the message. \textit{Information Distance}: The embedding distance between the conversations from current and previous round. \textit{Dialogue Acts}: The actions performed in the conversation, annotated with LLM. We extracted features from each round of communication and analyzed the trend of changes across different rounds, voting rules, and environments.

\begin{figure}[t]
    \centering
    \includegraphics[width=\linewidth]{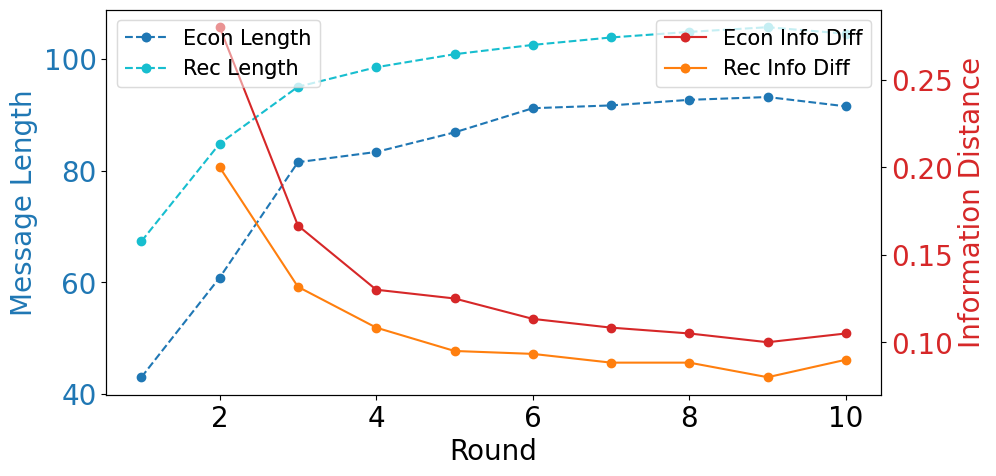}
    \caption{Line plots of average message length and information distance per round. Increasing message length and decreasing information distance indicate collaboration inefficiency in prolonged rounds.}
    \label{fig:language_feature}
\end{figure}

\label{sec:conversation_statistics}
\paragraph{Results}
Table \ref{tbl:heatmap} and Figure \ref{fig:language_feature} shows the averages of language features per round. There are two notable points: 1) Recommendation system is a harder environment than exchange economy, presented from overall higher message length and complexity. 2) Message length increases by 84\% over time, while information distance steadily decreases to 10\% in both environments. These phenomena indicate a convergence of topics and a decline in novel information as discussions progress, while the increasing message length adds additional cognitive and communication costs.

Dialogue acts annotation results are presented in Table \ref{tbl:dialogue_act}, and the transition graph between dialogue acts is in Figure \ref{fig:da_transition_econ}, \ref{fig:da_transition_recom}, \ref{fig:da_transition_econ_sc} and \ref{fig:da_transition_recom_sc}.\footnote{Detailed explanation of result can be found in Appendix \ref{app:language_analysis_result}.} In both environments, information exchange (\textit{Inform} and \textit{Request}) is the most frequent action, while direct disagreement (\textit{Defend} and \textit{Decline}) is very rare. Increasing usage of \textit{Confirm}, \textit{Summarize} and \textit{Compromise} also indicates the progress of collaboration. Dialogue acts also show differences across environments. Exchange economy shows more \textit{Evaluate}, \textit{Propose}, and strong loop between \textit{Request} and \textit{Propose}, showing how agents collaborate when the relationship is competitive. Meanwhile, recommendation system shows minimal compromise and accept, which are unnecessary in cooperate relationship. Instead, it has intensive use of request, indicating the focus on data exploration.  

\section{Early Stopping in MAS}
\begin{table}[t]
\centering
\resizebox{\linewidth}{!}{%
\begin{tabular}{lllllll}
\multicolumn{1}{c}{\textbf{}}& \multicolumn{1}{c}{\textbf{UNA.}} & \multicolumn{1}{c}{\textbf{MAJ.}} & \multicolumn{1}{c}{\textbf{PLU.}} & \multicolumn{1}{c}{\textbf{RAT.}} & \multicolumn{1}{c}{\textbf{RAN.}} & \multicolumn{1}{c}{\textbf{CUM.}} \\
\hline
\multicolumn{7}{c}{\textbf{RECOMMENDATION SYSTEM, in MAE($\downarrow$)}}\\
@10 (Baseline)&0.88&0.86&0.84&0.79&0.84&\textbf{0.76}\\
First Agreement&0.82&0.80&0.81&0.82&0.89&0.82\\
Consecutive Agreements&0.84&0.81&0.84&0.77&0.82&0.77\\
Validation Checkpoint&\textbf{0.81}&0.81&0.82&0.80&0.83&0.82\\
Information Distance&0.86&\textbf{0.78}&0.85&0.79&\textbf{0.78}&0.81\\
Dialogue Act&0.84&0.82&\textbf{0.79}&\textbf{0.76}&0.85&0.82\\
\cellcolor[RGB]{192,192,192}Oracle&\cellcolor[RGB]{192,192,192}0.73&\cellcolor[RGB]{192,192,192}0.63&\cellcolor[RGB]{192,192,192}0.59&\cellcolor[RGB]{192,192,192}0.62&\cellcolor[RGB]{192,192,192}0.69&\cellcolor[RGB]{192,192,192}0.67\\
\end{tabular}
}
\caption{Comparison between early stopping methods in recommendation system environment. The performance is compared with \textit{Oracle} and baseline \textit{@10} in MAE. ($\downarrow$) indicates better performance with lower values. Experiments are based on the results of 3 agents, gpt-4o-mini setting. Results are based on 5-fold cross validation. We observe that language-based methods performed well overall.}
\label{tbl:early_stopping_recom}
\end{table}

\paragraph{Method} 
While enforced 10 rounds of collaboration to study agent behaviors, we observed signals of ineffective collaboration from task performance and language feature analysis. To enhance collaboration efficiency, we compare five early stopping methods: \textit{First Agreement}, \textit{Consecutive Agreements}, \textit{Validation Checkpoint}, \textit{Information Distance}, and \textit{Dialogue Act}.\footnote{Detailed definitions of early stopping methods are listed in Appendix \ref{app:early_stopping_explanation}} We focus on the performance of \textit{Information Distance} and \textit{Dialogue Act} to examine whether language features help in identifying early stopping signals.

\paragraph{Results} 
We show comparison of early stopping methods in recommendation system in Table \ref{tbl:early_stopping_recom}, full results in Table \ref{tbl:early_stopping_all}, and their basic statistics are in Table \ref{tbl:early_stopping_statistics}. 

Almost all early stopping methods outperformed the non-stop baseline (\textit{@10}), indicating effectiveness of early termination in improving outcomes. Notably, methods leveraging language features (\textit{Information Distance} and \textit{Dialogue Act}) delivered better performance across most of voting rules and environments, each closed 12\% and 15\% of the gap between the non-stop and oracle approaches with fewer rounds. These findings highlight the advantage of incorporating language-based metrics in deciding when to stop, especially in complex multi-agent environments. Their results across both the exchange economy and recommendation system environments further supports the claim that language indicators can be powerful tools for optimizing collaboration outcomes, reducing cognitive load, and improving decision efficiency.

\section{Discussion}
\paragraph{Impact of Voting Rules on Collaborative Behavior}
This study demonstrates that voting rules, particularly their thresholds and formats, play a crucial role in multi-agent collaboration. Preference voting rules with a flexible format achieve higher performance and efficiency, especially in early rounds. In contrast, discrete voting rules struggle with rigidity, leading to lower initial performance. The strictness of voting thresholds, such as in unanimous voting, hampers collaboration due to its inability to accommodate dissent. Overall, preference voting rules, which allow for nuanced preference expression, facilitate dynamic exchanges and enhance decision quality, making a collaboration more effective for complex tasks.

Discrete voting rules also have unique advantages. First, the format of voting is simpler than preference voting rules, which led to fewer format errors during experiments from LLMs. Furthermore, unanimous voting showed the highest fairness and non-decreasing outcome over rounds when utility is quantifiable. These strengths suggest that voting rules in decentralized MAS should be thoughtfully chosen based on the specific task, or the primary metric being prioritized. Mixture of different methods in different stages of collaboration may also bring effectiveness, which we leave for future studies.

\paragraph{Language Indicators of Collaboration}
Language analysis of agent conversations reveals several key indicators of effective collaboration. The length and complexity of messages increase progressively across rounds, suggesting that deeper engagement and richer exchanges occur as the collaboration advanced. This is particularly evident in the recommendation system environment, where tasks require greater information exchange and negotiation. Complex tasks inherently demand more intricate language, as agents need to convey not just factual information but also interpret, analyze, and synthesize inputs from others. A key takeaway is the decline in collaboration efficiency over time, indicated by increasing message length and decreasing information distance between adjacent rounds.

Dialogue acts provide another layer of insight into collaborative behavior. We discover that dialogue acts comprehensively reflects agents' collaboration behaviors. When competitive relationship makes agents do more negotiation, cooperate relationship stimulate information exchange. This language pattern underscores that successful collaboration hinges on the flow of information, clarification, and ongoing evaluation of proposals. These indicators align with patterns seen in human collaboration, where information exchange and continuous feedback loops drive cooperative success.

\paragraph{Early Stopping}
As a part of decision-making mechanism, early stopping is essential in multi-agent collaboration because it prevents diminishing returns and inefficiency in decision-making. Our experiments show that while the initial rounds of collaboration led to significant improvements in group utility and decision quality, continuing beyond a certain point often results in stagnant or even declining performance. This highlights the need for mechanisms that can intelligently terminate the collaboration process once optimal solutions have been reached, avoiding unnecessary iterations.

Key insights into effective early stopping methods emerge from analyzing conversational patterns. Information distance suggests that as the rounds progress, the novelty of information decreases, with fewer new ideas being introduced. Transition graphs of dialogue acts reveal common collaborative patterns. These signals lead to language feature-based stopping methods, which halt collaboration when no meaningful new information is detected, or when dialogue act patterns indicate the need for termination. Our findings demonstrate that implementing early stopping methods based on language cues can enhance the efficiency and effectiveness of decentralized multi-agent collaboration by preventing unnecessary prolonged discussions.

\section{Conclusion}
This paper investigates the role of group decision-making mechanisms in decentralized MAS, emphasizing how different voting rules shape collaboration dynamics and efficiency. Through controlled experiments on RoundTable, we analyze the impact of decision thresholds and formats on performance, finding that adaptive voting rules lead to more effective outcomes. Our analysis of agent communication patterns reveals increasing message length and repetition over time, indicating inefficiencies in prolonged collaboration. Building on these insights, we demonstrate that language-based early stopping improves performance while reducing unnecessary rounds. Our study provides insights into decision-making and cross-agent communication, offering valuable implications for improving multi-agent collaboration and efficiency.

\section*{Limitations}
\paragraph{Platform Design}
In RoundTable, we fixed the phase order in our platform design to follow a structured sequence: message, propose, and voting. While this sequence represents a common collaboration flow, alternative action chains are possible, such as multi-round discussions before voting or iterative proposal refinements. However, we designed RoundTable to ensure that it captures the essential decision-making steps while maintaining clarity and reproducibility for experimental evaluation. By including both proposal and voting phases in each round, our design not only provides better transparency of agent intent but also allows researchers to collect granular collaboration data at multiple stages. This structured approach facilitates analysis of how agents form consensus and adapt strategies over time. Furthermore, repeating round-based collaboration will eventually enable agents to engage in multiple rounds of communication, proposing, and voting, giving them the chance to continue their discussion based on previous interactions. Although other collaboration models exist, we believe RoundTable represents a generalizable framework for studying the impact of decision-making mechanisms in decentralized MAS.

\paragraph{Environment Selection}
We conducted our experiments in two distinct environments: exchange economy and recommendation system. These environments were chosen because they inherently involve a balance between competition and cooperation, making them well-suited for studying collaborative decision-making. In exchange economy, agents must negotiate and trade resources while optimizing individual and collective goals, capturing the tension between self-interest and group benefit. In contrast, recommendation system simulates a setting where agents collaborate to improve recommendations based on user preferences, emphasizing mutual benefit and shared optimization. While our study focuses on these two environments, they represent a broad spectrum of real-world multi-agent interactions, spanning both competitive resource allocation and collaborative preference aggregation scenarios. As such, our findings offer insights applicable to a wide range of collaborative decision-making contexts.

\bibliography{custom}

\appendix
\section{Details on the Background of Decentralized MAS}
\label{app:background}

Agents powered by LLMs have demonstrated impressive problem-solving capabilities across a wide range of tasks. However, single-agent systems struggle with problems that are too large or complex, often leading to instability, misalignment with the intended request, and hallucination \citep{liu2024lost, kuhn2023semantic, lyu2023faithful}. To overcome these limitations, research has increasingly turned to Multi-Agent Systems (MAS). MAS leverage collective intelligence by enabling agents to specialize in distinct skills and collaborate effectively \citep{guo2024large}.

In MAS, agents often hold differing perspectives due to variations in persona roles, information access, and individual goals, making group decision-making inherently complex. Many LLM-driven MAS rely on centralized decision-making, where a manager agent or predefined power hierarchy resolves conflicts \citep{hong2023metagpt, qian2023communicative, dong2023self}. However, centralization introduces several challenges: 1) \textit{fairness}: Individual agent inputs may be misrepresented in the final decision \citep{jiang2019learning, owens2024multi}; 2) \textit{rigidity}: Fixed hierarchical structures limit adaptability, making systems prone to overfitting \citep{chen2023agentverse}; 3) \textit{information constraints}: A central authority cannot always dictate optimal strategies when agents possess private or partial information \citep{xu2023exploring}.

Decentralized group decision-making mitigates the limitations of centralized MAS by distributing decision power among agents, allowing each to actively participate in the process. This approach is particularly critical in dynamic and complex environments, where agents must operate independently with information asymmetry \citep{mandi2024roco, zhang2023building, xu2023exploring, park2023generative}. By removing reliance on a central authority, decentralized MAS enable greater adaptability and resilience. However, in the absence of a fixed decision-maker, agents must establish structured yet flexible coordination strategies to achieve coherent and effective group decisions. Identifying mechanisms that facilitate meaningful agreement in such systems remains a key challenge.

\section{Details of RoundTable}
    \subsection{RoundTable Platform Design}
    \label{app:roundtable_platform}
    In this section, we explain the details of the design in RoundTable by breaking it down to phases. 
    
    \textbf {Input and Initialization} defines each agent for collaboration. Each agent will be given a specific $\Omega_i$ and $u_i$, which include the same input user query, information of RoundTable, and the voting rule $F$.
    However, they may have different background, information access, or utility function. While $\Omega_i$ and $u_i$ can be defined automatically by agent recruiting or profiling \citep{chen2023agentverse}, we manually define them for simplicity. 
    
    \textbf{Message Phase} is the initial stage of the iterative collaboration process, where all agents participate in a simultaneous, open chat conversation. During this phase, agents can freely choose recipients based on their context to send messages. They can respond to others' questions, share insights, or ask their own questions. There are no restrictions on whom to communicate with or what to say. Once generated, all messages are updated to $\Omega_i$ for everyone, regardless of the original sender or intended recipients. This process guarantees that all messages are sent simultaneously, makes it has no specific order of communication. Since only one message can be sent per iteration, back-and-forth conversations can take place across multiple rounds.
    
    \textbf{Proposal Phase} allows agents to present a proposal $p_i$, offering a potential solution or response to the user query from their perspective, aiming to fulfil their individual objective or maximize the utility. Before generating a proposal, agents will be asked to do a step by step reasoning, which is by default invisible to other agents. Reasoning is designed to give agents a chance to summarize and analyze information in $\Omega_i$. The most recent proposal from each agent will be considered as a candidate for voting in the next phase. Like the Message Phase, proposals are made simultaneously and will be updated for $\Omega_i$ once everyone has generated a response. Agents also have the option to skip submitting a new proposal; in such cases, their previous proposal will automatically be included among the candidates.
    
    \textbf{Voting Phase} collects one latest proposal $p_i$ from each agent, additionally include the latest intermediate group decision $x^*$ from previous rounds to form a candidate proposal list. Agents in this round are asked to express preferences toward candidates. The preferences can be in various formats, including voting, rating and ranking, depends on what types of voting rule $F$ is used. Similar with the proposal phase, agents have option to not vote (give up). With all the collected proposals and preferences, voting rule selects one proposal as the intermediate group decision. Finally, voting details and the result are updated to all agents' context $\Omega_i$.
    
    \textbf{Output} The iteration will run for 10 rounds. After the 10th round, the latest intermediate group decision will be selected as the final output.

    \subsection{RoundTable Prompt Design}
    \label{app:roundtable_prompt}
        We present the prompt design for RoundTable in this section, which consists of initialization stage (Figure \ref{prompt:init}), message phase (Figure \ref{prompt:message}), proposal phase (Figure \ref{prompt:proposal}), and voting phase (Figure \ref{prompt:voting}). Italicized words represent inputs or variables within the prompt. 
        \begin{figure*}[!t]
\begin{tcolorbox}[colframe=blue!60, colback=blue!5!white, coltitle=white, title=Initialization Prompt, width=\textwidth, sharp corners, fonttitle=\bfseries, colbacktitle=blue!80!black, left=2mm, right=2mm, top=1mm, bottom=1mm]
\# Agent Initialization\\
You are \textit{{my\_name}}, an agent in a recurring collaboration environment designed to address and solve complex problems.\\

\# Task Description\\
\textit{{task\_description}}\\

\# Collaboration Rules\\
You start with nothing decided. The intermediate result will be decided by the voting rule at the end of each round.\\
In each round, the collaboration runs in 3 phases with the following order:\\
1. Message Phase: At the beginning of each round, you can send one message to a shared channel for either Talking to one or more agents. All agents will send messages simultaneously. You will be able to see all messages from all agents after the end of the message phase.\\
2. Proposal Phase: After the end of the message phase, you will have the opportunity to propose potential solution. If you don't propose in this phase, your latest proposal will be used for voting.\\
3. Voting Phase: At the end of the round, all agents' latest proposal will be voted. When agents didn't propose in this round, their latest proposal will be used for voting. All votes will be processed with the voting rule: \textit{{name\_of\_voting\_rule}}, where \textit{{explanation\_of\_voting\_rule}}. If the voting rule selects a proposal, the intermediate result will be updated accordingly. 
After each round, each agent will be able to see the result of the vote from the previous round and the conversation history from all rounds.\\

The collaboration will run for \textit{{max\_rounds}} rounds. After the last round, the latest result will be the final result.\\

\# Your Background\\
\textit{{my\_agent\_backround}}\\

\# Game History\\
\#\# Latest Candidates at Round \textit{latest\_candidates\_round}:\\
\textit{latest\_candidates}\\

\#\# Latest Voting Result at Round \textit{vote\_history\_length}:\\
\textit{latest\_vote\_history}\\

\#\# Latest Approved Proposal:\\
Proposal \textit{latest\_approved\_proposal\_id} from Round \textit{latest\_approved\_proposal\_round}.\\

Proposal \textit{latest\_approved\_proposal\_id} Detail:\\
\textit{latest\_approved\_proposal\_detail}\\

\#\# Conversation History until Round \textit{conversation\_history\_length}:\\
\textit{conversation\_history}
\end{tcolorbox}
\caption{Prompt used for initialization stage in RoundTable. Italicized words represent inputs or variables within the prompt. }
\label{prompt:init}
\end{figure*}
        \begin{figure*}[!t]
\begin{tcolorbox}[colframe=blue!60, colback=blue!5!white, coltitle=white, title=Message Phase Prompt, width=\textwidth, sharp corners, fonttitle=\bfseries, colbacktitle=blue!80!black, left=2mm, right=2mm, top=1mm, bottom=1mm]
You are \textit{my\_name}, currently in the message phase of round \textit{round\_num}. In this phase, you can:\\
1. Answer questions posed by others.\\
2. Share your findings or insights.\\
3. Ask questions to further the discussion.\\
You may engage in multiple activities using multiple sentences.\\

Please type your message in the following JSON format: \{``target":  \textless list of agent names\textgreater, ``message": \textless str, your message\textgreater\}\\
Don't generate anything except the JSON format.

\end{tcolorbox}
\caption{Prompt used for message phase in RoundTable. Italicized words represent inputs or variables within the prompt.}
\label{prompt:message}
\end{figure*}
        \begin{figure*}[!t]
\begin{tcolorbox}[colframe=blue!60, colback=blue!5!white, coltitle=white, title=Proposal Phase Prompt, width=\textwidth, sharp corners, fonttitle=\bfseries, colbacktitle=blue!80!black, left=2mm, right=2mm, top=1mm, bottom=1mm]
You are \textit{my\_name}, currently in the proposal phase of round \textit{round\_num}. You have an opportunity to make a proposal of the potential solution. Whether or not you submit a new proposal, your latest proposal will be considered as a candidate proposal for the voting phase.\\

You have two options:\\
1. Make a proposal:\\
    - You can propose a potential solution by the provided format.\\
2. Do not make a proposal:\\
    - If you do not want to propose a solution, you can return None as your proposal.\\

Please type your proposal in the following JSON format: \{``reason\_for\_decision": \textless your step by step reasoning for your decision\textgreater, ``proposal": \textit{proposal\_format\_text}, or None\}\\
Don't generate anything except the JSON format.\\

\end{tcolorbox}
\caption{Prompt used for proposal phase in RoundTable. Italicized words represent inputs or variables within the prompt.}
\label{prompt:proposal}
\end{figure*}
        \begin{figure*}[!t]
\begin{tcolorbox}[colframe=blue!60, colback=blue!5!white, coltitle=white, title=Voting Phase Prompt (for Discrete Voting Rules), width=\textwidth, sharp corners, fonttitle=\bfseries, colbacktitle=blue!80!black, left=2mm, right=2mm, top=1mm, bottom=1mm]
You are \textit{my\_name}, at the voting phase at round \textit{round\_num}. \\
In this phase, all agents' latest proposal will be voted by \textit{name\_of\_social\_choice}, where \textit{explanation\_of\_social\_choice}. If the voting rule selects a proposal, the intermediate result will be updated accordingly. \\

You have two actions to choose: vote or not vote.\\
1. For vote:\\
    - You can only vote for one of the proposals from the candidate list.\\
2. For not vote:\\
    - You should vote None.\\
    - If you do not want to vote for any of the proposals, you can vote None.\\
    - If there is no proposal, you vote None.\\

The same proposal proposed by multiple agents will be merged as one proposal.\\
If there are multiple proposals passed, none of the proposals will be selected.\\
If no proposals are passed, the current intermediate result will be kept.\\

The current candidate proposals are as follows:\\
\textit{proposal\_list}\\

What is your vote? Please answer in the following JSON format: \{``reason\_for\_decision": \textless your step by step reasoning for your decision\textgreater, ``decision": \textless id of the proposal from the candidates you want to vote, or None\textgreater\}\\
Don't generate anything except the JSON format.

\end{tcolorbox}
\caption{Prompt used for voting phase in RoundTable. Italicized words represent inputs or variables within the prompt.}
\label{prompt:voting}
\end{figure*}

\section{Details of Voting Rules}
    \label{app:social_choice}
    We present the definition and the prompt we use to define different voting rules, including unanimous voting (Figure \ref{prompt:unanimous}), majority voting (Figure \ref{prompt:majority}), plurality voting (Figure \ref{prompt:plurality}), rated voting (Figure \ref{prompt:rated}), ranked voting (Figure \ref{prompt:ranked}), and cumulative voting (Figure \ref{prompt:cumulative}). The formal definition of voting rules are as below: \\
    \noindent\textbf{Unanimous Voting} \citep{arrow2012social}: The proposal \( x^*_r \) that receives votes from all agents is selected:
\[
\hat{x^*}_r = \arg\max_{c \in C} \mathbbm{1} \left( \sum_{i \in \mathcal{I}} v_{i,c,r} = |\mathcal{I}| \right)
\]

\noindent\textbf{Majority Voting} \citep{arrow2012social}: The candidate $c$ that receives votes from more than half of all agents is selected:
\[
\hat{x^*}_r = \arg\max_{c \in C} \mathbbm{1} \left( \sum_{i \in \mathcal{I}} v_{i,c,r} > \frac{|\mathcal{I}|}{2} \right)
\]

\noindent\textbf{Plurality Voting} \citep{arrow2012social}: The candidate $c$ that receives the most votes is selected:
\[
\hat{x^*}_r = \arg\max_{c \in C} \sum_{i \in \mathcal{I}} v_{i,c,r}
\]

\noindent\textbf{Rated Voting} \citep{baujard2018voters}: Each agent rates proposals on a 5-point scale, and the candidate $c$ with the highest total score is selected:
\[
\hat{x^*}_r = \arg\max_{c \in C} \sum_{i \in \mathcal{I}} s_{i,c,r}, \quad s_{i,c,r} \in \{1,2,3,4,5\}
\]

\noindent\textbf{Ranked Voting} \citep{arrow2012social}: Agents rank proposals by preference, assigning points as \( 1, 1/2, 1/3, \) etc., and the candidate $c$ with the highest total score is selected:
\[
\hat{x^*}_r = \arg\max_{c \in C} \sum_{i \in \mathcal{I}} \frac{1}{rank_{i,c,r}}
\]
where \( rank_{i,c,r} \) is the rank of proposal \( c \) given by agent \( i \) (lower rank values indicate higher preference).

\noindent\textbf{Cumulative Voting} \citep{black1958theory}: Each agent distributes \( |\mathcal{I}| \) points among \( |\mathcal{I}| \) proposals, and the candidate $c$ with the highest total is selected:
\[
\hat{x^*}_r = \arg\max_{c \in C} \sum_{i \in \mathcal{I}} pt_{i,c,r}
\]
where \( pt_{i,c,r} \) represents the number of points agent \( i \) assigns to proposal \( c \) such that:
\[
\sum_{c \in C} p_{i,c,r} = |\mathcal{I}|
\]
    \begin{figure*}[!t]
        \begin{tcolorbox}[colframe=blue!60, colback=blue!5!white, coltitle=white, title=Unanimous Voting Description, width=\textwidth, sharp corners, fonttitle=\bfseries, colbacktitle=blue!80!black, left=2mm, right=2mm, top=1mm, bottom=1mm]
            The proposal that receives votes from all agents will be selected. If no proposal receives votes from all agents, no proposal will be selected.
        \end{tcolorbox}
        \caption{Prompt used for describing unanimous voting in RoundTable. }
        \label{prompt:unanimous}
    \end{figure*}
    \begin{figure*}[!t]
        \begin{tcolorbox}[colframe=blue!60, colback=blue!5!white, coltitle=white, title=Majority Voting Description, width=\textwidth, sharp corners, fonttitle=\bfseries, colbacktitle=blue!80!black, left=2mm, right=2mm, top=1mm, bottom=1mm]
            The proposal that receives votes from more than half of all agents will be selected. If no proposal meets this condition, none will be selected.
        \end{tcolorbox}
        \caption{Prompt used for describing majority voting in RoundTable.}
        \label{prompt:majority}
    \end{figure*}
    \begin{figure*}[!t]
        \begin{tcolorbox}[colframe=blue!60, colback=blue!5!white, coltitle=white, title=Plurality Voting Description, width=\textwidth, sharp corners, fonttitle=\bfseries, colbacktitle=blue!80!black, left=2mm, right=2mm, top=1mm, bottom=1mm]
            The proposal that receives the most votes will be selected.
        \end{tcolorbox}
        \caption{Prompt used for describing plurality voting in RoundTable.}
        \label{prompt:plurality}
    \end{figure*}
    \begin{figure*}[!t]
        \begin{tcolorbox}[colframe=blue!60, colback=blue!5!white, coltitle=white, title=Rated Voting Description, width=\textwidth, sharp corners, fonttitle=\bfseries, colbacktitle=blue!80!black, left=2mm, right=2mm, top=1mm, bottom=1mm]
            Each agent assigns ratings on a 5-point Likert scale to all candidate proposals, with 1 being the lowest and 5 being the highest. The proposal with the highest total score will be selected.
        \end{tcolorbox}
        \caption{Prompt used for describing rated voting in RoundTable.}
        \label{prompt:rated}
    \end{figure*}
    \begin{figure*}[!t]
        \begin{tcolorbox}[colframe=blue!60, colback=blue!5!white, coltitle=white, title=Ranked Voting Description, width=\textwidth, sharp corners, fonttitle=\bfseries, colbacktitle=blue!80!black, left=2mm, right=2mm, top=1mm, bottom=1mm]
            Each agent ranks all candidate proposals from the most preferred to the least preferred. 1, 1/2, 1/3... points will be assigned to the 1st, 2nd, 3rd... candidates on each ballot. The proposal with the highest total points will be selected.
        \end{tcolorbox}
        \caption{Prompt used for describing ranked voting in RoundTable.}
        \label{prompt:ranked}
    \end{figure*}
    \begin{figure*}[!t]
        \begin{tcolorbox}[colframe=blue!60, colback=blue!5!white, coltitle=white, title=Cumulative Voting Description, width=\textwidth, sharp corners, fonttitle=\bfseries, colbacktitle=blue!80!black, left=2mm, right=2mm, top=1mm, bottom=1mm]
            For X candidate proposals, each agent is given X points to distribute among the proposals as they see fit. The proposal with the highest total points will be selected.
        \end{tcolorbox}
        \caption{Prompt used for describing cumulative voting in RoundTable.}
        \label{prompt:cumulative}
    \end{figure*}

\section{Details of Experiments}
    \subsection{Exchange Economy}
    \label{app:exchange_economy}
    Exchange economy environment has multiple unique advantages. First, it is a plus-sum game. Economically, agents seek an equilibrium until no one can improve their utility without reducing others', implying the possibility of further collaboration exists if the allocation is not at equilibrium. Second, there are multiple possible equilibria exist in the market, allowing dynamic collaboration direction. Third, equilibria are not equal to reaching maximum total utility $U_{max}$, but $U_{max}$ must be lying in one of the equilibria. This can help us to measure if conflicts between agents harmful for the group to reach an ultimate goal.

    The task description we use is presented in Figure \ref{prompt:econ_task}. Prompt for agent's objective is shown in Figure \ref{prompt:econ_agent_goal}.

    \begin{figure*}[!t]
        \begin{tcolorbox}[colframe=blue!60, colback=blue!5!white, coltitle=white, title=Exchange Economy Task Description, width=\textwidth, sharp corners, fonttitle=\bfseries, colbacktitle=blue!80!black, left=2mm, right=2mm, top=1mm, bottom=1mm]
    You will collaborate with other agents in a recurring exchange market game.\\
    There are \textit{num\_of\_agents} agents in this market: \textit{list\_of\_agents}.\\
    There are \textit{num\_of\_goods} goods in the market: \textit{list\_of\_goods}. Total quantity of each good is as follows: \textit{total\_num\_of\_goods}.\\
    In this game, you will collaboratively decide how to distribute the goods among the agents. Your goal is to maximize your own utility function.
        \end{tcolorbox}
        \caption{Prompt used for describing the task in exchange economy environment in RoundTable. Italicized words represent inputs or variables within the prompt.}
        \label{prompt:econ_task}
    \end{figure*}

    \begin{figure*}[!t]
        \begin{tcolorbox}[colframe=blue!60, colback=blue!5!white, coltitle=white, title=Exchange Economy Agent Objective, width=\textwidth, sharp corners, fonttitle=\bfseries, colbacktitle=blue!80!black, left=2mm, right=2mm, top=1mm, bottom=1mm]
    Your goal is to maximize your individual utility function by communicating, proposing, and voting with other agents. Your utility function is \textit{util\_func}
        \end{tcolorbox}
        \caption{Prompt used for describing an agent's objective in exchange economy environment in RoundTable. Italicized words represent inputs or variables within the prompt.}
        \label{prompt:econ_agent_goal}
    \end{figure*}

    \subsection{Exchange Economy: Other Utility Sets}
    \label{app:other_utility_sets}
    We use the asymmetric scenario for the main result due to its similarity to real-world applications, here are other utility sets we examined.
    
    \textit{Symmetric} is the case where all agents prefer the same good, mirroring the scenario where agents collaborate with a common goal. The following utility function is applied to each agent: $u_i = a_1^{0.8}\prod_{k \neq 1} a_k^{\tilde{\theta}}, \tilde{\theta} = \frac{1-0.8}{|I|}$
    
    \textit{Uniform} is the case where all agents indifferently prefers all goods, using the utility function: $u_i = \prod a_k^{\tilde{\theta}}, \tilde{\theta} = \frac{1}{|I|}$

    We find these two sets of utility functions are not ideal for evaluating multi-agent collaboration. Since all the agents has the same utility function, the oracle allocation among agents to reach maximized group total utility is very close to even split. However, even split is the most frequent proposal from agents while they have no information for other agents' preferences, making most of the result almost perfect. This phenomenon is observed across all LLMs we've tested. 

    \subsection{Recommendation System}
    \label{app:recommendation_system}
    Recommendation system aims to reveal unique collaborative behaviors that exchange economic does not have. First, there exists strong information asymmetry between agents. Any piece of information is not sufficient to correctly predict the rating. Secondly, the group decision-making in these systems encourages diverse approaches to problem-solving. Agents are not only consumers of information but also contributors, creating a feedback loop where the quality of recommendations improves with increased participation and collaboration. Lastly, the value generated in such a system is not solely based on the final recommendation but also on the process of reaching that recommendation. The interactions, negotiations, and information exchanges that occur along the way contribute to the overall effectiveness and satisfaction of the system. We show the description of the environment in Figure \ref{prompt:recom_task}. In the recommendation system enviroment, all agents share the same objective, but has different background dataset. The prompt used for an agent's objective is in Figure \ref{prompt:recom_agent_goal}.
    
    \begin{figure*}[!t]
        \begin{tcolorbox}[colframe=blue!60, colback=blue!5!white, coltitle=white, title=Recommendation System Task Description, width=\textwidth, sharp corners, fonttitle=\bfseries, colbacktitle=blue!80!black, left=2mm, right=2mm, top=1mm, bottom=1mm]
    You will collaborate with other agents in a movie recommendation game.\\
    In this game, you will collaboratively predict the rating of a target movie (\textit{target\_movie\_title}) for a target user.\\
    There are 3 agents in this game: BasicInfo Agent, MovieHistory Agent, UserHistory Agent.\\
    1. BasicInfo Agent has access to the basic information of the target movie and target user. It has access to the data with the following schema:\\
    \textit{get\_schema(movie\_info)}\\
    \textit{get\_schema(user\_info)}\\
    2. MovieHistory Agent has access to the rating history of the target movie from other people. It has access to the data with the following schema:\\
    \textit{get\_schema(movie\_rating\_history)}\\
    3. UserHistory Agent has access to the rating history of the target user to other movies. It has access to the data with the following schema:\\
    \textit{get\_schema(user\_rating\_history)}\\
    
    You can't see other agents' information directly, but you can get information from other agents through communication.
    Your goal is to predict the rating a target user would give to \textit{target\_movie\_title}. Utilize all available information about both the user and the movie to make the most accurate prediction possible. You only have access to partial information, but you can communicate with other agents to get more information.
        \end{tcolorbox}
        \caption{Prompt used for describing the task in recommendation environment in RoundTable. Italicized words represent inputs or variables within the prompt.}
        \label{prompt:recom_task}
    \end{figure*}
    
    \begin{figure*}[!t]
        \begin{tcolorbox}[colframe=blue!60, colback=blue!5!white, coltitle=white, title=Recommendation System Agent Objective, width=\textwidth, sharp corners, fonttitle=\bfseries, colbacktitle=blue!80!black, left=2mm, right=2mm, top=1mm, bottom=1mm]
        Your goal is to predict the rating the target user would give to the target movie. Utilize all available information about both the user and the movie to make the most accurate prediction possible. You only have access to \textit{data\_access}, but you can communicate with other agents to get more information.\\
        
        \# Your Data:\\
        \textit{agent\_dataset}
        \end{tcolorbox}
        \caption{Prompt used for describing an agent's objective in recommendation environment in RoundTable. Italicized words represent inputs or variables within the prompt.}
        \label{prompt:recom_agent_goal}
    \end{figure*}

To ease the complexity of the data structure, we pre-processed the data into three parts: \textbf{basicInfo}, which contains the basic details of the target user and movie; \textbf{userHistory}, which includes the target user's ratings and basic information for other movies; and \textbf{movieHistory}, which organizes all ratings from other users for the target movie, ranked by a preference similarity score calculated using non-negative matrix factorization on the co-occurrence rating table between users and movies. In our experiment, each part of the dataset is assigned to a different agent, forming a 3-agent collaboration system. Examples of three datasets are in Table \ref{tbl:basicInfo}, \ref{tbl:userHistory} and \ref{tbl:movieHistory}.
    
\begin{table*}
    \centering
    \begin{tabular}{llll}
        \multicolumn{1}{c}{\textbf{movie\_id}} & \multicolumn{1}{c}{\textbf{movie\_title}} & \multicolumn{1}{c}{\textbf{release\_date}} & \multicolumn{1}{c}{\textbf{genre}} \\
        \hline
        231 & Batman Returns & 19920101 & ['Action', 'Adventure', 'Comedy', 'Crime'] \\
    \end{tabular}

    \vspace{0.5cm} 
    
    \begin{tabular}{lllll}
        \multicolumn{1}{c}{\textbf{user\_id}} & \multicolumn{1}{c}{\textbf{age}} & \multicolumn{1}{c}{\textbf{gender}} & \multicolumn{1}{c}{\textbf{occupation}} & \multicolumn{1}{c}{\textbf{state}}\\
        \hline
        7 & 29 & F & artist & NY \\
    \end{tabular}
    
    \caption{Example of \textit{basicInfo} dataset, which includes basic information about the target user and target movie to predict.}
    \label{tbl:basicInfo}
\end{table*}
\begin{table*}
    \centering
    \resizebox{\textwidth}{!}{%
    \begin{tabular}{llllll}
\multicolumn{1}{c}{\textbf{movie\_id}} & \multicolumn{1}{c}{\textbf{movie\_title}} & \multicolumn{1}{c}{\textbf{genre}} & \multicolumn{1}{c}{\textbf{release\_date}} & \multicolumn{1}{c}{\textbf{rating}} & \multicolumn{1}{c}{\textbf{rated\_date}} \\
\hline
1 & Toy Story & ['Animation', "Children's", 'Comedy'] & 19950101 & 3 & 19980331 \\
14 & Postino, Il & ['Drama', 'Romance'] & 19940101 & 3 & 19980331 \\
24 & Rumble in the Bronx & ['Action', 'Adventure', 'Crime'] & 19960223 & 3 & 19980331 \\
50 & Star Wars & ['Action', 'Adventure', 'Romance', 'Sci-Fi', 'War'] & 19770101 & 4 & 19980331 \\
109 & Mystery Science Theater 3000: The Movie & ['Comedy', 'Sci-Fi'] & 19960419 & 3 & 19980331 \\
\multicolumn{6}{c}{...}
    \end{tabular}
    }
    \caption{Example of \textit{userHistory} dataset, which includes rating history from the target user to other movies. Here only shows 5 rows for spacing.}
    \label{tbl:userHistory}
\end{table*}
\begin{table*}
    \centering
    \resizebox{\textwidth}{!}{%
    \begin{tabular}{lllllllll}
\multicolumn{1}{c}{\textbf{user\_id}} & \multicolumn{1}{c}{\textbf{user\_pref\_similarity}} & \multicolumn{1}{c}{\textbf{personal\_average\_score}} & \multicolumn{1}{c}{\textbf{age}} & \multicolumn{1}{c}{\textbf{gender}} & \multicolumn{1}{c}{\textbf{occupation}} & \multicolumn{1}{c}{\textbf{state}} & \multicolumn{1}{c}{\textbf{rated\_date}} & \multicolumn{1}{c}{\textbf{rating}} \\
\hline
343 & 0.98 & 3.99 & 43 & M & engineer & GA & 19971009 & 5 \\
806 & 0.98 & 3.64 & 27 & M & marketing & NY & 19971217 & 3 \\
773 & 0.98 & 3.28 & 20 & M & student & MN & 19980227 & 2 \\
805 & 0.98 & 3.35 & 27 & F & other & DC & 19971209 & 3 \\
447 & 0.97 & 3.57 & 30 & M & administrator & MN & 19971106 & 2 \\
\multicolumn{9}{c}{...}
    \end{tabular}
    }
    \caption{Example of \textit{movieHistory} dataset, which includes rating history of the target movie from other users. Here only shows 5 rows for spacing.}
    \label{tbl:movieHistory}
\end{table*}

    \subsection{Language Features}
    \label{app:language_features}
To understand agent collaboration and analyze their messages, we perform a language analysis focused on four key features. All observations are collected in 3 agents, gpt-4o-mini setting.\newline

    \noindent\textbf{Message Length} is a basic yet significant metric, reflecting the amount of information an agent conveys. We measure it using word count, as longer messages generally enrich the conversation.
    
    \noindent\textbf{Message Complexity} assesses how difficult a message is to understand, indicating the depth of the conversation. We use the Flesch-Kincaid grade level to calculate complexity, with higher scores representing more intricate messages \citep{klare1974assessing}.

    \noindent\textbf{Information Distance} measures how much new information is introduced. Effective collaboration should consistently bring new insights, while low information gain signals a stalled conversation. It is calculated as the average cosine distance between messages in the current round and the center embedding of the previous round.

    \noindent\textbf{Dialogue Acts} are communicative functions that capture actions, intents, and behaviors within messages as discrete states. We design a set of dialogue acts for multi-agent collaboration: \textit{Inform, Request, Confirm, Summarize, Evaluate, Propose, Compromise, Defend, Accept, Decline}, and use LLM-labeling to perform multi-label classification on a target message based on the previous round. Details of the dialogue act labels and prompt can be found in Appendix \ref{app:dialogue_act}.
    
    \subsection{Dialogue Act Labeling}
    \label{app:dialogue_act}
    We present the definition of each dialogue act we use in the experiments. The definitions are also used in LLM prompt for automatic annotation.
    
    Conversation Acts (Informational):
    \begin{itemize}
        \item \textit{Inform} - Shares new information that wasn't previously known.
        \item \textit{Request} - Asks for information that the speaker doesn't have.
        \item \textit{Confirm} - Asks to verify or validate shared information.
        \item \textit{Summarize} - Provides a brief overview of the main points.
        \item \textit{Evaluate} - Gives an opinion or judgment about the information.
    \end{itemize}
    
    Collaboration Acts (Decision-Making):
    \begin{itemize}
      \item \textit{Propose} - Introduce a new solution in the discussion.
      \item \textit{Compromise} - Offers a balanced solution that incorporates parts of different parties' preferences.
      \item \textit{Defend} - Maintain support for an idea or solution after consideration or challenge.
      \item \textit{Accept} - Agrees to or accept an idea or solution.
      \item \textit{Decline} - Refuses or disagrees with an idea or solution.
    \end{itemize}

    \subsection{Dialogue Act Transition Graph}
    \label{app:dialogue_act_transition_explanation}
    We constructed transition graphs to visualize the dynamics of dialogue act transitions across rounds, highlighting the most frequent transitions. Each node represents a dialogue act, and directed edges indicate the probability of transitions between acts from one round to the next. The figure displays only the most probable edge per node, excluding self-loops. 
    
    Here, we formally define dialogue act transition graph. Two nodes and the directed edge between them consist with a pair of dialogue acts and its transition probability. We denote a pair of dialogue acts as $A\rightarrow B$, where $A$ is a dialogue act observed from an agent's message in round $r-1$, and $B$ is the one from other speakers in round $r$. The transition probability, $p_{A\rightarrow B}$, is the probability of the existence of $A\rightarrow B$ among all observed $A$s To be more specific, a directed edge between dialogue act $A$ and $B$ is calculated by the following equation:

    \begin{equation}
        \resizebox{\columnwidth}{!}{$
        p_{A\rightarrow B} = \frac{|A\rightarrow B|}{|A|} = \frac{\sum_s\sum_{r=1}^{10}|\{(i,\neg i)|i\in\mathcal{I},A \in \mathcal{D}_{i,r-1}, B\in \mathcal{D}_{\neg i,r}\}|}{\sum_s\sum_{r=1}^{10}|\{i|i \in \mathcal{I}, A \in \mathcal{D}_{i,r-1}\}|}
        $}
    \end{equation}

    Where $s$ is the index of a collaboration, $\mathcal{D}_{i,r}$ is a set of dialogue acts from the message of the agent $i$ at round $r$, and $\mathcal{D}_{r=0}=\{Start\}, \mathcal{D}_{r=11}=\{End\}$. 
    \subsection{Early Stopping in MAS}
    \label{app:early_stopping_explanation}
    This section explains the detailed definitions of early stopping methods. In this experiment, we evaluate the following early stopping methods:\newline

    \noindent\textbf{@10} is our baseline, the final performance after 10 rounds without early stopping.
    
    \noindent\textbf{First Agreement} stops collaboration whenever a proposal passes the voting threshold.

    \noindent\textbf{Consecutive Agreements} is when no one make additional proposal after a proposal has been accepted in the previous round.

    \noindent\textbf{Validation Checkpoint} is the average number of rounds that produced the best outcomes in the train set, used as an early stopping criterion for all test sets. 

    \noindent\textbf{Information Distance} is the average embedding distance for messages between adjacent rounds captured at the round that produced the best outcomes in the train set, and stops collaborations in the test set when the distance became lower than the threshold. This idea is supported by the observation in Section \ref{sec:conversation_statistics}.

    \noindent\textbf{Dialogue Act} method utilizes pairs of dialogue acts, linking one from the previous round to another from the current round. We perform ordinary least squares (OLS) regression on all such pairs in relation to the performance. The regression coefficients indicate the most impactful pairs that act as stop signals. Further details of the algorithm are provided in Appendix \ref{app:dialogue_act_early_stopping}.\newline

We use 5-fold cross validation to compare the methods with \textbf{Oracle}: the oracle performance for early stopping, which reflects the best outcomes from each test simulation.

    \subsubsection{Dialogue Act Early Stopping Method}
    \label{app:dialogue_act_early_stopping}
    Details of \textbf{Dialogue Act} early stopping method is as follows. We first collect all pairs from dialogue acts as independent variable (with one-hot encoding), where one form round $r-1$ and another from $r$. Then make the group performance at round $r$ to match with dialogue act pairs as dependent variable. Next, we run OLS regression on this dataset, assigning coefficients and p-value to each pair. The strength of coefficient shows how each dialogue act pairs relate the performance, positively and negatively, and p-value shows the statistical significance of these relationships.

    To find the best early stopping rounds, we do a greedy search on training set for the following hyperparameters:
    \begin{itemize}
        \item \textit{top\_da}: the number of dialogue act pairs with the top coefficients to apply in candidates. 1 to 5, inclusive.
        \item \textit{p-value}: the threshold of p-value for a pair to be considered as candidates. 0.05, 0.1, 0.2 and None.
        \item \textit{count\_per\_round}: the threshold of how many occurrences of a candidate should exist in a round to pass. When passed, it gives +1 score for termination. 1 to 3, inclusive.
        \item \textit{score}: the threshold of how many scores a round needs to be terminated. 1 to value of top\_da, inclusive.
    \end{itemize}

    We examine all combinations of hyperparameters on the training data, and identifies the best sets for each voting rule and environment. Using the best hyperparameters, we evaluate the performance of early stopping on the test set. 

    \subsection{Models and Dataset}
        For all experiments, we by default use \textit{gpt-4o-mini-2024-07-18} \citep{openai2023gpt4omini} as the main LLM for agents. In ablation study, we use \textit{gpt-3.5-turbo-0125}, \textit{gpt-4o-2024-05-13}, \textit{Llama-3.1-70b-Instruct} and \textit{Llama-3.1-8b-Instruct} for comparison \citep{dubey2024llama}. To calculate sentence embeddings we used \textit{paraphrase-MiniLM-L6-v2} \citep{reimers-2019-sentence-bert}. For dialogue act labeling, we use \textit{Llama-3.1-8b-Instruct}. For all LLM inferences, we used \textit{temperature = 0}.

        MovieLens 100K dataset is used in recommendation system, which is under CC0: Public Domain license \cite{harper2015movielens}. Our use of the dataset is consistent with the intended use. MovieLens does not contain personally identifying info or offensive content.

\section{Details of Results}
    \begin{figure}[t]
        \centering
        \includegraphics[width=\linewidth]{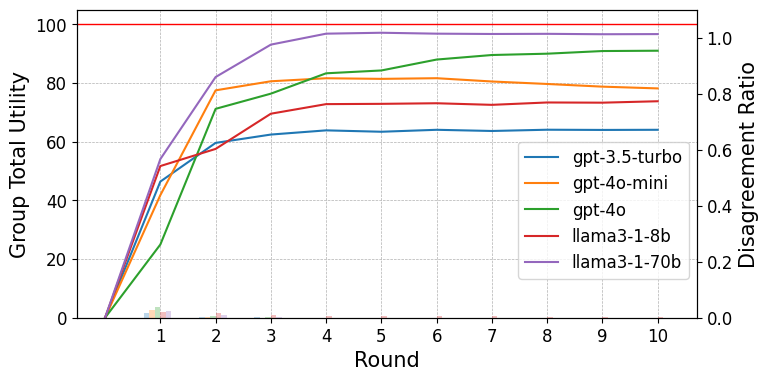}
        \caption{LLM comparison with group total utility in exchange economy, using 3 agent, asymmetric, majority voting setting. Line plots (left y-axis) show the group total utility achieved over rounds; bar plots (right y-axis) represent the ratio of cases where participants yet to reach an agreement until a certain round. The red horizontal line indicates the maximum achievable group total utility.}
        \label{fig:economics_model}
    \end{figure}

    \begin{figure}[t]
        \centering
        \includegraphics[width=\linewidth]{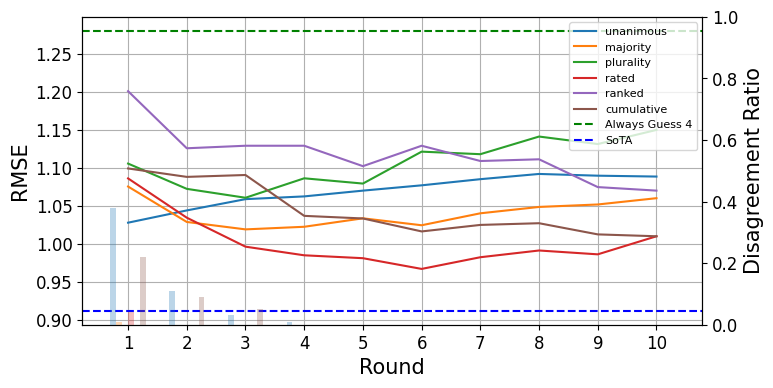}
        \caption{Comparison between voting rules in recommendation system environment. Line plots (left y-axis) show the comparison of RMSE achieved over rounds; bar plots (right y-axis) represent the ratio of cases where participants yet to reach an agreement until a certain round. The green horizontal line is baseline (\textit{Always Guess 4}), and the blue line is SoTA performance \cite{behera2023collaborative}.}
        \label{fig:recommendation_rmse}
    \end{figure}
    
    \subsection{Ablation Studies}
        \label{app:ablation}
        \begin{table*}[t]
\resizebox{\textwidth}{!}{%
\begin{tabular}{llllllllll}
\textbf{}  & \multicolumn{1}{c}{\textbf{\begin{tabular}[c]{@{}c@{}}GROUP\\TOTAL\\UTILITY@3\end{tabular}}} & \multicolumn{1}{c}{\textbf{\begin{tabular}[c]{@{}c@{}}GROUP\\TOTAL\\UTILITY@5\end{tabular}}} & \multicolumn{1}{c}{\textbf{\begin{tabular}[c]{@{}c@{}}GROUP\\TOTAL\\UTILITY@10\end{tabular}}} & \multicolumn{1}{c}{\textbf{AUC@3}} & \multicolumn{1}{c}{\textbf{AUC@5}} & \multicolumn{1}{c}{\textbf{AUC@10}} & \multicolumn{1}{c}{\textbf{RATIONALITY}} & \multicolumn{1}{c}{\textbf{FAIRNESS}} & \multicolumn{1}{c}{\textbf{RIGIDITY}} \\ 
\hline
\multicolumn{10}{c}{\textbf{Voting Rule: Majority, \# Agent: $\mathbf{K=3}$}}\\
gpt-3.5-turbo & \textbf{\textcolor{blue}{62.44 (2.24)}} & \textbf{\textcolor{blue}{63.40 (0.35)}} & \textbf{\textcolor{blue}{64.07 (0.61)}} & \textbf{\textcolor{blue}{56.15 (2.81)}} & \textbf{\textcolor{blue}{59.15 (1.69)}} & \textbf{\textcolor{blue}{61.56 (0.88)}} & 17.88 (2.11) & \textbf{\textcolor{red}{97.58 (0.91)}} & \textbf{\textcolor{blue}{47.00 (3.08)}} \\
gpt-4o-mini   & 80.61 (1.46) & 81.43 (1.32) & 78.18 (1.64) & 66.64 (3.07) & 72.59 (2.10) & 76.18 (1.49) & 25.89 (2.25) & \textbf{\textcolor{blue}{58.41 (6.14)}} & 70.00 (2.84) \\
gpt-4o        & 76.39 (4.26) & 84.29 (1.94) & 91.02 (1.15) & 57.51 (3.82) & 68.03 (2.34) & 78.97 (1.16) & \textbf{\textcolor{red}{38.16 (4.19)}} & 74.27 (3.44) & 73.00 (1.93) \\
llama3-1-8b   & 69.55 (5.87) & 72.91 (5.45) & 73.80 (5.39) & 59.60 (6.22) & 64.91 (5.55) & 69.07 (5.08) & \textbf{\textcolor{blue}{3.78 (0.71)}} & 77.66 (5.20) & \textbf{\textcolor{red}{86.00 (1.83)}} \\
llama3-1-70b  & \textbf{\textcolor{red}{93.10 (3.28)}} & \textbf{\textcolor{red}{97.15 (0.56)}} & \textbf{\textcolor{red}{96.69 (0.79)}} & \textbf{\textcolor{red}{76.41 (4.61)}} & \textbf{\textcolor{red}{84.64 (2.78)}} & \textbf{\textcolor{red}{90.69 (1.43)}} & 35.11 (4.03) & 87.66 (3.28) & 70.67 (2.67) \\
\hline
\multicolumn{10}{c}{\textbf{Voting Rule: Majority, Model: gpt-4o-mini}}\\
$K=3$ & \textbf{\textcolor{red}{79.88 (0.80)}} & \textbf{\textcolor{red}{81.33 (0.72)}} & \textbf{\textcolor{red}{79.61 (0.88)}} & \textbf{\textcolor{red}{64.08 (1.81)}} & \textbf{\textcolor{red}{70.91 (1.23)}} & \textbf{\textcolor{red}{75.67 (0.84)}} & \textbf{\textcolor{blue}{23.80 (0.94)}} & 64.00 (3.07) & \textbf{\textcolor{blue}{71.30 (1.32)}} \\
$K=4$ & \textbf{\textcolor{blue}{58.29 (2.39)}} & \textbf{\textcolor{blue}{63.10 (2.12)}} & \textbf{\textcolor{blue}{64.99 (2.12)}} & 43.64 (2.04) & \textbf{\textcolor{blue}{51.05 (1.87)}} & \textbf{\textcolor{blue}{57.66 (1.89)}} & 34.25 (1.39) & \textbf{\textcolor{red}{64.54 (3.14)}} & \textbf{\textcolor{red}{80.30 (0.96)}} \\
$K=5$ & 63.67 (2.72) & 72.42 (2.33) & 74.56 (2.18) & \textbf{\textcolor{blue}{40.23 (1.90)}} & 52.17 (1.88) & 63.13 (1.94) & \textbf{\textcolor{red}{36.26 (1.36)}} & \textbf{\textcolor{blue}{57.54 (3.28)}} & 73.50 (1.36)
\end{tabular}
}
\caption{Ablation studies on exchange economics environment. The reported numbers are an average of all simulations, and the numbers in parentheses are standard errors. The smallest and largest value in a category is colored in \textcolor{blue}{blue} and \textcolor{red}{red}.}
\label{tbl:economics_ablation}
\end{table*}
        The result of ablation studies is shown in Table \ref{tbl:economics_ablation}. We further compare the performance of multi-agent collaboration with different LLMs in Figure \ref{fig:economics_model}. For simplicity, we only compare $K=3$ agent system in majority voting setting with 30 collaborations. The results of \textit{gpt-3.5-turbo-0125}, \textit{gpt-4o-2024-05-13}, \textit{Llama-3.1-8b-Instruct} and \textit{Llama-3.1-70b-Instruct} in Table \ref{tbl:economics} show that multi-agent collaboration works with different LLMs. The stronger (larger) the model, the better performance observed in collaboration, both in end round performance and also efficiency in reaching to an agreement. Notably, the Rationality of the weaker model, \textit{gpt-3.5-turbo-0125} and \textit{Llama-3.1-8b-Instruct}, is only 17.88\% and 3.78\%, indicates that single agent will largely fail due to the complexity of the task. Nonetheless, the final round performance has reached 64\% and 80\%, shows the effectiveness of multi-agent collaboration with weak models.
        
        Scaling on MAS is evaluated with different number of agents. Due to the nature of decentralized collaboration, the more the agents, the harder to reach an agreement. Furthermore, the additional amount of historical context brought by increasing number of agents can also be a burden to the agent's performance. Comparing $K=3,4,5$ settings in Table \ref{tbl:economics}, the quality and efficiency of collaboration generally drops with the increasing number of agents. Exceptionally, performance in $K=4$ setting is worse than that of $K=5$. This is because majority voting is harder to achieve an agreement with, even number of participants. With 4 agents, a proposal needs at least 3 votes to get accepted; 5 agent setting also requires 3 votes, but they can ignore up to 2 agents' preferences or agreements.
    \subsection{Recommendation System}
        We show the detailed result between voting rules in the recommendation system environment in Table \ref{tbl:recommendation}.
        \begin{table*}[t]
\centering
\resizebox{\textwidth}{!}{%
\begin{tabular}{lllllllllll}
 & \textbf{MAE@1} & \textbf{MAE@2} & \textbf{MAE@3} & \textbf{MAE@4} & \textbf{MAE@5} & \textbf{MAE@6} & \textbf{MAE@7} & \textbf{MAE@8} & \textbf{MAE@9} & \textbf{MAE@10} \\ \hline
Unanimous & 0.85 & \textbf{\textcolor{red}{0.84}} & 0.84 & \textbf{\textcolor{red}{0.86}} & \textbf{\textcolor{red}{0.86}} & \textbf{\textcolor{red}{0.87}} & \textbf{\textcolor{red}{0.88}} & \textbf{\textcolor{red}{0.89}} & \textbf{\textcolor{red}{0.89}} & \textbf{\textcolor{red}{0.88}} \\
Majority & \textbf{\textcolor{blue}{0.81}} & 0.81 & 0.81 & 0.81 & 0.84 & 0.82 & 0.84 & 0.84 & 0.85 & 0.86 \\
Plurality & \textbf{\textcolor{blue}{0.81}} & 0.82 & 0.82 & 0.83 & 0.82 & 0.83 & 0.83 & 0.84 & 0.83 & 0.84 \\
Rated & 0.83 & \textbf{\textcolor{blue}{0.80}} & \textbf{\textcolor{blue}{0.77}} & \textbf{\textcolor{blue}{0.77}} & \textbf{\textcolor{blue}{0.77}} & \textbf{\textcolor{blue}{0.75}} & \textbf{\textcolor{blue}{0.76}} & \textbf{\textcolor{blue}{0.77}} & \textbf{\textcolor{blue}{0.76}} & 0.79 \\
Ranked & \textbf{\textcolor{red}{0.90}} & 0.83 & \textbf{\textcolor{red}{0.86}} & \textbf{\textcolor{red}{0.86}} & 0.84 & \textbf{\textcolor{red}{0.87}} & 0.86 & 0.86 & 0.84 & 0.84 \\
Cumulative & 0.83 & 0.83 & 0.83 & 0.78 & 0.78 & 0.77 & 0.78 & 0.79 & 0.77 & \textbf{\textcolor{blue}{0.76}} \\
\\
 & \textbf{RMSE@1} & \textbf{RMSE@2} & \textbf{RMSE@3} & \textbf{RMSE@4} & \textbf{RMSE@5} & \textbf{RMSE@6} & \textbf{RMSE@7} & \textbf{RMSE@8} & \textbf{RMSE@9} & \textbf{RMSE@10} \\
 \hline
 Unanimous & \textbf{\textcolor{blue}{1.03}} & 1.04 & 1.06 & 1.06 & 1.07 & 1.08 & 1.09 & 1.09 & 1.09 & 1.09 \\
Majority & 1.08 & \textbf{\textcolor{blue}{1.03}} & 1.02 & 1.02 & 1.03 & 1.02 & 1.04 & 1.05 & 1.05 & 1.06 \\
Plurality & 1.11 & 1.07 & 1.06 & 1.09 & 1.08 & 1.12 & \textbf{\textcolor{red}{1.12}} & \textbf{\textcolor{red}{1.14}} & \textbf{\textcolor{red}{1.13}} & \textbf{\textcolor{red}{1.15}} \\
Rated & 1.09 & \textbf{\textcolor{blue}{1.03}} & \textbf{\textcolor{blue}{1.00}} & \textbf{\textcolor{blue}{0.98}} & \textbf{\textcolor{blue}{0.98}} & \textbf{\textcolor{blue}{0.97}} & \textbf{\textcolor{blue}{0.98}} & \textbf{\textcolor{blue}{0.99}} & \textbf{\textcolor{blue}{0.99}} & \textbf{\textcolor{blue}{1.01}} \\
Ranked & \textbf{\textcolor{red}{1.20}} & \textbf{\textcolor{red}{1.13}} & \textbf{\textcolor{red}{1.13}} & \textbf{\textcolor{red}{1.13}} & \textbf{\textcolor{red}{1.10}} & \textbf{\textcolor{red}{1.13}} & 1.11 & 1.11 & 1.07 & 1.07 \\
Cumulative & 1.10 & 1.09 & 1.09 & 1.04 & 1.03 & 1.02 & 1.02 & 1.03 & 1.01 & \textbf{\textcolor{blue}{1.01}} \\
\end{tabular}
}
\caption{Recommendation system environment results comparison between voting rules. The smallest and largest value in a category is colored in \textcolor{blue}{blue} and \textcolor{red}{red}. Experiments are made with gpt-4o-mini.}
\label{tbl:recommendation}
\end{table*}

    \subsection{Cross-Agent Communication Analysis}
            \label{app:language_analysis_result}
            In this section, we show how language features are presented in different environments and voting rules. Table \ref{tbl:heatmap} is heatmaps for average message length, message complexity, and information distance observed in agent conversation. Message length increases over time in both environments, with shorter messages in early rounds and longer ones later. The recommendation system generally has longer messages due to its need for detailed explanations. Message complexity in the exchange economy starts low, peaks early, dips, and then gradually increases toward the final round, while the recommendation system shows consistently higher and increasing complexity throughout. Information distance steadily decreases in both environments, indicating a convergence of topics and less novel information as the discussions progress.
            \begin{table*}[t]
\centering
\tiny
\resizebox{\textwidth}{!}{%
\begin{tabular}{llllllllllllllllllllll}
\multicolumn{1}{l}{}& \multicolumn{10}{c}{\textbf{EXCHANGE ECONOMY}} & \multicolumn{1}{l}{}&\multicolumn{10}{c}{\textbf{RECOMMENDATION SYSTEM}}\\
\textbf{} & \textbf{1} & \textbf{2} & \textbf{3} & \textbf{4} & \textbf{5} & \textbf{6} & \textbf{7} & \textbf{8} & \textbf{9} & \textbf{10} & &\textbf{1} & \textbf{2} & \textbf{3} & \textbf{4} & \textbf{5} & \textbf{6} & \textbf{7} & \textbf{8} & \textbf{9} & \textbf{10}\\
\hline
\multicolumn{22}{c}{\textbf{AVERAGE MESSAGE LENGTH}}\\
Unanimous & \cellcolor[RGB]{232,135,126}46 & \cellcolor[RGB]{240,179,174}62 & \cellcolor[RGB]{251,238,237}83 & \cellcolor[RGB]{252,243,243}85 & 89 & \cellcolor[RGB]{150,212,181}94 & \cellcolor[RGB]{108,195,152}96 & \cellcolor[RGB]{108,195,152}96 & \cellcolor[RGB]{87,187,138}97 & \cellcolor[RGB]{150,212,181}94  & & \cellcolor[RGB]{230,127,119}68 & \cellcolor[RGB]{243,192,188}85 & \cellcolor[RGB]{249,226,224}94 & \cellcolor[RGB]{251,234,232}96 & \cellcolor[RGB]{253,249,248}100 & \cellcolor[RGB]{254,253,252}101 & \cellcolor[RGB]{209,236,223}103 & \cellcolor[RGB]{178,224,201}104 & \cellcolor[RGB]{148,211,180}105 & \cellcolor[RGB]{178,224,201}104  \\

Majority & \cellcolor[RGB]{230,126,117}43 & \cellcolor[RGB]{240,179,174}62 & \cellcolor[RGB]{251,235,234}82 & \cellcolor[RGB]{252,241,240}84 & \cellcolor[RGB]{253,249,249}87 & \cellcolor[RGB]{213,238,225}91 & \cellcolor[RGB]{213,238,225}91 & \cellcolor[RGB]{171,221,196}93 & \cellcolor[RGB]{213,238,225}91 & \cellcolor[RGB]{213,238,225}91  & & \cellcolor[RGB]{230,124,115}67 & \cellcolor[RGB]{243,192,188}85 & \cellcolor[RGB]{248,222,220}93 & \cellcolor[RGB]{253,245,244}99 & \cellcolor[RGB]{254,253,252}101 & \cellcolor[RGB]{209,236,223}103 & \cellcolor[RGB]{148,211,180}105 & \cellcolor[RGB]{148,211,180}105 & \cellcolor[RGB]{117,199,159}106 & \cellcolor[RGB]{148,211,180}105  \\

Plurality & \cellcolor[RGB]{230,124,115}42 & \cellcolor[RGB]{239,171,165}59 & \cellcolor[RGB]{250,232,231}81 & \cellcolor[RGB]{251,235,234}82 & \cellcolor[RGB]{252,241,240}84 & \cellcolor[RGB]{234,246,240}90 & 89 & \cellcolor[RGB]{213,238,225}91 & \cellcolor[RGB]{171,221,196}93 & \cellcolor[RGB]{234,246,240}90  & & \cellcolor[RGB]{230,124,115}67 & \cellcolor[RGB]{242,188,183}84 & \cellcolor[RGB]{249,226,224}94 & \cellcolor[RGB]{253,245,244}99 & \cellcolor[RGB]{254,253,252}101 & \cellcolor[RGB]{178,224,201}104 & \cellcolor[RGB]{148,211,180}105 & \cellcolor[RGB]{117,199,159}106 & \cellcolor[RGB]{87,187,138}107 & \cellcolor[RGB]{117,199,159}106  \\

Rated & \cellcolor[RGB]{230,124,115}42 & \cellcolor[RGB]{239,174,168}60 & \cellcolor[RGB]{251,235,234}82 & \cellcolor[RGB]{251,238,237}83 & 89 & \cellcolor[RGB]{213,238,225}91 & \cellcolor[RGB]{192,229,211}92 & \cellcolor[RGB]{150,212,181}94 & \cellcolor[RGB]{150,212,181}94 & \cellcolor[RGB]{171,221,196}93  & & \cellcolor[RGB]{230,124,115}67 & \cellcolor[RGB]{243,196,192}86 & \cellcolor[RGB]{251,234,232}96 & \cellcolor[RGB]{253,249,248}100 & \cellcolor[RGB]{254,253,252}101 & \cellcolor[RGB]{209,236,223}103 & \cellcolor[RGB]{178,224,201}104 & \cellcolor[RGB]{117,199,159}106 & \cellcolor[RGB]{87,187,138}107 & \cellcolor[RGB]{148,211,180}105  \\

Ranked & \cellcolor[RGB]{230,126,117}43 & \cellcolor[RGB]{240,176,171}61 & \cellcolor[RGB]{250,229,228}80 & \cellcolor[RGB]{251,238,237}83 & \cellcolor[RGB]{253,246,246}86 & \cellcolor[RGB]{234,246,240}90 & \cellcolor[RGB]{213,238,225}91 & \cellcolor[RGB]{234,246,240}90 & \cellcolor[RGB]{213,238,225}91 & \cellcolor[RGB]{234,246,240}90  & & \cellcolor[RGB]{230,124,115}67 & \cellcolor[RGB]{242,188,183}84 & \cellcolor[RGB]{252,241,240}98 & \cellcolor[RGB]{253,245,244}99 & \cellcolor[RGB]{239,248,244}102 & \cellcolor[RGB]{239,248,244}102 & \cellcolor[RGB]{209,236,223}103 & \cellcolor[RGB]{178,224,201}104 & \cellcolor[RGB]{148,211,180}105 & \cellcolor[RGB]{178,224,201}104  \\

Cumulative & \cellcolor[RGB]{230,124,115}42 & \cellcolor[RGB]{240,176,171}61 & \cellcolor[RGB]{250,232,231}81 & \cellcolor[RGB]{251,238,237}83 & \cellcolor[RGB]{253,246,246}86 & \cellcolor[RGB]{213,238,225}91 & \cellcolor[RGB]{213,238,225}91 & \cellcolor[RGB]{192,229,211}92 & \cellcolor[RGB]{171,221,196}93 & \cellcolor[RGB]{213,238,225}91  & & \cellcolor[RGB]{230,127,119}68 & \cellcolor[RGB]{243,192,188}85 & \cellcolor[RGB]{250,230,228}95 & \cellcolor[RGB]{252,241,240}98 & \cellcolor[RGB]{253,249,248}100 & \cellcolor[RGB]{239,248,244}102 & \cellcolor[RGB]{209,236,223}103 & \cellcolor[RGB]{178,224,201}104 & \cellcolor[RGB]{178,224,201}104 & \cellcolor[RGB]{209,236,223}103  \\
\\
\multicolumn{22}{c}{\textbf{AVERAGE MESSAGE COMPLEXITY}}\\
Unanimous & \cellcolor[RGB]{239,171,165}7.1 & \cellcolor[RGB]{143,209,177}10.2 & \cellcolor[RGB]{254,251,251}9.3 & \cellcolor[RGB]{252,240,239}9.0 & \cellcolor[RGB]{252,240,239}9.0 & \cellcolor[RGB]{252,244,243}9.1 & \cellcolor[RGB]{253,247,247}9.2 & \cellcolor[RGB]{241,249,245}9.5 & \cellcolor[RGB]{241,249,245}9.5 & \cellcolor[RGB]{227,243,235}9.6 & & \cellcolor[RGB]{231,132,123}10.4 & \cellcolor[RGB]{236,156,149}10.7 & \cellcolor[RGB]{248,222,219}11.5 & \cellcolor[RGB]{251,238,237}11.7 & 11.9 & \cellcolor[RGB]{213,238,225}12.0 & \cellcolor[RGB]{213,238,225}12.0 & \cellcolor[RGB]{171,221,196}12.1 & \cellcolor[RGB]{87,187,138}12.3 & \cellcolor[RGB]{129,204,167}12.2  \\

Majority & \cellcolor[RGB]{242,189,185}7.6 & \cellcolor[RGB]{114,198,157}10.4 & \cellcolor[RGB]{241,249,245}9.5 & 9.4 & \cellcolor[RGB]{253,247,247}9.2 & \cellcolor[RGB]{253,247,247}9.2 & 9.4 & \cellcolor[RGB]{241,249,245}9.5 & \cellcolor[RGB]{227,243,235}9.6 & \cellcolor[RGB]{213,238,225}9.7 & & \cellcolor[RGB]{230,124,115}10.3 & \cellcolor[RGB]{236,156,149}10.7 & \cellcolor[RGB]{248,222,219}11.5 & \cellcolor[RGB]{251,238,237}11.7 & 11.9 & \cellcolor[RGB]{213,238,225}12.0 & \cellcolor[RGB]{171,221,196}12.1 & \cellcolor[RGB]{171,221,196}12.1 & \cellcolor[RGB]{171,221,196}12.1 & \cellcolor[RGB]{171,221,196}12.1  \\

Plurality & \cellcolor[RGB]{236,160,153}6.8 & \cellcolor[RGB]{100,192,147}10.5 & 9.4 & \cellcolor[RGB]{252,244,243}9.1 & \cellcolor[RGB]{251,236,235}8.9 & \cellcolor[RGB]{252,244,243}9.1 & \cellcolor[RGB]{253,247,247}9.2 & \cellcolor[RGB]{254,251,251}9.3 & 9.4 & 9.4 & & \cellcolor[RGB]{230,124,115}10.3 & \cellcolor[RGB]{234,148,141}10.6 & \cellcolor[RGB]{248,222,219}11.5 & \cellcolor[RGB]{251,238,237}11.7 & 11.9 & \cellcolor[RGB]{213,238,225}12.0 & \cellcolor[RGB]{213,238,225}12.0 & \cellcolor[RGB]{171,221,196}12.1 & \cellcolor[RGB]{171,221,196}12.1 & \cellcolor[RGB]{171,221,196}12.1  \\

Rated & \cellcolor[RGB]{230,124,115}5.8 & \cellcolor[RGB]{114,198,157}10.4 & 9.4 & \cellcolor[RGB]{252,244,243}9.1 & \cellcolor[RGB]{252,244,243}9.1 & \cellcolor[RGB]{253,247,247}9.2 & 9.4 & \cellcolor[RGB]{254,251,251}9.3 & \cellcolor[RGB]{241,249,245}9.5 & \cellcolor[RGB]{227,243,235}9.6 & & \cellcolor[RGB]{230,124,115}10.3 & \cellcolor[RGB]{236,156,149}10.7 & \cellcolor[RGB]{250,230,228}11.6 & \cellcolor[RGB]{253,246,246}11.8 & \cellcolor[RGB]{213,238,225}12.0 & \cellcolor[RGB]{213,238,225}12.0 & \cellcolor[RGB]{129,204,167}12.2 & \cellcolor[RGB]{129,204,167}12.2 & \cellcolor[RGB]{87,187,138}12.3 & \cellcolor[RGB]{87,187,138}12.3  \\

Ranked & \cellcolor[RGB]{236,156,150}6.7 & \cellcolor[RGB]{143,209,177}10.2 & \cellcolor[RGB]{241,249,245}9.5 & \cellcolor[RGB]{253,247,247}9.2 & \cellcolor[RGB]{252,244,243}9.1 & \cellcolor[RGB]{252,244,243}9.1 & 9.4 & 9.4 & \cellcolor[RGB]{241,249,245}9.5 & \cellcolor[RGB]{227,243,235}9.6 & & \cellcolor[RGB]{230,124,115}10.3 & \cellcolor[RGB]{236,156,149}10.7 & \cellcolor[RGB]{250,230,228}11.6 & \cellcolor[RGB]{251,238,237}11.7 & 11.9 & \cellcolor[RGB]{171,221,196}12.1 & \cellcolor[RGB]{171,221,196}12.1 & \cellcolor[RGB]{129,204,167}12.2 & \cellcolor[RGB]{129,204,167}12.2 & \cellcolor[RGB]{129,204,167}12.2  \\

Cumulative & \cellcolor[RGB]{237,164,157}6.9 & \cellcolor[RGB]{87,187,138}10.6 & 9.4 & \cellcolor[RGB]{254,251,251}9.3 & \cellcolor[RGB]{252,244,243}9.1 & \cellcolor[RGB]{253,247,247}9.2 & 9.4 & \cellcolor[RGB]{241,249,245}9.5 & \cellcolor[RGB]{227,243,235}9.6 & \cellcolor[RGB]{213,238,225}9.7 & & \cellcolor[RGB]{230,124,115}10.3 & \cellcolor[RGB]{234,148,141}10.6 & \cellcolor[RGB]{248,222,219}11.5 & \cellcolor[RGB]{251,238,237}11.7 & \cellcolor[RGB]{251,238,237}11.7 & 11.9 & 11.9 & \cellcolor[RGB]{213,238,225}12.0 & \cellcolor[RGB]{171,221,196}12.1 & \cellcolor[RGB]{171,221,196}12.1  \\
\\
\multicolumn{22}{c}{\textbf{AVERAGE INFORMATION DISTANCE}}\\

Unanimous & \cellcolor[RGB]{192,192,192} &\cellcolor[RGB]{163,217,190}0.23 & \cellcolor[RGB]{219,240,230}0.16 & \cellcolor[RGB]{241,249,245}0.13 & \cellcolor[RGB]{238,248,243}0.13 & \cellcolor[RGB]{250,253,252}0.12 & \cellcolor[RGB]{252,242,241}0.11 & \cellcolor[RGB]{245,205,202}0.11 & \cellcolor[RGB]{230,124,115}0.10 & \cellcolor[RGB]{237,162,156}0.10  & & \cellcolor[RGB]{192,192,192}& \cellcolor[RGB]{100,192,147}0.20 & \cellcolor[RGB]{194,230,213}0.13 & \cellcolor[RGB]{234,246,240}0.11 & \cellcolor[RGB]{252,253,253}0.09 & \cellcolor[RGB]{251,237,235}0.09 & \cellcolor[RGB]{235,151,144}0.08 & \cellcolor[RGB]{234,148,141}0.08 & \cellcolor[RGB]{230,124,115}0.08 & \cellcolor[RGB]{246,210,207}0.09  \\

Majority & \cellcolor[RGB]{192,192,192}&\cellcolor[RGB]{141,208,175}0.26 & \cellcolor[RGB]{214,238,227}0.17 & \cellcolor[RGB]{242,249,246}0.13 & \cellcolor[RGB]{246,251,248}0.12 & \cellcolor[RGB]{253,254,254}0.12 & \cellcolor[RGB]{240,177,172}0.11 & \cellcolor[RGB]{238,166,160}0.10 & \cellcolor[RGB]{238,170,164}0.10 & \cellcolor[RGB]{245,203,199}0.11  & & \cellcolor[RGB]{192,192,192}& \cellcolor[RGB]{104,193,150}0.19 & \cellcolor[RGB]{196,231,214}0.13 & \cellcolor[RGB]{229,244,237}0.11 & \cellcolor[RGB]{250,253,251}0.10 & \cellcolor[RGB]{248,252,250}0.10 & \cellcolor[RGB]{250,230,228}0.09 & \cellcolor[RGB]{243,195,191}0.09 & \cellcolor[RGB]{235,154,147}0.08 & \cellcolor[RGB]{252,239,238}0.09  \\

Plurality & \cellcolor[RGB]{192,192,192}&\cellcolor[RGB]{124,202,164}0.28 & \cellcolor[RGB]{220,241,231}0.16 & \cellcolor[RGB]{238,248,243}0.13 & \cellcolor[RGB]{238,248,243}0.14 & \cellcolor[RGB]{253,254,254}0.12 & \cellcolor[RGB]{244,201,197}0.11 & \cellcolor[RGB]{240,178,173}0.11 & \cellcolor[RGB]{238,168,162}0.10 & \cellcolor[RGB]{245,203,200}0.11  & & \cellcolor[RGB]{192,192,192}& \cellcolor[RGB]{87,187,138}0.21 & \cellcolor[RGB]{191,229,210}0.14 & \cellcolor[RGB]{233,246,240}0.11 & \cellcolor[RGB]{251,253,252}0.10 & \cellcolor[RGB]{254,254,254}0.09 & \cellcolor[RGB]{245,205,202}0.09 & \cellcolor[RGB]{243,193,189}0.09 & \cellcolor[RGB]{236,155,148}0.08 & \cellcolor[RGB]{246,210,206}0.09  \\

Rated & \cellcolor[RGB]{192,192,192}&\cellcolor[RGB]{87,187,138}0.33 & \cellcolor[RGB]{219,240,230}0.16 & \cellcolor[RGB]{244,250,247}0.13 & \cellcolor[RGB]{248,252,250}0.12 & \cellcolor[RGB]{239,171,165}0.10 & \cellcolor[RGB]{242,187,182}0.11 & \cellcolor[RGB]{239,174,169}0.10 & \cellcolor[RGB]{230,126,118}0.10 & \cellcolor[RGB]{241,186,181}0.11  & & \cellcolor[RGB]{192,192,192}& \cellcolor[RGB]{93,189,142}0.20 & \cellcolor[RGB]{200,233,217}0.13 & \cellcolor[RGB]{233,246,239}0.11 & \cellcolor[RGB]{247,251,249}0.10 & \cellcolor[RGB]{251,253,252}0.10 & \cellcolor[RGB]{244,201,197}0.09 & \cellcolor[RGB]{237,163,157}0.09 & \cellcolor[RGB]{233,140,132}0.08 & \cellcolor[RGB]{247,213,210}0.09  \\

Ranked & \cellcolor[RGB]{192,192,192}&\cellcolor[RGB]{128,203,166}0.28 & \cellcolor[RGB]{209,236,223}0.17 & \cellcolor[RGB]{246,251,248}0.13 & \cellcolor[RGB]{250,253,251}0.12 & \cellcolor[RGB]{245,203,200}0.11 & \cellcolor[RGB]{230,128,119}0.10 & \cellcolor[RGB]{230,127,119}0.10 & \cellcolor[RGB]{234,145,137}0.10 & \cellcolor[RGB]{236,160,153}0.10  & & \cellcolor[RGB]{192,192,192}& \cellcolor[RGB]{92,189,142}0.20 & \cellcolor[RGB]{193,230,212}0.13 & \cellcolor[RGB]{235,247,241}0.11 & \cellcolor[RGB]{252,253,253}0.09 & \cellcolor[RGB]{251,234,232}0.09 & \cellcolor[RGB]{248,220,218}0.09 & \cellcolor[RGB]{247,216,213}0.09 & \cellcolor[RGB]{232,138,130}0.08 & \cellcolor[RGB]{249,225,223}0.09  \\

Cumulative & \cellcolor[RGB]{192,192,192}&\cellcolor[RGB]{113,197,156}0.30 & \cellcolor[RGB]{207,235,222}0.18 & \cellcolor[RGB]{243,250,247}0.13 & \cellcolor[RGB]{248,252,250}0.12 & \cellcolor[RGB]{243,196,192}0.11 & \cellcolor[RGB]{242,188,183}0.11 & \cellcolor[RGB]{243,194,189}0.11 & \cellcolor[RGB]{230,124,115}0.10 & \cellcolor[RGB]{232,135,127}0.10  & & \cellcolor[RGB]{192,192,192}& \cellcolor[RGB]{98,191,146}0.20 & \cellcolor[RGB]{195,230,213}0.13 & \cellcolor[RGB]{240,249,245}0.10 & \cellcolor[RGB]{253,254,254}0.09 & \cellcolor[RGB]{254,252,252}0.09 & \cellcolor[RGB]{239,171,166}0.09 & \cellcolor[RGB]{238,169,163}0.09 & \cellcolor[RGB]{233,143,135}0.08 & \cellcolor[RGB]{249,228,226}0.09  \\

\end{tabular}
}
\caption{Heatmaps of average message length, message complexity, and information distance in each round. Analysis is made in 3 agents, gpt-4o-mini setting. Color gradient is calculated with green as maximum, red as minimum, and white as median value in each table.}
\label{tbl:heatmap}
\end{table*}

            The dialogue act annotation results are presented in Table \ref{tbl:dialogue_act}. The dialogue act annotation results in both environments show that \textit{Inform} and \textit{Request} acts dominate, indicating frequent information sharing and input requests critical for task progression. The \textit{Confirm} act rises sharply after round 3 in the recommendation system and more gradually in the exchange economy, showing validation is more prominent in recommendation tasks. \textit{Summarize} acts increase slightly in later rounds, consolidating information, while \textit{Evaluate} acts remain consistently high, reflecting ongoing assessment. \textit{Propose} acts surge early in the exchange economy and peak later in the recommendation system, suggesting early proposals are vital in negotiations. Acts like \textit{Compromise}, \textit{Defend}, \textit{Accept}, and \textit{Decline} are rare, reflecting minimal adversarial behavior, while \textit{Others} remain low, indicating the high quality of dialogue act definitions.

\begin{figure}[t]
    \centering
    \subfigure[Dialogue act transition graph for exchange economy.]{
        \includegraphics[width=\linewidth]{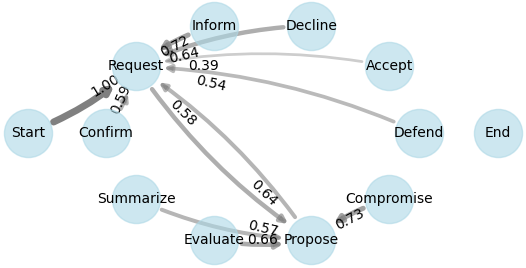}
        \label{fig:da_transition_econ}
    }
    \hspace{0.02\textwidth} 
    \subfigure[Dialogue act transition graph for recommendation system.]{
        \includegraphics[width=\linewidth]{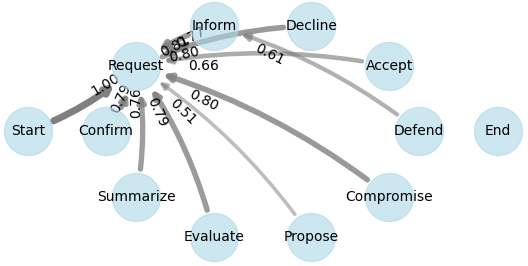}
        \label{fig:da_transition_recom}
    }
    \caption{Dialogue act transition graph for exchange economy and recommendation system environments. Only the most probable outgoing edge is presented. Self-loops are excluded.}
\end{figure}
\begin{figure*}[t]
                \centering
                \includegraphics[width=\textwidth]{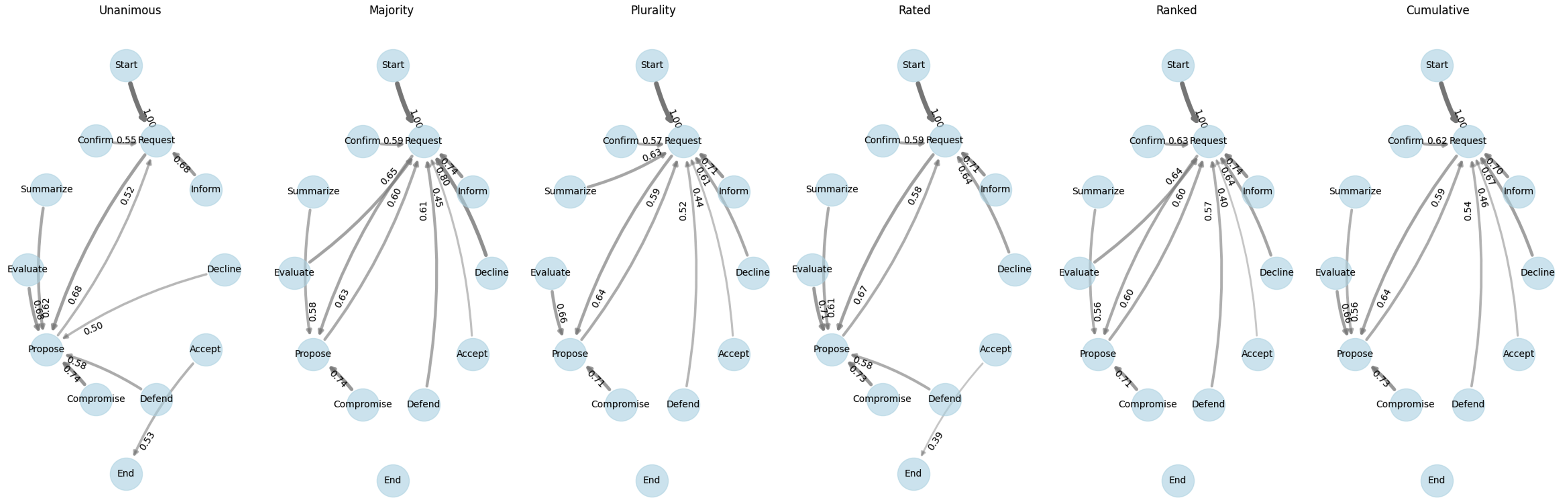}
                \caption{Dialogue act transition graph for different voting rules in exchange economics environment.}
                \label{fig:da_transition_econ_sc}
            \end{figure*}
\begin{figure*}[t]
                \centering
                \includegraphics[width=\textwidth]{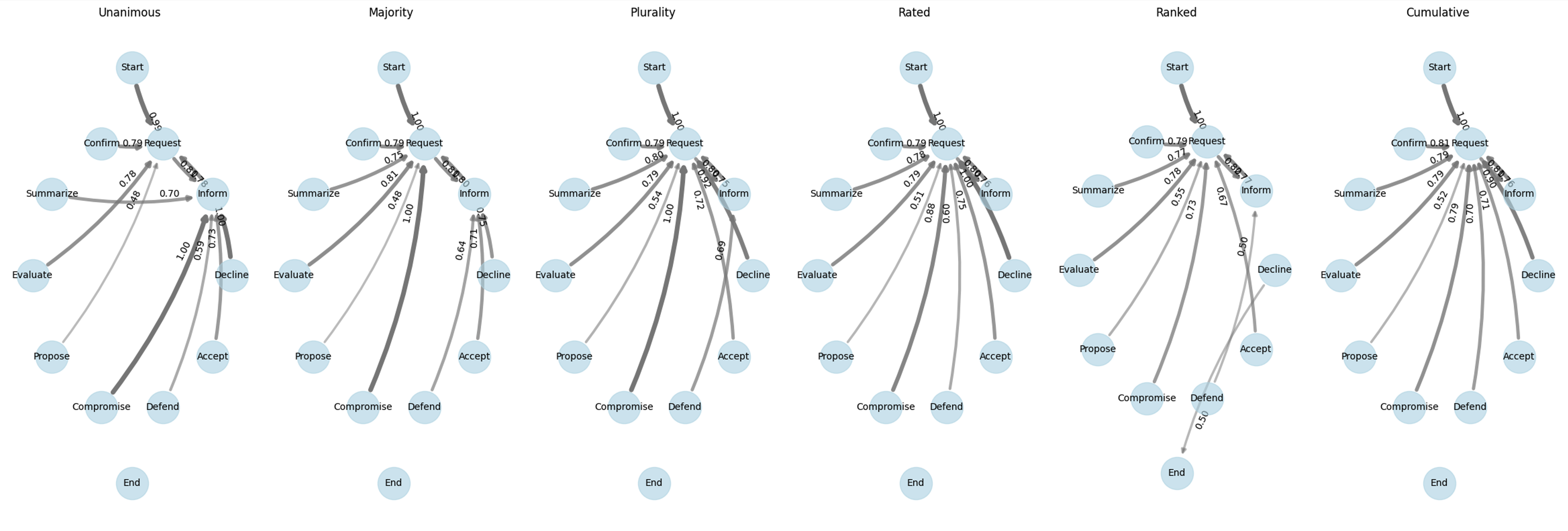}
                \caption{Dialogue act transition graph for different voting rules in recommendation system environment.}
                \label{fig:da_transition_recom_sc}
            \end{figure*}

            The most probable transition graphs in both environments are shown in Figure \ref{fig:da_transition_econ} and \ref{fig:da_transition_recom}. The breakdown of voting rules are displayed in Figure \ref{fig:da_transition_econ_sc} and \ref{fig:da_transition_recom_sc}. The transitions reveal a structured progression from requesting information to proposing and evaluating solutions, followed by resolution or negotiation. In the exchange economy, collaboration centers around a loop between \textit{Request} and \textit{Propose}, reflecting the cooperative decision-making process. Key transitions include \textit{Inform} to \textit{Request}, \textit{Confirm} to \textit{Request}, and \textit{Compromise} to \textit{Propose}, showing how agents share knowledge, confirm information, and adopt ideas. \textit{Accept} transitions to \textit{End}, signaling negotiation closure. In contrast, the recommendation system centers on \textit{Request}, indicating a more cooperative, exploratory process with incomplete information.

            \label{app:dialogue_act_statistics}
            \begin{table*}[t]
\centering
\tiny
\resizebox{\textwidth}{!}{%
\begin{tabular}{llllllllllllllllllllll}
\multicolumn{1}{l}{}& \multicolumn{10}{c}{\textbf{EXCHANGE ECONOMY}} & \multicolumn{1}{l}{}&\multicolumn{10}{c}{\textbf{RECOMMENDATION SYSTEM}}\\
\textbf{} & \textbf{1} & \textbf{2} & \textbf{3} & \textbf{4} & \textbf{5} & \textbf{6} & \textbf{7} & \textbf{8} & \textbf{9} & \textbf{10} & &\textbf{1} & \textbf{2} & \textbf{3} & \textbf{4} & \textbf{5} & \textbf{6} & \textbf{7} & \textbf{8} & \textbf{9} & \textbf{10}\\
\hline
\multicolumn{22}{c}{\textbf{UNANIMOUS}}\\
Inform & \cellcolor[RGB]{87,187,138}1.00 & \cellcolor[RGB]{88,187,139}0.99 & \cellcolor[RGB]{107,195,152}0.88 & \cellcolor[RGB]{155,214,185}0.59 & \cellcolor[RGB]{139,208,174}0.69 & \cellcolor[RGB]{140,208,175}0.68 & \cellcolor[RGB]{150,212,182}0.62 & \cellcolor[RGB]{144,210,177}0.66 & \cellcolor[RGB]{150,212,182}0.62 & \cellcolor[RGB]{164,218,191}0.54  & & \cellcolor[RGB]{87,187,138}1.00 & \cellcolor[RGB]{90,188,140}0.98 & \cellcolor[RGB]{88,187,139}0.99 & \cellcolor[RGB]{87,187,138}1.00 & \cellcolor[RGB]{87,187,138}1.00 & \cellcolor[RGB]{87,187,138}1.00 & \cellcolor[RGB]{88,187,139}0.99 & \cellcolor[RGB]{87,187,138}1.00 & \cellcolor[RGB]{87,187,138}1.00 & \cellcolor[RGB]{87,187,138}1.00  \\
Request & \cellcolor[RGB]{87,187,138}1.00 & \cellcolor[RGB]{87,187,138}1.00 & \cellcolor[RGB]{93,189,142}0.96 & \cellcolor[RGB]{97,191,145}0.94 & \cellcolor[RGB]{105,194,150}0.89 & \cellcolor[RGB]{102,193,148}0.91 & \cellcolor[RGB]{105,194,150}0.89 & \cellcolor[RGB]{105,194,150}0.89 & \cellcolor[RGB]{112,197,155}0.85 & \cellcolor[RGB]{117,199,159}0.82  & & \cellcolor[RGB]{87,187,138}1.00 & \cellcolor[RGB]{87,187,138}1.00 & \cellcolor[RGB]{87,187,138}1.00 & \cellcolor[RGB]{87,187,138}1.00 & \cellcolor[RGB]{87,187,138}1.00 & \cellcolor[RGB]{87,187,138}1.00 & \cellcolor[RGB]{88,187,139}0.99 & \cellcolor[RGB]{87,187,138}1.00 & \cellcolor[RGB]{87,187,138}1.00 & \cellcolor[RGB]{90,188,140}0.98  \\
Confirm & \cellcolor[RGB]{241,249,245}0.08 & \cellcolor[RGB]{197,231,215}0.34 & \cellcolor[RGB]{207,235,222}0.28 & \cellcolor[RGB]{226,243,235}0.17 & \cellcolor[RGB]{209,236,223}0.27 & \cellcolor[RGB]{207,235,222}0.28 & \cellcolor[RGB]{199,232,216}0.33 & \cellcolor[RGB]{192,229,211}0.37 & \cellcolor[RGB]{186,227,207}0.41 & \cellcolor[RGB]{165,218,192}0.53  & & \cellcolor[RGB]{206,235,221}0.29 & \cellcolor[RGB]{172,221,197}0.49 & \cellcolor[RGB]{149,212,181}0.63 & \cellcolor[RGB]{144,210,177}0.66 & \cellcolor[RGB]{122,201,162}0.79 & \cellcolor[RGB]{140,208,175}0.68 & \cellcolor[RGB]{130,204,168}0.74 & \cellcolor[RGB]{139,208,174}0.69 & \cellcolor[RGB]{134,206,170}0.72 & \cellcolor[RGB]{147,211,180}0.64  \\
Summarize & \cellcolor[RGB]{228,244,236}0.16 & \cellcolor[RGB]{251,253,252}0.02 & \cellcolor[RGB]{229,244,237}0.15 & \cellcolor[RGB]{226,243,235}0.17 & \cellcolor[RGB]{228,244,236}0.16 & \cellcolor[RGB]{223,242,232}0.19 & \cellcolor[RGB]{233,246,239}0.13 & \cellcolor[RGB]{233,246,239}0.13 & \cellcolor[RGB]{224,242,233}0.18 & \cellcolor[RGB]{201,233,217}0.32  & & \cellcolor[RGB]{243,250,246}0.07 & \cellcolor[RGB]{236,247,242}0.11 & \cellcolor[RGB]{241,249,245}0.08 & \cellcolor[RGB]{228,244,236}0.16 & \cellcolor[RGB]{228,244,236}0.16 & \cellcolor[RGB]{224,242,233}0.18 & \cellcolor[RGB]{228,244,236}0.16 & \cellcolor[RGB]{228,244,236}0.16 & \cellcolor[RGB]{223,242,232}0.19 & \cellcolor[RGB]{214,238,226}0.24  \\
Evaluate & \cellcolor[RGB]{90,188,140}0.98 & \cellcolor[RGB]{102,193,148}0.91 & \cellcolor[RGB]{157,215,187}0.58 & \cellcolor[RGB]{120,200,161}0.80 & \cellcolor[RGB]{103,193,149}0.90 & \cellcolor[RGB]{97,191,145}0.94 & \cellcolor[RGB]{102,193,148}0.91 & \cellcolor[RGB]{98,191,146}0.93 & \cellcolor[RGB]{102,193,148}0.91 & \cellcolor[RGB]{100,192,147}0.92  & & \cellcolor[RGB]{253,254,253}0.01 & \cellcolor[RGB]{157,215,187}0.58 & \cellcolor[RGB]{157,215,187}0.58 & \cellcolor[RGB]{150,212,182}0.62 & \cellcolor[RGB]{154,214,184}0.60 & \cellcolor[RGB]{155,214,185}0.59 & \cellcolor[RGB]{147,211,180}0.64 & \cellcolor[RGB]{150,212,182}0.62 & \cellcolor[RGB]{147,211,180}0.64 & \cellcolor[RGB]{142,209,176}0.67  \\
Propose & \cellcolor[RGB]{244,250,247}0.06 & \cellcolor[RGB]{176,223,200}0.47 & \cellcolor[RGB]{88,187,139}0.99 & \cellcolor[RGB]{88,187,139}0.99 & \cellcolor[RGB]{88,187,139}0.99 & \cellcolor[RGB]{87,187,138}1.00 & \cellcolor[RGB]{90,188,140}0.98 & \cellcolor[RGB]{90,188,140}0.98 & \cellcolor[RGB]{100,192,147}0.92 & \cellcolor[RGB]{92,189,141}0.97  & & \cellcolor[RGB]{238,248,243}0.10 & \cellcolor[RGB]{243,250,246}0.07 & \cellcolor[RGB]{221,241,231}0.20 & \cellcolor[RGB]{221,241,231}0.20 & \cellcolor[RGB]{229,244,237}0.15 & \cellcolor[RGB]{219,240,230}0.21 & \cellcolor[RGB]{204,234,219}0.30 & \cellcolor[RGB]{207,235,222}0.28 & \cellcolor[RGB]{207,235,222}0.28 & \cellcolor[RGB]{105,194,150}0.89  \\
Compromise & \cellcolor[RGB]{255,255,255}0.00 & \cellcolor[RGB]{239,248,244}0.09 & \cellcolor[RGB]{182,225,204}0.43 & \cellcolor[RGB]{233,246,239}0.13 & \cellcolor[RGB]{214,238,226}0.24 & \cellcolor[RGB]{189,228,209}0.39 & \cellcolor[RGB]{184,226,205}0.42 & \cellcolor[RGB]{192,229,211}0.37 & \cellcolor[RGB]{189,228,209}0.39 & \cellcolor[RGB]{187,227,208}0.40  & & \cellcolor[RGB]{253,254,253}0.01 & \cellcolor[RGB]{255,255,255}0.00 & \cellcolor[RGB]{255,255,255}0.00 & \cellcolor[RGB]{255,255,255}0.00 & \cellcolor[RGB]{253,254,253}0.01 & \cellcolor[RGB]{255,255,255}0.00 & \cellcolor[RGB]{255,255,255}0.00 & \cellcolor[RGB]{255,255,255}0.00 & \cellcolor[RGB]{255,255,255}0.00 & \cellcolor[RGB]{255,255,255}0.00  \\
Defend & \cellcolor[RGB]{255,255,255}0.00 & \cellcolor[RGB]{255,255,255}0.00 & \cellcolor[RGB]{248,252,250}0.04 & \cellcolor[RGB]{255,255,255}0.00 & \cellcolor[RGB]{251,253,252}0.02 & \cellcolor[RGB]{249,252,251}0.03 & \cellcolor[RGB]{251,253,252}0.02 & \cellcolor[RGB]{253,254,253}0.01 & \cellcolor[RGB]{249,252,251}0.03 & \cellcolor[RGB]{241,249,245}0.08  & & \cellcolor[RGB]{255,255,255}0.00 & \cellcolor[RGB]{255,255,255}0.00 & \cellcolor[RGB]{255,255,255}0.00 & \cellcolor[RGB]{253,254,253}0.01 & \cellcolor[RGB]{253,254,253}0.01 & \cellcolor[RGB]{251,253,252}0.02 & \cellcolor[RGB]{251,253,252}0.02 & \cellcolor[RGB]{253,254,253}0.01 & \cellcolor[RGB]{249,252,251}0.03 & \cellcolor[RGB]{246,251,249}0.05  \\
Accept & \cellcolor[RGB]{255,255,255}0.00 & \cellcolor[RGB]{255,255,255}0.00 & \cellcolor[RGB]{253,254,253}0.01 & \cellcolor[RGB]{255,255,255}0.00 & \cellcolor[RGB]{255,255,255}0.00 & \cellcolor[RGB]{251,253,252}0.02 & \cellcolor[RGB]{244,250,247}0.06 & \cellcolor[RGB]{241,249,245}0.08 & \cellcolor[RGB]{234,246,240}0.12 & \cellcolor[RGB]{209,236,223}0.27  & & \cellcolor[RGB]{255,255,255}0.00 & \cellcolor[RGB]{255,255,255}0.00 & \cellcolor[RGB]{255,255,255}0.00 & \cellcolor[RGB]{253,254,253}0.01 & \cellcolor[RGB]{255,255,255}0.00 & \cellcolor[RGB]{251,253,252}0.02 & \cellcolor[RGB]{253,254,253}0.01 & \cellcolor[RGB]{253,254,253}0.01 & \cellcolor[RGB]{248,252,250}0.04 & \cellcolor[RGB]{251,253,252}0.02  \\
Decline & \cellcolor[RGB]{255,255,255}0.00 & \cellcolor[RGB]{255,255,255}0.00 & \cellcolor[RGB]{253,254,253}0.01 & \cellcolor[RGB]{255,255,255}0.00 & \cellcolor[RGB]{255,255,255}0.00 & \cellcolor[RGB]{253,254,253}0.01 & \cellcolor[RGB]{253,254,253}0.01 & \cellcolor[RGB]{255,255,255}0.00 & \cellcolor[RGB]{251,253,252}0.02 & \cellcolor[RGB]{249,252,251}0.03  & & \cellcolor[RGB]{255,255,255}0.00 & \cellcolor[RGB]{255,255,255}0.00 & \cellcolor[RGB]{255,255,255}0.00 & \cellcolor[RGB]{255,255,255}0.00 & \cellcolor[RGB]{255,255,255}0.00 & \cellcolor[RGB]{255,255,255}0.00 & \cellcolor[RGB]{255,255,255}0.00 & \cellcolor[RGB]{253,254,253}0.01 & \cellcolor[RGB]{253,254,253}0.01 & \cellcolor[RGB]{255,255,255}0.00  \\
Others & \cellcolor[RGB]{255,255,255}0.00 & \cellcolor[RGB]{255,255,255}0.00 & \cellcolor[RGB]{251,253,252}0.02 & \cellcolor[RGB]{244,250,247}0.06 & \cellcolor[RGB]{248,252,250}0.04 & \cellcolor[RGB]{253,254,253}0.01 & \cellcolor[RGB]{244,250,247}0.06 & \cellcolor[RGB]{248,252,250}0.04 & \cellcolor[RGB]{248,252,250}0.04 & \cellcolor[RGB]{248,252,250}0.04  & & \cellcolor[RGB]{249,252,251}0.03 & \cellcolor[RGB]{248,252,250}0.04 & \cellcolor[RGB]{255,255,255}0.00 & \cellcolor[RGB]{249,252,251}0.03 & \cellcolor[RGB]{253,254,253}0.01 & \cellcolor[RGB]{251,253,252}0.02 & \cellcolor[RGB]{248,252,250}0.04 & \cellcolor[RGB]{253,254,253}0.01 & \cellcolor[RGB]{249,252,251}0.03 & \cellcolor[RGB]{253,254,253}0.01  \\
\\
\multicolumn{22}{c}{\textbf{MAJORITY}}\\
Inform & \cellcolor[RGB]{87,187,138}1.00 & \cellcolor[RGB]{87,187,138}1.00 & \cellcolor[RGB]{93,189,142}0.96 & \cellcolor[RGB]{134,206,170}0.72 & \cellcolor[RGB]{134,206,170}0.72 & \cellcolor[RGB]{144,210,177}0.66 & \cellcolor[RGB]{145,210,178}0.65 & \cellcolor[RGB]{125,202,164}0.77 & \cellcolor[RGB]{144,210,177}0.66 & \cellcolor[RGB]{147,211,180}0.64  & & \cellcolor[RGB]{87,187,138}1.00 & \cellcolor[RGB]{87,187,138}1.00 & \cellcolor[RGB]{87,187,138}1.00 & \cellcolor[RGB]{87,187,138}1.00 & \cellcolor[RGB]{87,187,138}1.00 & \cellcolor[RGB]{87,187,138}1.00 & \cellcolor[RGB]{87,187,138}1.00 & \cellcolor[RGB]{87,187,138}1.00 & \cellcolor[RGB]{88,187,139}0.99 & \cellcolor[RGB]{87,187,138}1.00  \\
Request & \cellcolor[RGB]{87,187,138}1.00 & \cellcolor[RGB]{87,187,138}1.00 & \cellcolor[RGB]{87,187,138}1.00 & \cellcolor[RGB]{93,189,142}0.96 & \cellcolor[RGB]{95,190,143}0.95 & \cellcolor[RGB]{93,189,142}0.96 & \cellcolor[RGB]{107,195,152}0.88 & \cellcolor[RGB]{93,189,142}0.96 & \cellcolor[RGB]{103,193,149}0.90 & \cellcolor[RGB]{112,197,155}0.85  & & \cellcolor[RGB]{87,187,138}1.00 & \cellcolor[RGB]{87,187,138}1.00 & \cellcolor[RGB]{87,187,138}1.00 & \cellcolor[RGB]{87,187,138}1.00 & \cellcolor[RGB]{87,187,138}1.00 & \cellcolor[RGB]{87,187,138}1.00 & \cellcolor[RGB]{87,187,138}1.00 & \cellcolor[RGB]{88,187,139}0.99 & \cellcolor[RGB]{87,187,138}1.00 & \cellcolor[RGB]{100,192,147}0.92  \\
Confirm & \cellcolor[RGB]{228,244,236}0.16 & \cellcolor[RGB]{213,238,225}0.25 & \cellcolor[RGB]{197,231,215}0.34 & \cellcolor[RGB]{202,233,218}0.31 & \cellcolor[RGB]{181,225,203}0.44 & \cellcolor[RGB]{194,230,212}0.36 & \cellcolor[RGB]{186,227,207}0.41 & \cellcolor[RGB]{186,227,207}0.41 & \cellcolor[RGB]{165,218,192}0.53 & \cellcolor[RGB]{167,219,194}0.52  & & \cellcolor[RGB]{213,238,225}0.25 & \cellcolor[RGB]{174,222,198}0.48 & \cellcolor[RGB]{160,216,189}0.56 & \cellcolor[RGB]{145,210,178}0.65 & \cellcolor[RGB]{135,206,171}0.71 & \cellcolor[RGB]{135,206,171}0.71 & \cellcolor[RGB]{130,204,168}0.74 & \cellcolor[RGB]{135,206,171}0.71 & \cellcolor[RGB]{118,199,160}0.81 & \cellcolor[RGB]{145,210,178}0.65  \\
Summarize & \cellcolor[RGB]{229,244,237}0.15 & \cellcolor[RGB]{251,253,252}0.02 & \cellcolor[RGB]{229,244,237}0.15 & \cellcolor[RGB]{233,246,239}0.13 & \cellcolor[RGB]{223,242,232}0.19 & \cellcolor[RGB]{219,240,230}0.21 & \cellcolor[RGB]{221,241,231}0.20 & \cellcolor[RGB]{224,242,233}0.18 & \cellcolor[RGB]{199,232,216}0.33 & \cellcolor[RGB]{204,234,219}0.30  & & \cellcolor[RGB]{253,254,253}0.01 & \cellcolor[RGB]{229,244,237}0.15 & \cellcolor[RGB]{243,250,246}0.07 & \cellcolor[RGB]{224,242,233}0.18 & \cellcolor[RGB]{236,247,242}0.11 & \cellcolor[RGB]{226,243,235}0.17 & \cellcolor[RGB]{223,242,232}0.19 & \cellcolor[RGB]{228,244,236}0.16 & \cellcolor[RGB]{224,242,233}0.18 & \cellcolor[RGB]{213,238,225}0.25  \\
Evaluate & \cellcolor[RGB]{98,191,146}0.93 & \cellcolor[RGB]{102,193,148}0.91 & \cellcolor[RGB]{142,209,176}0.67 & \cellcolor[RGB]{123,201,163}0.78 & \cellcolor[RGB]{110,196,154}0.86 & \cellcolor[RGB]{107,195,152}0.88 & \cellcolor[RGB]{107,195,152}0.88 & \cellcolor[RGB]{110,196,154}0.86 & \cellcolor[RGB]{105,194,150}0.89 & \cellcolor[RGB]{108,195,153}0.87  & & \cellcolor[RGB]{246,251,249}0.05 & \cellcolor[RGB]{152,213,183}0.61 & \cellcolor[RGB]{167,219,194}0.52 & \cellcolor[RGB]{172,221,197}0.49 & \cellcolor[RGB]{157,215,187}0.58 & \cellcolor[RGB]{167,219,194}0.52 & \cellcolor[RGB]{152,213,183}0.61 & \cellcolor[RGB]{164,218,191}0.54 & \cellcolor[RGB]{157,215,187}0.58 & \cellcolor[RGB]{160,216,189}0.56  \\
Propose & \cellcolor[RGB]{192,229,211}0.37 & \cellcolor[RGB]{186,227,207}0.41 & \cellcolor[RGB]{92,189,141}0.97 & \cellcolor[RGB]{88,187,139}0.99 & \cellcolor[RGB]{95,190,143}0.95 & \cellcolor[RGB]{92,189,141}0.97 & \cellcolor[RGB]{88,187,139}0.99 & \cellcolor[RGB]{93,189,142}0.96 & \cellcolor[RGB]{98,191,146}0.93 & \cellcolor[RGB]{103,193,149}0.90  & & \cellcolor[RGB]{246,251,249}0.05 & \cellcolor[RGB]{241,249,245}0.08 & \cellcolor[RGB]{206,235,221}0.29 & \cellcolor[RGB]{226,243,235}0.17 & \cellcolor[RGB]{218,240,229}0.22 & \cellcolor[RGB]{229,244,237}0.15 & \cellcolor[RGB]{213,238,225}0.25 & \cellcolor[RGB]{219,240,230}0.21 & \cellcolor[RGB]{211,237,224}0.26 & \cellcolor[RGB]{105,194,150}0.89  \\
Compromise & \cellcolor[RGB]{255,255,255}0.00 & \cellcolor[RGB]{249,252,251}0.03 & \cellcolor[RGB]{191,229,210}0.38 & \cellcolor[RGB]{231,245,238}0.14 & \cellcolor[RGB]{213,238,225}0.25 & \cellcolor[RGB]{186,227,207}0.41 & \cellcolor[RGB]{189,228,209}0.39 & \cellcolor[RGB]{202,233,218}0.31 & \cellcolor[RGB]{194,230,212}0.36 & \cellcolor[RGB]{204,234,219}0.30  & & \cellcolor[RGB]{255,255,255}0.00 & \cellcolor[RGB]{255,255,255}0.00 & \cellcolor[RGB]{251,253,252}0.02 & \cellcolor[RGB]{255,255,255}0.00 & \cellcolor[RGB]{255,255,255}0.00 & \cellcolor[RGB]{255,255,255}0.00 & \cellcolor[RGB]{255,255,255}0.00 & \cellcolor[RGB]{255,255,255}0.00 & \cellcolor[RGB]{255,255,255}0.00 & \cellcolor[RGB]{255,255,255}0.00  \\
Defend & \cellcolor[RGB]{255,255,255}0.00 & \cellcolor[RGB]{255,255,255}0.00 & \cellcolor[RGB]{248,252,250}0.04 & \cellcolor[RGB]{253,254,253}0.01 & \cellcolor[RGB]{248,252,250}0.04 & \cellcolor[RGB]{244,250,247}0.06 & \cellcolor[RGB]{243,250,246}0.07 & \cellcolor[RGB]{246,251,249}0.05 & \cellcolor[RGB]{239,248,244}0.09 & \cellcolor[RGB]{243,250,246}0.07  & & \cellcolor[RGB]{255,255,255}0.00 & \cellcolor[RGB]{253,254,253}0.01 & \cellcolor[RGB]{253,254,253}0.01 & \cellcolor[RGB]{255,255,255}0.00 & \cellcolor[RGB]{251,253,252}0.02 & \cellcolor[RGB]{253,254,253}0.01 & \cellcolor[RGB]{251,253,252}0.02 & \cellcolor[RGB]{255,255,255}0.00 & \cellcolor[RGB]{249,252,251}0.03 & \cellcolor[RGB]{248,252,250}0.04  \\
Accept & \cellcolor[RGB]{255,255,255}0.00 & \cellcolor[RGB]{255,255,255}0.00 & \cellcolor[RGB]{251,253,252}0.02 & \cellcolor[RGB]{249,252,251}0.03 & \cellcolor[RGB]{248,252,250}0.04 & \cellcolor[RGB]{236,247,242}0.11 & \cellcolor[RGB]{241,249,245}0.08 & \cellcolor[RGB]{234,246,240}0.12 & \cellcolor[RGB]{221,241,231}0.20 & \cellcolor[RGB]{221,241,231}0.20  & & \cellcolor[RGB]{255,255,255}0.00 & \cellcolor[RGB]{251,253,252}0.02 & \cellcolor[RGB]{251,253,252}0.02 & \cellcolor[RGB]{253,254,253}0.01 & \cellcolor[RGB]{253,254,253}0.01 & \cellcolor[RGB]{255,255,255}0.00 & \cellcolor[RGB]{251,253,252}0.02 & \cellcolor[RGB]{253,254,253}0.01 & \cellcolor[RGB]{251,253,252}0.02 & \cellcolor[RGB]{249,252,251}0.03  \\
Decline & \cellcolor[RGB]{255,255,255}0.00 & \cellcolor[RGB]{255,255,255}0.00 & \cellcolor[RGB]{251,253,252}0.02 & \cellcolor[RGB]{251,253,252}0.02 & \cellcolor[RGB]{249,252,251}0.03 & \cellcolor[RGB]{251,253,252}0.02 & \cellcolor[RGB]{248,252,250}0.04 & \cellcolor[RGB]{249,252,251}0.03 & \cellcolor[RGB]{248,252,250}0.04 & \cellcolor[RGB]{255,255,255}0.00  & & \cellcolor[RGB]{255,255,255}0.00 & \cellcolor[RGB]{253,254,253}0.01 & \cellcolor[RGB]{251,253,252}0.02 & \cellcolor[RGB]{255,255,255}0.00 & \cellcolor[RGB]{255,255,255}0.00 & \cellcolor[RGB]{255,255,255}0.00 & \cellcolor[RGB]{253,254,253}0.01 & \cellcolor[RGB]{255,255,255}0.00 & \cellcolor[RGB]{253,254,253}0.01 & \cellcolor[RGB]{253,254,253}0.01  \\
Others & \cellcolor[RGB]{255,255,255}0.00 & \cellcolor[RGB]{253,254,253}0.01 & \cellcolor[RGB]{253,254,253}0.01 & \cellcolor[RGB]{241,249,245}0.08 & \cellcolor[RGB]{241,249,245}0.08 & \cellcolor[RGB]{241,249,245}0.08 & \cellcolor[RGB]{241,249,245}0.08 & \cellcolor[RGB]{241,249,245}0.08 & \cellcolor[RGB]{243,250,246}0.07 & \cellcolor[RGB]{244,250,247}0.06  & & \cellcolor[RGB]{251,253,252}0.02 & \cellcolor[RGB]{255,255,255}0.00 & \cellcolor[RGB]{249,252,251}0.03 & \cellcolor[RGB]{249,252,251}0.03 & \cellcolor[RGB]{253,254,253}0.01 & \cellcolor[RGB]{251,253,252}0.02 & \cellcolor[RGB]{251,253,252}0.02 & \cellcolor[RGB]{253,254,253}0.01 & \cellcolor[RGB]{253,254,253}0.01 & \cellcolor[RGB]{253,254,253}0.01  \\
\\
\multicolumn{22}{c}{\textbf{PLURALITY}}\\
Inform & \cellcolor[RGB]{87,187,138}1.00 & \cellcolor[RGB]{90,188,140}0.98 & \cellcolor[RGB]{98,191,146}0.93 & \cellcolor[RGB]{130,204,168}0.74 & \cellcolor[RGB]{139,208,174}0.69 & \cellcolor[RGB]{140,208,175}0.68 & \cellcolor[RGB]{152,213,183}0.61 & \cellcolor[RGB]{147,211,180}0.64 & \cellcolor[RGB]{144,210,177}0.66 & \cellcolor[RGB]{147,211,180}0.64  & & \cellcolor[RGB]{87,187,138}1.00 & \cellcolor[RGB]{87,187,138}1.00 & \cellcolor[RGB]{90,188,140}0.98 & \cellcolor[RGB]{88,187,139}0.99 & \cellcolor[RGB]{88,187,139}0.99 & \cellcolor[RGB]{88,187,139}0.99 & \cellcolor[RGB]{87,187,138}1.00 & \cellcolor[RGB]{90,188,140}0.98 & \cellcolor[RGB]{87,187,138}1.00 & \cellcolor[RGB]{87,187,138}1.00  \\
Request & \cellcolor[RGB]{87,187,138}1.00 & \cellcolor[RGB]{87,187,138}1.00 & \cellcolor[RGB]{95,190,143}0.95 & \cellcolor[RGB]{92,189,141}0.97 & \cellcolor[RGB]{87,187,138}1.00 & \cellcolor[RGB]{98,191,146}0.93 & \cellcolor[RGB]{102,193,148}0.91 & \cellcolor[RGB]{103,193,149}0.90 & \cellcolor[RGB]{93,189,142}0.96 & \cellcolor[RGB]{112,197,155}0.85  & & \cellcolor[RGB]{87,187,138}1.00 & \cellcolor[RGB]{87,187,138}1.00 & \cellcolor[RGB]{87,187,138}1.00 & \cellcolor[RGB]{87,187,138}1.00 & \cellcolor[RGB]{87,187,138}1.00 & \cellcolor[RGB]{87,187,138}1.00 & \cellcolor[RGB]{88,187,139}0.99 & \cellcolor[RGB]{87,187,138}1.00 & \cellcolor[RGB]{87,187,138}1.00 & \cellcolor[RGB]{98,191,146}0.93  \\
Confirm & \cellcolor[RGB]{253,254,253}0.01 & \cellcolor[RGB]{206,235,221}0.29 & \cellcolor[RGB]{228,244,236}0.16 & \cellcolor[RGB]{211,237,224}0.26 & \cellcolor[RGB]{187,227,208}0.40 & \cellcolor[RGB]{184,226,205}0.42 & \cellcolor[RGB]{182,225,204}0.43 & \cellcolor[RGB]{177,223,201}0.46 & \cellcolor[RGB]{167,219,194}0.52 & \cellcolor[RGB]{162,217,190}0.55  & & \cellcolor[RGB]{206,235,221}0.29 & \cellcolor[RGB]{181,225,203}0.44 & \cellcolor[RGB]{152,213,183}0.61 & \cellcolor[RGB]{137,207,173}0.70 & \cellcolor[RGB]{132,205,169}0.73 & \cellcolor[RGB]{123,201,163}0.78 & \cellcolor[RGB]{127,203,166}0.76 & \cellcolor[RGB]{123,201,163}0.78 & \cellcolor[RGB]{125,202,164}0.77 & \cellcolor[RGB]{152,213,183}0.61  \\
Summarize & \cellcolor[RGB]{189,228,209}0.39 & \cellcolor[RGB]{253,254,253}0.01 & \cellcolor[RGB]{236,247,242}0.11 & \cellcolor[RGB]{226,243,235}0.17 & \cellcolor[RGB]{226,243,235}0.17 & \cellcolor[RGB]{218,240,229}0.22 & \cellcolor[RGB]{221,241,231}0.20 & \cellcolor[RGB]{218,240,229}0.22 & \cellcolor[RGB]{224,242,233}0.18 & \cellcolor[RGB]{213,238,225}0.25  & & \cellcolor[RGB]{246,251,249}0.05 & \cellcolor[RGB]{233,246,239}0.13 & \cellcolor[RGB]{236,247,242}0.11 & \cellcolor[RGB]{216,239,228}0.23 & \cellcolor[RGB]{224,242,233}0.18 & \cellcolor[RGB]{216,239,228}0.23 & \cellcolor[RGB]{223,242,232}0.19 & \cellcolor[RGB]{226,243,235}0.17 & \cellcolor[RGB]{219,240,230}0.21 & \cellcolor[RGB]{223,242,232}0.19  \\
Evaluate & \cellcolor[RGB]{112,197,155}0.85 & \cellcolor[RGB]{93,189,142}0.96 & \cellcolor[RGB]{150,212,182}0.62 & \cellcolor[RGB]{132,205,169}0.73 & \cellcolor[RGB]{110,196,154}0.86 & \cellcolor[RGB]{112,197,155}0.85 & \cellcolor[RGB]{102,193,148}0.91 & \cellcolor[RGB]{105,194,150}0.89 & \cellcolor[RGB]{115,198,157}0.83 & \cellcolor[RGB]{110,196,154}0.86  & & \cellcolor[RGB]{244,250,247}0.06 & \cellcolor[RGB]{145,210,178}0.65 & \cellcolor[RGB]{155,214,185}0.59 & \cellcolor[RGB]{152,213,183}0.61 & \cellcolor[RGB]{139,208,174}0.69 & \cellcolor[RGB]{157,215,187}0.58 & \cellcolor[RGB]{150,212,182}0.62 & \cellcolor[RGB]{142,209,176}0.67 & \cellcolor[RGB]{145,210,178}0.65 & \cellcolor[RGB]{152,213,183}0.61  \\
Propose & \cellcolor[RGB]{223,242,232}0.19 & \cellcolor[RGB]{172,221,197}0.49 & \cellcolor[RGB]{88,187,139}0.99 & \cellcolor[RGB]{87,187,138}1.00 & \cellcolor[RGB]{90,188,140}0.98 & \cellcolor[RGB]{92,189,141}0.97 & \cellcolor[RGB]{92,189,141}0.97 & \cellcolor[RGB]{95,190,143}0.95 & \cellcolor[RGB]{98,191,146}0.93 & \cellcolor[RGB]{108,195,153}0.87  & & \cellcolor[RGB]{239,248,244}0.09 & \cellcolor[RGB]{244,250,247}0.06 & \cellcolor[RGB]{202,233,218}0.31 & \cellcolor[RGB]{202,233,218}0.31 & \cellcolor[RGB]{204,234,219}0.30 & \cellcolor[RGB]{201,233,217}0.32 & \cellcolor[RGB]{201,233,217}0.32 & \cellcolor[RGB]{197,231,215}0.34 & \cellcolor[RGB]{194,230,212}0.36 & \cellcolor[RGB]{112,197,155}0.85  \\
Compromise & \cellcolor[RGB]{255,255,255}0.00 & \cellcolor[RGB]{251,253,252}0.02 & \cellcolor[RGB]{165,218,192}0.53 & \cellcolor[RGB]{231,245,238}0.14 & \cellcolor[RGB]{231,245,238}0.14 & \cellcolor[RGB]{191,229,210}0.38 & \cellcolor[RGB]{192,229,211}0.37 & \cellcolor[RGB]{199,232,216}0.33 & \cellcolor[RGB]{201,233,217}0.32 & \cellcolor[RGB]{201,233,217}0.32  & & \cellcolor[RGB]{255,255,255}0.00 & \cellcolor[RGB]{253,254,253}0.01 & \cellcolor[RGB]{255,255,255}0.00 & \cellcolor[RGB]{253,254,253}0.01 & \cellcolor[RGB]{251,253,252}0.02 & \cellcolor[RGB]{255,255,255}0.00 & \cellcolor[RGB]{255,255,255}0.00 & \cellcolor[RGB]{255,255,255}0.00 & \cellcolor[RGB]{255,255,255}0.00 & \cellcolor[RGB]{255,255,255}0.00  \\
Defend & \cellcolor[RGB]{255,255,255}0.00 & \cellcolor[RGB]{255,255,255}0.00 & \cellcolor[RGB]{255,255,255}0.00 & \cellcolor[RGB]{251,253,252}0.02 & \cellcolor[RGB]{255,255,255}0.00 & \cellcolor[RGB]{246,251,249}0.05 & \cellcolor[RGB]{244,250,247}0.06 & \cellcolor[RGB]{244,250,247}0.06 & \cellcolor[RGB]{243,250,246}0.07 & \cellcolor[RGB]{243,250,246}0.07  & & \cellcolor[RGB]{255,255,255}0.00 & \cellcolor[RGB]{255,255,255}0.00 & \cellcolor[RGB]{253,254,253}0.01 & \cellcolor[RGB]{249,252,251}0.03 & \cellcolor[RGB]{251,253,252}0.02 & \cellcolor[RGB]{248,252,250}0.04 & \cellcolor[RGB]{248,252,250}0.04 & \cellcolor[RGB]{251,253,252}0.02 & \cellcolor[RGB]{251,253,252}0.02 & \cellcolor[RGB]{249,252,251}0.03  \\
Accept & \cellcolor[RGB]{255,255,255}0.00 & \cellcolor[RGB]{255,255,255}0.00 & \cellcolor[RGB]{253,254,253}0.01 & \cellcolor[RGB]{251,253,252}0.02 & \cellcolor[RGB]{255,255,255}0.00 & \cellcolor[RGB]{241,249,245}0.08 & \cellcolor[RGB]{229,244,237}0.15 & \cellcolor[RGB]{238,248,243}0.10 & \cellcolor[RGB]{233,246,239}0.13 & \cellcolor[RGB]{213,238,225}0.25  & & \cellcolor[RGB]{255,255,255}0.00 & \cellcolor[RGB]{255,255,255}0.00 & \cellcolor[RGB]{253,254,253}0.01 & \cellcolor[RGB]{249,252,251}0.03 & \cellcolor[RGB]{253,254,253}0.01 & \cellcolor[RGB]{251,253,252}0.02 & \cellcolor[RGB]{248,252,250}0.04 & \cellcolor[RGB]{251,253,252}0.02 & \cellcolor[RGB]{253,254,253}0.01 & \cellcolor[RGB]{251,253,252}0.02  \\
Decline & \cellcolor[RGB]{255,255,255}0.00 & \cellcolor[RGB]{255,255,255}0.00 & \cellcolor[RGB]{253,254,253}0.01 & \cellcolor[RGB]{251,253,252}0.02 & \cellcolor[RGB]{255,255,255}0.00 & \cellcolor[RGB]{246,251,249}0.05 & \cellcolor[RGB]{246,251,249}0.05 & \cellcolor[RGB]{253,254,253}0.01 & \cellcolor[RGB]{251,253,252}0.02 & \cellcolor[RGB]{251,253,252}0.02  & & \cellcolor[RGB]{255,255,255}0.00 & \cellcolor[RGB]{255,255,255}0.00 & \cellcolor[RGB]{253,254,253}0.01 & \cellcolor[RGB]{253,254,253}0.01 & \cellcolor[RGB]{253,254,253}0.01 & \cellcolor[RGB]{253,254,253}0.01 & \cellcolor[RGB]{253,254,253}0.01 & \cellcolor[RGB]{253,254,253}0.01 & \cellcolor[RGB]{255,255,255}0.00 & \cellcolor[RGB]{255,255,255}0.00  \\
Others & \cellcolor[RGB]{255,255,255}0.00 & \cellcolor[RGB]{255,255,255}0.00 & \cellcolor[RGB]{251,253,252}0.02 & \cellcolor[RGB]{238,248,243}0.10 & \cellcolor[RGB]{244,250,247}0.06 & \cellcolor[RGB]{241,249,245}0.08 & \cellcolor[RGB]{246,251,249}0.05 & \cellcolor[RGB]{239,248,244}0.09 & \cellcolor[RGB]{243,250,246}0.07 & \cellcolor[RGB]{248,252,250}0.04  & & \cellcolor[RGB]{253,254,253}0.01 & \cellcolor[RGB]{253,254,253}0.01 & \cellcolor[RGB]{251,253,252}0.02 & \cellcolor[RGB]{253,254,253}0.01 & \cellcolor[RGB]{244,250,247}0.06 & \cellcolor[RGB]{251,253,252}0.02 & \cellcolor[RGB]{243,250,246}0.07 & \cellcolor[RGB]{255,255,255}0.00 & \cellcolor[RGB]{249,252,251}0.03 & \cellcolor[RGB]{253,254,253}0.01  \\
\\
\multicolumn{22}{c}{\textbf{RATED}}\\
Inform & \cellcolor[RGB]{87,187,138}1.00 & \cellcolor[RGB]{97,191,145}0.94 & \cellcolor[RGB]{92,189,141}0.97 & \cellcolor[RGB]{123,201,163}0.78 & \cellcolor[RGB]{139,208,174}0.69 & \cellcolor[RGB]{160,216,189}0.56 & \cellcolor[RGB]{155,214,185}0.59 & \cellcolor[RGB]{145,210,178}0.65 & \cellcolor[RGB]{145,210,178}0.65 & \cellcolor[RGB]{152,213,183}0.61  & & \cellcolor[RGB]{87,187,138}1.00 & \cellcolor[RGB]{87,187,138}1.00 & \cellcolor[RGB]{90,188,140}0.98 & \cellcolor[RGB]{88,187,139}0.99 & \cellcolor[RGB]{88,187,139}0.99 & \cellcolor[RGB]{87,187,138}1.00 & \cellcolor[RGB]{87,187,138}1.00 & \cellcolor[RGB]{87,187,138}1.00 & \cellcolor[RGB]{87,187,138}1.00 & \cellcolor[RGB]{88,187,139}0.99  \\
Request & \cellcolor[RGB]{87,187,138}1.00 & \cellcolor[RGB]{87,187,138}1.00 & \cellcolor[RGB]{87,187,138}1.00 & \cellcolor[RGB]{92,189,141}0.97 & \cellcolor[RGB]{93,189,142}0.96 & \cellcolor[RGB]{100,192,147}0.92 & \cellcolor[RGB]{98,191,146}0.93 & \cellcolor[RGB]{102,193,148}0.91 & \cellcolor[RGB]{103,193,149}0.90 & \cellcolor[RGB]{118,199,160}0.81  & & \cellcolor[RGB]{87,187,138}1.00 & \cellcolor[RGB]{87,187,138}1.00 & \cellcolor[RGB]{87,187,138}1.00 & \cellcolor[RGB]{87,187,138}1.00 & \cellcolor[RGB]{87,187,138}1.00 & \cellcolor[RGB]{87,187,138}1.00 & \cellcolor[RGB]{87,187,138}1.00 & \cellcolor[RGB]{88,187,139}0.99 & \cellcolor[RGB]{87,187,138}1.00 & \cellcolor[RGB]{97,191,145}0.94  \\
Confirm & \cellcolor[RGB]{255,255,255}0.00 & \cellcolor[RGB]{176,223,200}0.47 & \cellcolor[RGB]{199,232,216}0.33 & \cellcolor[RGB]{202,233,218}0.31 & \cellcolor[RGB]{196,231,214}0.35 & \cellcolor[RGB]{186,227,207}0.41 & \cellcolor[RGB]{199,232,216}0.33 & \cellcolor[RGB]{187,227,208}0.40 & \cellcolor[RGB]{182,225,204}0.43 & \cellcolor[RGB]{174,222,198}0.48  & & \cellcolor[RGB]{209,236,223}0.27 & \cellcolor[RGB]{174,222,198}0.48 & \cellcolor[RGB]{142,209,176}0.67 & \cellcolor[RGB]{137,207,173}0.70 & \cellcolor[RGB]{127,203,166}0.76 & \cellcolor[RGB]{145,210,178}0.65 & \cellcolor[RGB]{142,209,176}0.67 & \cellcolor[RGB]{135,206,171}0.71 & \cellcolor[RGB]{139,208,174}0.69 & \cellcolor[RGB]{157,215,187}0.58  \\
Summarize & \cellcolor[RGB]{228,244,236}0.16 & \cellcolor[RGB]{251,253,252}0.02 & \cellcolor[RGB]{226,243,235}0.17 & \cellcolor[RGB]{233,246,239}0.13 & \cellcolor[RGB]{226,243,235}0.17 & \cellcolor[RGB]{223,242,232}0.19 & \cellcolor[RGB]{216,239,228}0.23 & \cellcolor[RGB]{216,239,228}0.23 & \cellcolor[RGB]{221,241,231}0.20 & \cellcolor[RGB]{213,238,225}0.25  & & \cellcolor[RGB]{239,248,244}0.09 & \cellcolor[RGB]{233,246,239}0.13 & \cellcolor[RGB]{231,245,238}0.14 & \cellcolor[RGB]{228,244,236}0.16 & \cellcolor[RGB]{226,243,235}0.17 & \cellcolor[RGB]{218,240,229}0.22 & \cellcolor[RGB]{209,236,223}0.27 & \cellcolor[RGB]{219,240,230}0.21 & \cellcolor[RGB]{224,242,233}0.18 & \cellcolor[RGB]{213,238,225}0.25  \\
Evaluate & \cellcolor[RGB]{187,227,208}0.40 & \cellcolor[RGB]{87,187,138}1.00 & \cellcolor[RGB]{159,216,188}0.57 & \cellcolor[RGB]{139,208,174}0.69 & \cellcolor[RGB]{108,195,153}0.87 & \cellcolor[RGB]{112,197,155}0.85 & \cellcolor[RGB]{112,197,155}0.85 & \cellcolor[RGB]{113,197,156}0.84 & \cellcolor[RGB]{100,192,147}0.92 & \cellcolor[RGB]{112,197,155}0.85  & & \cellcolor[RGB]{239,248,244}0.09 & \cellcolor[RGB]{154,214,184}0.60 & \cellcolor[RGB]{150,212,182}0.62 & \cellcolor[RGB]{144,210,177}0.66 & \cellcolor[RGB]{157,215,187}0.58 & \cellcolor[RGB]{144,210,177}0.66 & \cellcolor[RGB]{157,215,187}0.58 & \cellcolor[RGB]{169,220,195}0.51 & \cellcolor[RGB]{162,217,190}0.55 & \cellcolor[RGB]{135,206,171}0.71  \\
Propose & \cellcolor[RGB]{236,247,242}0.11 & \cellcolor[RGB]{144,210,177}0.66 & \cellcolor[RGB]{92,189,141}0.97 & \cellcolor[RGB]{88,187,139}0.99 & \cellcolor[RGB]{90,188,140}0.98 & \cellcolor[RGB]{88,187,139}0.99 & \cellcolor[RGB]{87,187,138}1.00 & \cellcolor[RGB]{92,189,141}0.97 & \cellcolor[RGB]{95,190,143}0.95 & \cellcolor[RGB]{97,191,145}0.94  & & \cellcolor[RGB]{228,244,236}0.16 & \cellcolor[RGB]{246,251,249}0.05 & \cellcolor[RGB]{211,237,224}0.26 & \cellcolor[RGB]{221,241,231}0.20 & \cellcolor[RGB]{219,240,230}0.21 & \cellcolor[RGB]{216,239,228}0.23 & \cellcolor[RGB]{209,236,223}0.27 & \cellcolor[RGB]{202,233,218}0.31 & \cellcolor[RGB]{194,230,212}0.36 & \cellcolor[RGB]{107,195,152}0.88  \\
Compromise & \cellcolor[RGB]{255,255,255}0.00 & \cellcolor[RGB]{249,252,251}0.03 & \cellcolor[RGB]{174,222,198}0.48 & \cellcolor[RGB]{234,246,240}0.12 & \cellcolor[RGB]{214,238,226}0.24 & \cellcolor[RGB]{187,227,208}0.40 & \cellcolor[RGB]{201,233,217}0.32 & \cellcolor[RGB]{197,231,215}0.34 & \cellcolor[RGB]{191,229,210}0.38 & \cellcolor[RGB]{201,233,217}0.32  & & \cellcolor[RGB]{255,255,255}0.00 & \cellcolor[RGB]{255,255,255}0.00 & \cellcolor[RGB]{253,254,253}0.01 & \cellcolor[RGB]{253,254,253}0.01 & \cellcolor[RGB]{253,254,253}0.01 & \cellcolor[RGB]{255,255,255}0.00 & \cellcolor[RGB]{255,255,255}0.00 & \cellcolor[RGB]{255,255,255}0.00 & \cellcolor[RGB]{253,254,253}0.01 & \cellcolor[RGB]{255,255,255}0.00  \\
Defend & \cellcolor[RGB]{255,255,255}0.00 & \cellcolor[RGB]{255,255,255}0.00 & \cellcolor[RGB]{249,252,251}0.03 & \cellcolor[RGB]{253,254,253}0.01 & \cellcolor[RGB]{251,253,252}0.02 & \cellcolor[RGB]{249,252,251}0.03 & \cellcolor[RGB]{243,250,246}0.07 & \cellcolor[RGB]{249,252,251}0.03 & \cellcolor[RGB]{243,250,246}0.07 & \cellcolor[RGB]{248,252,250}0.04  & & \cellcolor[RGB]{255,255,255}0.00 & \cellcolor[RGB]{255,255,255}0.00 & \cellcolor[RGB]{255,255,255}0.00 & \cellcolor[RGB]{253,254,253}0.01 & \cellcolor[RGB]{253,254,253}0.01 & \cellcolor[RGB]{255,255,255}0.00 & \cellcolor[RGB]{255,255,255}0.00 & \cellcolor[RGB]{253,254,253}0.01 & \cellcolor[RGB]{255,255,255}0.00 & \cellcolor[RGB]{251,253,252}0.02  \\
Accept & \cellcolor[RGB]{255,255,255}0.00 & \cellcolor[RGB]{255,255,255}0.00 & \cellcolor[RGB]{249,252,251}0.03 & \cellcolor[RGB]{251,253,252}0.02 & \cellcolor[RGB]{246,251,249}0.05 & \cellcolor[RGB]{251,253,252}0.02 & \cellcolor[RGB]{243,250,246}0.07 & \cellcolor[RGB]{238,248,243}0.10 & \cellcolor[RGB]{236,247,242}0.11 & \cellcolor[RGB]{218,240,229}0.22  & & \cellcolor[RGB]{255,255,255}0.00 & \cellcolor[RGB]{255,255,255}0.00 & \cellcolor[RGB]{255,255,255}0.00 & \cellcolor[RGB]{255,255,255}0.00 & \cellcolor[RGB]{255,255,255}0.00 & \cellcolor[RGB]{255,255,255}0.00 & \cellcolor[RGB]{253,254,253}0.01 & \cellcolor[RGB]{251,253,252}0.02 & \cellcolor[RGB]{255,255,255}0.00 & \cellcolor[RGB]{253,254,253}0.01  \\
Decline & \cellcolor[RGB]{255,255,255}0.00 & \cellcolor[RGB]{255,255,255}0.00 & \cellcolor[RGB]{251,253,252}0.02 & \cellcolor[RGB]{253,254,253}0.01 & \cellcolor[RGB]{253,254,253}0.01 & \cellcolor[RGB]{255,255,255}0.00 & \cellcolor[RGB]{253,254,253}0.01 & \cellcolor[RGB]{249,252,251}0.03 & \cellcolor[RGB]{249,252,251}0.03 & \cellcolor[RGB]{255,255,255}0.00  & & \cellcolor[RGB]{255,255,255}0.00 & \cellcolor[RGB]{255,255,255}0.00 & \cellcolor[RGB]{255,255,255}0.00 & \cellcolor[RGB]{255,255,255}0.00 & \cellcolor[RGB]{255,255,255}0.00 & \cellcolor[RGB]{255,255,255}0.00 & \cellcolor[RGB]{253,254,253}0.01 & \cellcolor[RGB]{255,255,255}0.00 & \cellcolor[RGB]{255,255,255}0.00 & \cellcolor[RGB]{255,255,255}0.00  \\
Others & \cellcolor[RGB]{251,253,252}0.02 & \cellcolor[RGB]{255,255,255}0.00 & \cellcolor[RGB]{251,253,252}0.02 & \cellcolor[RGB]{238,248,243}0.10 & \cellcolor[RGB]{241,249,245}0.08 & \cellcolor[RGB]{241,249,245}0.08 & \cellcolor[RGB]{244,250,247}0.06 & \cellcolor[RGB]{243,250,246}0.07 & \cellcolor[RGB]{246,251,249}0.05 & \cellcolor[RGB]{241,249,245}0.08  & & \cellcolor[RGB]{249,252,251}0.03 & \cellcolor[RGB]{251,253,252}0.02 & \cellcolor[RGB]{244,250,247}0.06 & \cellcolor[RGB]{248,252,250}0.04 & \cellcolor[RGB]{249,252,251}0.03 & \cellcolor[RGB]{251,253,252}0.02 & \cellcolor[RGB]{251,253,252}0.02 & \cellcolor[RGB]{246,251,249}0.05 & \cellcolor[RGB]{246,251,249}0.05 & \cellcolor[RGB]{244,250,247}0.06  \\
\\
\multicolumn{22}{c}{\textbf{RANKED}}\\
Inform & \cellcolor[RGB]{87,187,138}1.00 & \cellcolor[RGB]{93,189,142}0.96 & \cellcolor[RGB]{93,189,142}0.96 & \cellcolor[RGB]{123,201,163}0.78 & \cellcolor[RGB]{122,201,162}0.79 & \cellcolor[RGB]{144,210,177}0.66 & \cellcolor[RGB]{139,208,174}0.69 & \cellcolor[RGB]{135,206,171}0.71 & \cellcolor[RGB]{134,206,170}0.72 & \cellcolor[RGB]{129,203,167}0.75  & & \cellcolor[RGB]{87,187,138}1.00 & \cellcolor[RGB]{87,187,138}1.00 & \cellcolor[RGB]{87,187,138}1.00 & \cellcolor[RGB]{87,187,138}1.00 & \cellcolor[RGB]{88,187,139}0.99 & \cellcolor[RGB]{88,187,139}0.99 & \cellcolor[RGB]{87,187,138}1.00 & \cellcolor[RGB]{87,187,138}1.00 & \cellcolor[RGB]{87,187,138}1.00 & \cellcolor[RGB]{88,187,139}0.99  \\
Request & \cellcolor[RGB]{87,187,138}1.00 & \cellcolor[RGB]{87,187,138}1.00 & \cellcolor[RGB]{88,187,139}0.99 & \cellcolor[RGB]{90,188,140}0.98 & \cellcolor[RGB]{93,189,142}0.96 & \cellcolor[RGB]{95,190,143}0.95 & \cellcolor[RGB]{95,190,143}0.95 & \cellcolor[RGB]{93,189,142}0.96 & \cellcolor[RGB]{102,193,148}0.91 & \cellcolor[RGB]{105,194,150}0.89  & & \cellcolor[RGB]{87,187,138}1.00 & \cellcolor[RGB]{87,187,138}1.00 & \cellcolor[RGB]{87,187,138}1.00 & \cellcolor[RGB]{87,187,138}1.00 & \cellcolor[RGB]{87,187,138}1.00 & \cellcolor[RGB]{87,187,138}1.00 & \cellcolor[RGB]{87,187,138}1.00 & \cellcolor[RGB]{87,187,138}1.00 & \cellcolor[RGB]{87,187,138}1.00 & \cellcolor[RGB]{90,188,140}0.98  \\
Confirm & \cellcolor[RGB]{249,252,251}0.03 & \cellcolor[RGB]{169,220,195}0.51 & \cellcolor[RGB]{172,221,197}0.49 & \cellcolor[RGB]{199,232,216}0.33 & \cellcolor[RGB]{187,227,208}0.40 & \cellcolor[RGB]{196,231,214}0.35 & \cellcolor[RGB]{179,224,202}0.45 & \cellcolor[RGB]{167,219,194}0.52 & \cellcolor[RGB]{174,222,198}0.48 & \cellcolor[RGB]{164,218,191}0.54  & & \cellcolor[RGB]{219,240,230}0.21 & \cellcolor[RGB]{182,225,204}0.43 & \cellcolor[RGB]{165,218,192}0.53 & \cellcolor[RGB]{150,212,182}0.62 & \cellcolor[RGB]{132,205,169}0.73 & \cellcolor[RGB]{129,203,167}0.75 & \cellcolor[RGB]{123,201,163}0.78 & \cellcolor[RGB]{140,208,175}0.68 & \cellcolor[RGB]{137,207,173}0.70 & \cellcolor[RGB]{137,207,173}0.70  \\
Summarize & \cellcolor[RGB]{251,253,252}0.02 & \cellcolor[RGB]{249,252,251}0.03 & \cellcolor[RGB]{239,248,244}0.09 & \cellcolor[RGB]{228,244,236}0.16 & \cellcolor[RGB]{216,239,228}0.23 & \cellcolor[RGB]{221,241,231}0.20 & \cellcolor[RGB]{221,241,231}0.20 & \cellcolor[RGB]{228,244,236}0.16 & \cellcolor[RGB]{209,236,223}0.27 & \cellcolor[RGB]{201,233,217}0.32  & & \cellcolor[RGB]{239,248,244}0.09 & \cellcolor[RGB]{231,245,238}0.14 & \cellcolor[RGB]{231,245,238}0.14 & \cellcolor[RGB]{219,240,230}0.21 & \cellcolor[RGB]{214,238,226}0.24 & \cellcolor[RGB]{216,239,228}0.23 & \cellcolor[RGB]{219,240,230}0.21 & \cellcolor[RGB]{211,237,224}0.26 & \cellcolor[RGB]{226,243,235}0.17 & \cellcolor[RGB]{214,238,226}0.24  \\
Evaluate & \cellcolor[RGB]{100,192,147}0.92 & \cellcolor[RGB]{95,190,143}0.95 & \cellcolor[RGB]{152,213,183}0.61 & \cellcolor[RGB]{145,210,178}0.65 & \cellcolor[RGB]{122,201,162}0.79 & \cellcolor[RGB]{110,196,154}0.86 & \cellcolor[RGB]{108,195,153}0.87 & \cellcolor[RGB]{110,196,154}0.86 & \cellcolor[RGB]{115,198,157}0.83 & \cellcolor[RGB]{105,194,150}0.89  & & \cellcolor[RGB]{244,250,247}0.06 & \cellcolor[RGB]{167,219,194}0.52 & \cellcolor[RGB]{147,211,180}0.64 & \cellcolor[RGB]{144,210,177}0.66 & \cellcolor[RGB]{149,212,181}0.63 & \cellcolor[RGB]{157,215,187}0.58 & \cellcolor[RGB]{155,214,185}0.59 & \cellcolor[RGB]{154,214,184}0.60 & \cellcolor[RGB]{139,208,174}0.69 & \cellcolor[RGB]{125,202,164}0.77  \\
Propose & \cellcolor[RGB]{229,244,237}0.15 & \cellcolor[RGB]{174,222,198}0.48 & \cellcolor[RGB]{100,192,147}0.92 & \cellcolor[RGB]{93,189,142}0.96 & \cellcolor[RGB]{88,187,139}0.99 & \cellcolor[RGB]{90,188,140}0.98 & \cellcolor[RGB]{92,189,141}0.97 & \cellcolor[RGB]{98,191,146}0.93 & \cellcolor[RGB]{103,193,149}0.90 & \cellcolor[RGB]{108,195,153}0.87  & & \cellcolor[RGB]{239,248,244}0.09 & \cellcolor[RGB]{248,252,250}0.04 & \cellcolor[RGB]{209,236,223}0.27 & \cellcolor[RGB]{228,244,236}0.16 & \cellcolor[RGB]{206,235,221}0.29 & \cellcolor[RGB]{223,242,232}0.19 & \cellcolor[RGB]{214,238,226}0.24 & \cellcolor[RGB]{207,235,222}0.28 & \cellcolor[RGB]{201,233,217}0.32 & \cellcolor[RGB]{112,197,155}0.85  \\
Compromise & \cellcolor[RGB]{255,255,255}0.00 & \cellcolor[RGB]{253,254,253}0.01 & \cellcolor[RGB]{196,231,214}0.35 & \cellcolor[RGB]{231,245,238}0.14 & \cellcolor[RGB]{219,240,230}0.21 & \cellcolor[RGB]{201,233,217}0.32 & \cellcolor[RGB]{196,231,214}0.35 & \cellcolor[RGB]{194,230,212}0.36 & \cellcolor[RGB]{192,229,211}0.37 & \cellcolor[RGB]{219,240,230}0.21  & & \cellcolor[RGB]{255,255,255}0.00 & \cellcolor[RGB]{253,254,253}0.01 & \cellcolor[RGB]{253,254,253}0.01 & \cellcolor[RGB]{251,253,252}0.02 & \cellcolor[RGB]{255,255,255}0.00 & \cellcolor[RGB]{255,255,255}0.00 & \cellcolor[RGB]{253,254,253}0.01 & \cellcolor[RGB]{249,252,251}0.03 & \cellcolor[RGB]{255,255,255}0.00 & \cellcolor[RGB]{249,252,251}0.03  \\
Defend & \cellcolor[RGB]{255,255,255}0.00 & \cellcolor[RGB]{255,255,255}0.00 & \cellcolor[RGB]{246,251,249}0.05 & \cellcolor[RGB]{251,253,252}0.02 & \cellcolor[RGB]{249,252,251}0.03 & \cellcolor[RGB]{244,250,247}0.06 & \cellcolor[RGB]{249,252,251}0.03 & \cellcolor[RGB]{251,253,252}0.02 & \cellcolor[RGB]{246,251,249}0.05 & \cellcolor[RGB]{239,248,244}0.09  & & \cellcolor[RGB]{255,255,255}0.00 & \cellcolor[RGB]{255,255,255}0.00 & \cellcolor[RGB]{255,255,255}0.00 & \cellcolor[RGB]{255,255,255}0.00 & \cellcolor[RGB]{255,255,255}0.00 & \cellcolor[RGB]{253,254,253}0.01 & \cellcolor[RGB]{253,254,253}0.01 & \cellcolor[RGB]{253,254,253}0.01 & \cellcolor[RGB]{251,253,252}0.02 & \cellcolor[RGB]{251,253,252}0.02  \\
Accept & \cellcolor[RGB]{255,255,255}0.00 & \cellcolor[RGB]{255,255,255}0.00 & \cellcolor[RGB]{246,251,249}0.05 & \cellcolor[RGB]{249,252,251}0.03 & \cellcolor[RGB]{248,252,250}0.04 & \cellcolor[RGB]{246,251,249}0.05 & \cellcolor[RGB]{234,246,240}0.12 & \cellcolor[RGB]{233,246,239}0.13 & \cellcolor[RGB]{236,247,242}0.11 & \cellcolor[RGB]{209,236,223}0.27  & & \cellcolor[RGB]{255,255,255}0.00 & \cellcolor[RGB]{255,255,255}0.00 & \cellcolor[RGB]{251,253,252}0.02 & \cellcolor[RGB]{253,254,253}0.01 & \cellcolor[RGB]{251,253,252}0.02 & \cellcolor[RGB]{253,254,253}0.01 & \cellcolor[RGB]{253,254,253}0.01 & \cellcolor[RGB]{251,253,252}0.02 & \cellcolor[RGB]{255,255,255}0.00 & \cellcolor[RGB]{249,252,251}0.03  \\
Decline & \cellcolor[RGB]{255,255,255}0.00 & \cellcolor[RGB]{255,255,255}0.00 & \cellcolor[RGB]{249,252,251}0.03 & \cellcolor[RGB]{253,254,253}0.01 & \cellcolor[RGB]{253,254,253}0.01 & \cellcolor[RGB]{249,252,251}0.03 & \cellcolor[RGB]{253,254,253}0.01 & \cellcolor[RGB]{255,255,255}0.00 & \cellcolor[RGB]{255,255,255}0.00 & \cellcolor[RGB]{251,253,252}0.02  & & \cellcolor[RGB]{255,255,255}0.00 & \cellcolor[RGB]{255,255,255}0.00 & \cellcolor[RGB]{255,255,255}0.00 & \cellcolor[RGB]{255,255,255}0.00 & \cellcolor[RGB]{255,255,255}0.00 & \cellcolor[RGB]{255,255,255}0.00 & \cellcolor[RGB]{253,254,253}0.01 & \cellcolor[RGB]{255,255,255}0.00 & \cellcolor[RGB]{255,255,255}0.00 & \cellcolor[RGB]{253,254,253}0.01  \\
Others & \cellcolor[RGB]{253,254,253}0.01 & \cellcolor[RGB]{251,253,252}0.02 & \cellcolor[RGB]{251,253,252}0.02 & \cellcolor[RGB]{246,251,249}0.05 & \cellcolor[RGB]{244,250,247}0.06 & \cellcolor[RGB]{233,246,239}0.13 & \cellcolor[RGB]{236,247,242}0.11 & \cellcolor[RGB]{248,252,250}0.04 & \cellcolor[RGB]{249,252,251}0.03 & \cellcolor[RGB]{253,254,253}0.01  & & \cellcolor[RGB]{253,254,253}0.01 & \cellcolor[RGB]{251,253,252}0.02 & \cellcolor[RGB]{241,249,245}0.08 & \cellcolor[RGB]{249,252,251}0.03 & \cellcolor[RGB]{253,254,253}0.01 & \cellcolor[RGB]{249,252,251}0.03 & \cellcolor[RGB]{253,254,253}0.01 & \cellcolor[RGB]{249,252,251}0.03 & \cellcolor[RGB]{251,253,252}0.02 & \cellcolor[RGB]{244,250,247}0.06  \\
\\
\multicolumn{22}{c}{\textbf{CUMULATIVE}}\\
Inform & \cellcolor[RGB]{87,187,138}1.00 & \cellcolor[RGB]{97,191,145}0.94 & \cellcolor[RGB]{88,187,139}0.99 & \cellcolor[RGB]{117,199,159}0.82 & \cellcolor[RGB]{134,206,170}0.72 & \cellcolor[RGB]{140,208,175}0.68 & \cellcolor[RGB]{154,214,184}0.60 & \cellcolor[RGB]{142,209,176}0.67 & \cellcolor[RGB]{134,206,170}0.72 & \cellcolor[RGB]{130,204,168}0.74  & & \cellcolor[RGB]{87,187,138}1.00 & \cellcolor[RGB]{87,187,138}1.00 & \cellcolor[RGB]{87,187,138}1.00 & \cellcolor[RGB]{90,188,140}0.98 & \cellcolor[RGB]{92,189,141}0.97 & \cellcolor[RGB]{90,188,140}0.98 & \cellcolor[RGB]{87,187,138}1.00 & \cellcolor[RGB]{87,187,138}1.00 & \cellcolor[RGB]{87,187,138}1.00 & \cellcolor[RGB]{87,187,138}1.00  \\
Request & \cellcolor[RGB]{87,187,138}1.00 & \cellcolor[RGB]{87,187,138}1.00 & \cellcolor[RGB]{90,188,140}0.98 & \cellcolor[RGB]{92,189,141}0.97 & \cellcolor[RGB]{93,189,142}0.96 & \cellcolor[RGB]{95,190,143}0.95 & \cellcolor[RGB]{97,191,145}0.94 & \cellcolor[RGB]{100,192,147}0.92 & \cellcolor[RGB]{95,190,143}0.95 & \cellcolor[RGB]{110,196,154}0.86  & & \cellcolor[RGB]{87,187,138}1.00 & \cellcolor[RGB]{87,187,138}1.00 & \cellcolor[RGB]{87,187,138}1.00 & \cellcolor[RGB]{87,187,138}1.00 & \cellcolor[RGB]{87,187,138}1.00 & \cellcolor[RGB]{88,187,139}0.99 & \cellcolor[RGB]{87,187,138}1.00 & \cellcolor[RGB]{87,187,138}1.00 & \cellcolor[RGB]{87,187,138}1.00 & \cellcolor[RGB]{93,189,142}0.96  \\
Confirm & \cellcolor[RGB]{249,252,251}0.03 & \cellcolor[RGB]{201,233,217}0.32 & \cellcolor[RGB]{204,234,219}0.30 & \cellcolor[RGB]{206,235,221}0.29 & \cellcolor[RGB]{206,235,221}0.29 & \cellcolor[RGB]{199,232,216}0.33 & \cellcolor[RGB]{186,227,207}0.41 & \cellcolor[RGB]{177,223,201}0.46 & \cellcolor[RGB]{179,224,202}0.45 & \cellcolor[RGB]{162,217,190}0.55  & & \cellcolor[RGB]{213,238,225}0.25 & \cellcolor[RGB]{181,225,203}0.44 & \cellcolor[RGB]{144,210,177}0.66 & \cellcolor[RGB]{134,206,170}0.72 & \cellcolor[RGB]{123,201,163}0.78 & \cellcolor[RGB]{132,205,169}0.73 & \cellcolor[RGB]{132,205,169}0.73 & \cellcolor[RGB]{127,203,166}0.76 & \cellcolor[RGB]{135,206,171}0.71 & \cellcolor[RGB]{150,212,182}0.62  \\
Summarize & \cellcolor[RGB]{231,245,238}0.14 & \cellcolor[RGB]{253,254,253}0.01 & \cellcolor[RGB]{231,245,238}0.14 & \cellcolor[RGB]{228,244,236}0.16 & \cellcolor[RGB]{239,248,244}0.09 & \cellcolor[RGB]{226,243,235}0.17 & \cellcolor[RGB]{213,238,225}0.25 & \cellcolor[RGB]{216,239,228}0.23 & \cellcolor[RGB]{219,240,230}0.21 & \cellcolor[RGB]{201,233,217}0.32  & & \cellcolor[RGB]{238,248,243}0.10 & \cellcolor[RGB]{234,246,240}0.12 & \cellcolor[RGB]{234,246,240}0.12 & \cellcolor[RGB]{218,240,229}0.22 & \cellcolor[RGB]{218,240,229}0.22 & \cellcolor[RGB]{218,240,229}0.22 & \cellcolor[RGB]{226,243,235}0.17 & \cellcolor[RGB]{206,235,221}0.29 & \cellcolor[RGB]{221,241,231}0.20 & \cellcolor[RGB]{211,237,224}0.26  \\
Evaluate & \cellcolor[RGB]{118,199,160}0.81 & \cellcolor[RGB]{90,188,140}0.98 & \cellcolor[RGB]{157,215,187}0.58 & \cellcolor[RGB]{130,204,168}0.74 & \cellcolor[RGB]{110,196,154}0.86 & \cellcolor[RGB]{103,193,149}0.90 & \cellcolor[RGB]{100,192,147}0.92 & \cellcolor[RGB]{103,193,149}0.90 & \cellcolor[RGB]{103,193,149}0.90 & \cellcolor[RGB]{110,196,154}0.86  & & \cellcolor[RGB]{248,252,250}0.04 & \cellcolor[RGB]{145,210,178}0.65 & \cellcolor[RGB]{149,212,181}0.63 & \cellcolor[RGB]{169,220,195}0.51 & \cellcolor[RGB]{152,213,183}0.61 & \cellcolor[RGB]{150,212,182}0.62 & \cellcolor[RGB]{155,214,185}0.59 & \cellcolor[RGB]{160,216,189}0.56 & \cellcolor[RGB]{154,214,184}0.60 & \cellcolor[RGB]{134,206,170}0.72  \\
Propose & \cellcolor[RGB]{226,243,235}0.17 & \cellcolor[RGB]{157,215,187}0.58 & \cellcolor[RGB]{90,188,140}0.98 & \cellcolor[RGB]{92,189,141}0.97 & \cellcolor[RGB]{88,187,139}0.99 & \cellcolor[RGB]{90,188,140}0.98 & \cellcolor[RGB]{95,190,143}0.95 & \cellcolor[RGB]{105,194,150}0.89 & \cellcolor[RGB]{105,194,150}0.89 & \cellcolor[RGB]{108,195,153}0.87  & & \cellcolor[RGB]{241,249,245}0.08 & \cellcolor[RGB]{249,252,251}0.03 & \cellcolor[RGB]{214,238,226}0.24 & \cellcolor[RGB]{211,237,224}0.26 & \cellcolor[RGB]{218,240,229}0.22 & \cellcolor[RGB]{218,240,229}0.22 & \cellcolor[RGB]{209,236,223}0.27 & \cellcolor[RGB]{207,235,222}0.28 & \cellcolor[RGB]{201,233,217}0.32 & \cellcolor[RGB]{113,197,156}0.84  \\
Compromise & \cellcolor[RGB]{255,255,255}0.00 & \cellcolor[RGB]{246,251,249}0.05 & \cellcolor[RGB]{164,218,191}0.54 & \cellcolor[RGB]{231,245,238}0.14 & \cellcolor[RGB]{223,242,232}0.19 & \cellcolor[RGB]{184,226,205}0.42 & \cellcolor[RGB]{204,234,219}0.30 & \cellcolor[RGB]{196,231,214}0.35 & \cellcolor[RGB]{207,235,222}0.28 & \cellcolor[RGB]{209,236,223}0.27  & & \cellcolor[RGB]{253,254,253}0.01 & \cellcolor[RGB]{255,255,255}0.00 & \cellcolor[RGB]{255,255,255}0.00 & \cellcolor[RGB]{251,253,252}0.02 & \cellcolor[RGB]{255,255,255}0.00 & \cellcolor[RGB]{255,255,255}0.00 & \cellcolor[RGB]{251,253,252}0.02 & \cellcolor[RGB]{253,254,253}0.01 & \cellcolor[RGB]{255,255,255}0.00 & \cellcolor[RGB]{253,254,253}0.01  \\
Defend & \cellcolor[RGB]{255,255,255}0.00 & \cellcolor[RGB]{255,255,255}0.00 & \cellcolor[RGB]{253,254,253}0.01 & \cellcolor[RGB]{255,255,255}0.00 & \cellcolor[RGB]{249,252,251}0.03 & \cellcolor[RGB]{246,251,249}0.05 & \cellcolor[RGB]{246,251,249}0.05 & \cellcolor[RGB]{251,253,252}0.02 & \cellcolor[RGB]{246,251,249}0.05 & \cellcolor[RGB]{244,250,247}0.06  & & \cellcolor[RGB]{253,254,253}0.01 & \cellcolor[RGB]{253,254,253}0.01 & \cellcolor[RGB]{253,254,253}0.01 & \cellcolor[RGB]{253,254,253}0.01 & \cellcolor[RGB]{253,254,253}0.01 & \cellcolor[RGB]{255,255,255}0.00 & \cellcolor[RGB]{251,253,252}0.02 & \cellcolor[RGB]{249,252,251}0.03 & \cellcolor[RGB]{251,253,252}0.02 & \cellcolor[RGB]{249,252,251}0.03  \\
Accept & \cellcolor[RGB]{255,255,255}0.00 & \cellcolor[RGB]{255,255,255}0.00 & \cellcolor[RGB]{253,254,253}0.01 & \cellcolor[RGB]{255,255,255}0.00 & \cellcolor[RGB]{253,254,253}0.01 & \cellcolor[RGB]{239,248,244}0.09 & \cellcolor[RGB]{238,248,243}0.10 & \cellcolor[RGB]{238,248,243}0.10 & \cellcolor[RGB]{219,240,230}0.21 & \cellcolor[RGB]{219,240,230}0.21  & & \cellcolor[RGB]{253,254,253}0.01 & \cellcolor[RGB]{253,254,253}0.01 & \cellcolor[RGB]{255,255,255}0.00 & \cellcolor[RGB]{255,255,255}0.00 & \cellcolor[RGB]{253,254,253}0.01 & \cellcolor[RGB]{249,252,251}0.03 & \cellcolor[RGB]{253,254,253}0.01 & \cellcolor[RGB]{249,252,251}0.03 & \cellcolor[RGB]{251,253,252}0.02 & \cellcolor[RGB]{248,252,250}0.04  \\
Decline & \cellcolor[RGB]{255,255,255}0.00 & \cellcolor[RGB]{255,255,255}0.00 & \cellcolor[RGB]{255,255,255}0.00 & \cellcolor[RGB]{255,255,255}0.00 & \cellcolor[RGB]{255,255,255}0.00 & \cellcolor[RGB]{248,252,250}0.04 & \cellcolor[RGB]{253,254,253}0.01 & \cellcolor[RGB]{255,255,255}0.00 & \cellcolor[RGB]{255,255,255}0.00 & \cellcolor[RGB]{253,254,253}0.01  & & \cellcolor[RGB]{253,254,253}0.01 & \cellcolor[RGB]{253,254,253}0.01 & \cellcolor[RGB]{255,255,255}0.00 & \cellcolor[RGB]{255,255,255}0.00 & \cellcolor[RGB]{255,255,255}0.00 & \cellcolor[RGB]{255,255,255}0.00 & \cellcolor[RGB]{255,255,255}0.00 & \cellcolor[RGB]{251,253,252}0.02 & \cellcolor[RGB]{253,254,253}0.01 & \cellcolor[RGB]{255,255,255}0.00  \\
Others & \cellcolor[RGB]{255,255,255}0.00 & \cellcolor[RGB]{255,255,255}0.00 & \cellcolor[RGB]{255,255,255}0.00 & \cellcolor[RGB]{244,250,247}0.06 & \cellcolor[RGB]{244,250,247}0.06 & \cellcolor[RGB]{243,250,246}0.07 & \cellcolor[RGB]{246,251,249}0.05 & \cellcolor[RGB]{244,250,247}0.06 & \cellcolor[RGB]{244,250,247}0.06 & \cellcolor[RGB]{249,252,251}0.03  & & \cellcolor[RGB]{253,254,253}0.01 & \cellcolor[RGB]{249,252,251}0.03 & \cellcolor[RGB]{248,252,250}0.04 & \cellcolor[RGB]{249,252,251}0.03 & \cellcolor[RGB]{253,254,253}0.01 & \cellcolor[RGB]{243,250,246}0.07 & \cellcolor[RGB]{253,254,253}0.01 & \cellcolor[RGB]{249,252,251}0.03 & \cellcolor[RGB]{248,252,250}0.04 & \cellcolor[RGB]{241,249,245}0.08  \\

\end{tabular}
}

\caption{Heatmaps of ratio of dialogue acts in each round. Color gradient is calculated with green as maximum, and white as minimum value in each table.}
\label{tbl:dialogue_act}
\end{table*}

    \subsection{Early Stopping}
        \label{app:early_stopping}
        \begin{table*}[t]
\centering
\resizebox{\textwidth}{!}{%
\begin{tabular}{lllllll}
\multicolumn{1}{c}{\textbf{EARLY STOPPING METHOD}}& \multicolumn{1}{c}{\textbf{UNANIMOUS}} & \multicolumn{1}{c}{\textbf{MAJORITY}} & \multicolumn{1}{c}{\textbf{PLURALIRY}} & \multicolumn{1}{c}{\textbf{RATED}} & \multicolumn{1}{c}{\textbf{RANKED}} & \multicolumn{1}{c}{\textbf{CUMULATIVE}} \\
\hline
\multicolumn{7}{c}{\textbf{EXCHANGE ECONOMY, in Group Total Utility($\uparrow$)}}\\
@10 (Baseline)&\textbf{0.48}&0.80&0.77&0.80&0.78&0.78\\
First Agreement&0.48&0.78&0.76&0.74&0.77&0.74\\
Consecutive Agreements&\textbf{0.48}&0.80&0.77&0.80&0.78&0.79\\
Validation Checkpoint&0.37&0.81&0.80&0.81&0.81&0.81\\
Information Distance&0.39&\textbf{0.81}&\textbf{0.80}&\textbf{0.81}&\textbf{0.81}&\textbf{0.82}\\
Dialogue Act&0.42&0.81&0.80&0.81&0.81&0.81\\
\cellcolor[RGB]{192,192,192}Oracle&\cellcolor[RGB]{192,192,192}0.48&\cellcolor[RGB]{192,192,192}0.84&\cellcolor[RGB]{192,192,192}0.82&\cellcolor[RGB]{192,192,192}0.83&\cellcolor[RGB]{192,192,192}0.84&\cellcolor[RGB]{192,192,192}0.84\\
\\
\multicolumn{7}{c}{\textbf{RECOMMENDATION SYSTEM, in MAE($\downarrow$)}}\\
@10 (Baseline)&0.88&0.86&0.84&0.79&0.84&\textbf{0.76}\\
First Agreement&0.82&0.80&0.81&0.82&0.89&0.82\\
Consecutive Agreements&0.84&0.81&0.84&0.77&0.82&0.77\\
Validation Checkpoint&\textbf{0.81}&0.81&0.82&0.80&0.83&0.82\\
Information Distance&0.86&\textbf{0.78}&0.85&0.79&\textbf{0.78}&0.81\\
Dialogue Act&0.84&0.82&\textbf{0.79}&\textbf{0.76}&0.85&0.82\\
\cellcolor[RGB]{192,192,192}Oracle&\cellcolor[RGB]{192,192,192}0.73&\cellcolor[RGB]{192,192,192}0.63&\cellcolor[RGB]{192,192,192}0.59&\cellcolor[RGB]{192,192,192}0.62&\cellcolor[RGB]{192,192,192}0.69&\cellcolor[RGB]{192,192,192}0.67\\
\end{tabular}
}

\caption{Comparison between early stopping methods in different voting rules. The performance is shown with group total utility and MAE. ($\uparrow$) and ($\downarrow$) indicates better performance with higher and lower values, respectively. Experiments are based on the results of 3 agents, gpt-4o-mini setting. Results are based on 5-fold cross validation. We observe that language-based methods performed well overall.}
\label{tbl:early_stopping_all}
\end{table*}
        \begin{table*}[!t]
\centering
\resizebox{\textwidth}{!}{%
\begin{tabular}{lllllllll}
 \textbf{STATISTICS}&\textbf{EARLY STOPPING METHOD}& \textbf{UNANIMOUS} & \textbf{MAJORITY} & \textbf{PLURALITY} & \textbf{RATED} & \textbf{RANKED} & \textbf{CUMULATIVE} & \textbf{AVERAGE} \\
 \hline
 \multicolumn{9}{c}{\textbf{EXCHANGE ECONOMY}}\\
\multirow{7}{*}{Early Stopped Round} & Oracle & 3.50 & 3.61 & 3.84 & 3.81 & 3.49 & 4.11 & 3.73 \\
& First Agreement & 3.00 & 1.57 & 1.54 & 1.15 & 1.07 & 1.33 & 1.61 \\
& Consecutive Agreement & - & 8.00 & 6.50 & 8.00 & 10.00 & 7.00 & 7.90 \\
& Validation Checkpoint & 2.60 & 2.80 & 3.00 & 3.00 & 2.60 & 3.00 & 2.83 \\
& Information Distance & 3.21 & 3.32 & 3.28 & 3.09 & 3.31 & 3.66 & 3.31 \\\
& Dialogue Act & 6.23 & 7.09 & 8.33 & 7.64 & 5.47 & 7.40 & 7.03 \\
 \hline
\multirow{4}{*}{Effective Ratio} & First Agreement & 0.62 & 1.00 & 1.00 & 1.00 & 1.00 & 1.00 & 0.94 \\
& Consecutive Agreement & 0.00 & 0.03 & 0.03 & 0.02 & 0.00 & 0.01 & 0.02 \\
& Information Distance & 1.00 & 1.00 & 1.00 & 1.00 & 1.00 & 1.00 & 1.00 \\
& Dialogue Act & 0.62 & 0.34 & 0.07 & 0.19 & 0.63 & 0.31 & 0.36 \\
 \hline
Threshold& Information Distance & 0.17 & 0.17 & 0.17 & 0.20 & 0.18 & 0.17 & 0.18 \\
\\
\multicolumn{9}{c}{\textbf{RECOMMENDATION SYSTEM}}\\
\multirow{7}{*}{Early Stopped Round}& Oracle & 2.40 & 2.10 & 2.38 & 2.15 & 2.25 & 2.44 & 2.29 \\
& First Agreement & 1.53 & 1.01 & 1.00 & 1.04 & 1.00 & 1.36 & 1.16 \\
& Consecutive Agreement & 3.34 & 3.31 & 3.15 & 3.61 & 3.46 & 3.87 & 3.46 \\
& Validation Checkpoint & 1.20 & 1.00 & 1.20 & 1.00 & 1.00 & 1.00 & 1.07 \\
& Information Distance & 3.12 & 3.67 & 4.30 & 3.50 & 3.97 & 3.34 & 3.65 \\
& Dialogue Act & 7.58 & 6.98 & 6.02 & 5.22 & 5.37 & 6.53 & 6.28 \\
\hline
\multirow{4}{*}{Effective Ratio}& First Agreement & 1.00 & 1.00 & 1.00 & 1.00 & 1.00 & 1.00 & 1.00 \\
& Consecutive Agreement & 0.82 & 0.92 & 0.91 & 0.77 & 0.62 & 0.78 & 0.80 \\
& Information Distance & 1.00 & 1.00 & 1.00 & 1.00 & 0.99 & 1.00 & 1.00 \\
& Dialogue Act & 0.06 & 0.24 & 0.52 & 0.58 & 0.20 & 0.30 & 0.32 \\
\hline
Threshold& Information Distance & 0.15 & 0.13 & 0.12 & 0.14 & 0.12 & 0.14 & 0.13 \\
\end{tabular}
}
\caption{Basic statistics of different early stopping methods.}
\label{tbl:early_stopping_statistics}
\end{table*}
        We share full result of the early stopping experiment in Table \ref{tbl:early_stopping_all}, and basic statistics of early stopping methods, including early-stopped round, effective ratio, and information distance threshold in Table \ref{tbl:early_stopping_statistics}. The effective ratio stands for the chance that a certain rule can be applied in the test set. The threshold is the embedding distance threshold used for early stopping. 

        Although language feature based methods, \textit{Information Distance} and \textit{Dialogue Act}, outperforms other early stopping methods and baseline in exchange economy environment, the difference is small. This is due to the small room for improvement, identified between baseline (\textit{@10}) and \textit{Oracle} performance.


\end{document}